\documentclass[fleqn,10pt]{wlscirep}
\usepackage[utf8]{inputenc}
\usepackage[T1]{fontenc}
\usepackage{amsmath, bm}
\usepackage{braket}
\usepackage{color}
\usepackage{soul, mdframed} 
\usepackage{adjustbox}
\usepackage{array, makecell}
\DeclareMathAlphabet{\mathcal}{OMS}{cmsy}{m}{n}

\title{Seeking a quantum advantage with trapped-ion quantum simulations of condensed-phase chemical dynamics}

\author[1,2,*]{Mingyu Kang}
\author[3]{Hanggai Nuomin}
\author[3]{Sutirtha N. Chowdhury}
\author[4]{Jonathon L. Yuly}
\author[1,2]{Ke Sun}
\author[1,5]{Jacob Whitlow}
\author[6, 7, 8]{Jesús Valdiviezo}
\author[3]{Zhendian Zhang}
\author[3]{Peng Zhang}
\author[2,3,9,$\dag$]{David N. Beratan}
\author[1,2,3,5,$\ddag$]{Kenneth R. Brown}

\affil[1]{Duke Quantum Center, Duke University, Durham, NC, USA}
\affil[2]{Department of Physics, Duke University, Durham, NC, USA}
\affil[3]{Department of Chemistry, Duke University, Durham, NC, USA}
\affil[4]{Lewis-Sigler Institute for Integrative Genomics, Princeton University, Princeton, NJ, USA}
\affil[5]{Department of Electrical and Computer Engineering, Duke University, Durham, NC, USA}
\affil[6]{Kenneth S. Pitzer Theory Center, University of California, Berkeley, CA, USA}
\affil[7]{Department of Chemistry, University of California, Berkeley, CA, USA}
\affil[8]{Departamento de Ciencias, Sección Química, Pontificia Universidad Católica del Perú, Lima, Perú}
\affil[9]{Department of Biochemistry, Duke University, Durham, NC, USA}

\affil[*]{e-mail: mingyu.kang@duke.edu}
\affil[$\dag$]{e-mail: david.beratan@duke.edu}
\affil[$\ddag$]{e-mail: ken.brown@duke.edu}

\begin{abstract}
Simulating the quantum dynamics of molecules in the condensed phase represents a longstanding challenge in chemistry. Trapped-ion quantum systems may serve as a platform for the analog-quantum simulation of chemical dynamics that is beyond the reach of current classical-digital simulation. To identify a `quantum advantage’ for these simulations, performance analysis of both analog-quantum simulation on noisy hardware and classical-digital algorithms is needed. In this Review, we make a comparison between a noisy analog trapped-ion simulator and a few choice classical-digital methods on simulating the dynamics of a model molecular Hamiltonian with linear vibronic coupling. We describe several simple Hamiltonians that are commonly used to model molecular systems, which can be simulated with existing or emerging trapped-ion hardware. These Hamiltonians may serve as stepping stones toward the use of trapped-ion simulators for systems beyond the reach of classical-digital methods. Finally, we identify dynamical regimes where classical-digital simulations seem to have the weakest performance compared to analog-quantum simulations. These regimes may provide the lowest hanging fruit to exploit potential quantum advantages.
\end{abstract}

\begin{document}

\flushbottom
\maketitle

\thispagestyle{empty}

%%%%%%%%%%%%%%%%%%%%%%%%%%%%%%%%%%%%%%%%
%%%%%%%%%%%%%%%%%%%%%%%%%%%%%%%%%%%%%%%%
%% Section
%%%%%%%%%%%%%%%%%%%%%%%%%%%%%%%%%%%%%%%%
%%%%%%%%%%%%%%%%%%%%%%%%%%%%%%%%%%%%%%%%
\section*{Introduction} %: the goal and challenge of quantum advantage

Quantum chemistry explores how chemical and physical properties emerge from the quantum mechanical equations of motion. Two ubiquitous challenges face quantum chemistry: (i) characterizing the molecular electronic structure and the corresponding properties of molecules at equilibrium, and (ii) describing the time-evolving dynamics of molecules, often in the condensed phase. For both tasks, the highest accuracy is achieved using the full quantum theory, but the computational cost grows exponentially with system size for the most reliable descriptions. As such, high-accuracy simulations of even modest-sized molecules (tens of atoms) are challenging, even on the world's finest supercomputers \cite{park2020multireference,park2017fly,larsson2022chromium}. The most common approach to addressing quantum problems on classical computers is to use approximations. While many practical approximations are available, there is generally a trade-off between accuracy and computational cost.

It was suggested that quantum computers/simulators might provide an advantage over classical-digital simulation for problems faced in quantum chemistry, since quantum properties may be best explored using computational resources that, themselves, are intrinsically quantum mechanical~\cite{Cirac12, Georgescu14, Daley22}. A quantum advantage could drive scientific progress by providing understanding of quantum phenomena that is not achievable by solely classical computation, thus opening up new opportunities in science and engineering. Yet, progress on this path remains an outstanding challenge, not only because of the daunting scientific and engineering challenges of building scalable quantum computers/simulators, but also because of the challenge of discovering how best to exploit the capabilities of (both classical and quantum) simulation tools~\cite{Alexeev21}. A clear ``quantum advantage'' for a specific quantum computation/simulation over classical-digital simulations for solving a practical problem in quantum chemistry remains to be proven~\cite{lee2023evaluating}. 

This Review motivates the search for a quantum advantage in the area of molecular quantum dynamics~\cite{Kassal08, Sawaya20, Jahangiri20, MacDonell21, Saha21}, which has received less attention than time-independent (stationary state) quantum chemistry~\cite{Kitaev96, Aspuru05, Lanyon10, Whitfield11, Peruzzo14, OMalley16, Kandala17, Nam20}. Box 1 overviews the classical and quantum approaches to simulating quantum dynamics as well as offers a definition for quantum advantage. Even if the full spectrum of electronic and vibrational energies is known, classical-digital simulation of quantum dynamics with high accuracy remains computationally difficult, as the number of evolving phase factors between quantum states scales exponentially with the system size. Describing the quantum dynamics of molecules, especially for structures in the condensed phase, has the potential to produce valuable insight into a wide variety of chemical, biological, and material phenomena of fundamental and practical value. Advances in understanding the physical underpinnings of photosynthesis~\cite{Wang19, Cao20} and protein function~\cite{hammes2008proton, hammes2015proton} could result from advances in quantum computing and simulations.

Finding quantum advantages requires making detailed comparisons between the strengths and weaknesses of classical-digital and quantum simulation methods. Classical-digital methods, of course, face limitations defined by state-of-art algorithms as well as hardware constraints. Finding opportunities where the performance of quantum simulations may surpass those of classical-digital methods is anticipated to be challenging, as over 80 years of progress in quantum calculations using classical-digital computers must be confronted by an emerging quantum-simulation technology.

At least two quantum-simulation strategies exist: fault-tolerant digital computation and analog simulation. While fault-tolerant digital quantum computation may efficiently solve many problems that are believed to be intractable with classical computers, including quantum molecular dynamics, those solutions require a tremendously large number of operations on a large number of quantum bits (qubits)~\cite{Reiher17, Babbush18, Su21, Kim22}.  These demands are far beyond the capabilities of state-of-art quantum-computing hardware. In the current era of small- or moderate-scale noisy quantum-computing hardware, analog quantum simulation, which exploits the fact that the hardware's native operations are particularly well suited to simulate specific kinds of Hamiltonians, may be considered. The weakness of analog quantum simulation is its inaccuracy, especially when noise is intrinsic to the computing hardware. 

Analog quantum simulations have been performed on several platforms, including superconducting structures~\cite{Houck12, Hartmann16}, neutral-atom assemblies~\cite{Greiner02, Gross17}, photonic architectures~\cite{Aspuru12}, and trapped-ion systems~\cite{Blatt12, Monroe21}. We focus our attention on trapped-ion systems, which have been used widely for quantum information processing, because of the excellent coherence of their atomic states and the availability of high-fidelity state preparation, manipulation, and measurement~\cite{Brown16, Bruzewicz19}. As well, significant progress has been made to control the motional degrees of freedom of the ion chain~\cite{Kienzler16, Um16, Zhang18, Fluhmann19, Neeve22, Jia22}. The atomic and motional states of trapped ions can be mapped onto the electronic and nuclear degrees of freedom of molecular Hamiltonians~\cite{MacDonell21, Gorman18}. This makes trapped-ion systems natural candidates for analog quantum simulators of quantum molecular dynamics. Box 2 overviews typical trapped-ion system. 

We suggest a path toward identifying models for quantum molecular-dynamics systems that are challenging to address using classical-digital methods, but could be solved using analog simulations on trapped-ion hardware. In this Review, we first provide an overview of classical-digital simulation methods and analog trapped-ion simulation approaches for computing the quantum dynamics of molecular Hamiltonians. Then, we make a comparison between the performance of a few classical-digital methods and noisy trapped-ion simulators in solving a model problem in quantum dynamics. We discuss parameter regimes where the trapped-ion simulation may provide an advantage. Then, we introduce candidate models of chemical dynamics to explore in the near term using trapped-ion simulations. Finally, we discuss the outlook for using trapped-ion simulators to address less well-studied challenges in quantum chemistry, and we conclude with a summary of future prospects.

%%%%%%%%%%%%%%%%%%%%%%%%%%%%%%%BOX 1%%%%%%%%%%%%%%%%%%%%%%%%%%%%%%%%%%%%%
\begin{mdframed}
\section*{Box 1 | Definitions}
The table below summarizes the four different approaches to solving a computation/simulation problem. In the description of analog-quantum methods, qubit (qudit) is the unit component of a quantum system, each consisting of two ($d$) states. This Review focuses on comparing the analog-quantum and digital-classical methods for simulating quantum molecular dynamics. We expect that analog-quantum simulators will be actively used for solving various problems in chemistry, physics, and material science~\cite{Daley22}, at least until fault-tolerant digital-quantum computers emerge, similarly to how analog-classical computers had been used for, for example, solving differential equations~\cite{karplus1959analog, jackson1960analog}, before fault-tolerant digital-classical computers were developed. 
\begin{center}
\includegraphics[width=\linewidth]{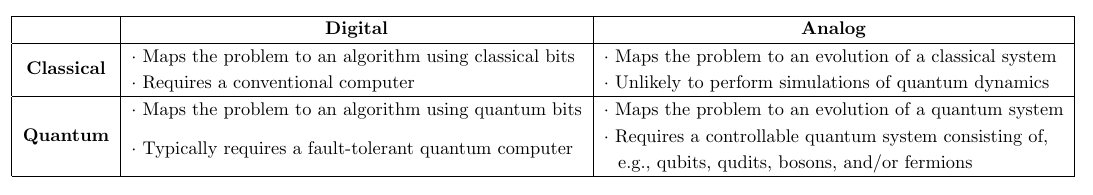}
\end{center} 

For the problem of simulating quantum dynamics, we are interested in finding expectation values of an observable $\hat{O}$ up to time $\tau$ and accuracy $\epsilon$, or estimating quantities related to molecular behavior that can be derived from these expectation values. We assume that the parameters that define the initial state and the time evolution are given. For example, given the initial state $\rho_0$ and the Hamiltonian $\hat{H}$, the problem is to find $O(t) := \text{Tr} \{\hat{O} e^{-i\hat{H}t} \rho_0 e^{i\hat{H}t}\}$ for $t = t_1, t_2, ..., \tau$ up to accuracy $\epsilon$, or to extract quantities of interest (rate constants, branching ratios, etc.) from these $O(t)$ values. A more general form of evolution, such as the Lindblad master equation~\cite{lindblad1976generators}, may be considered for dissipative dynamics. 

A model for quantum dynamics will always be an approximation of the dynamics of an actual molecular system. Nonetheless, it is valuable to solve approximate models accurately in order to improve the understanding of the quantum molecular phenomena. For example, solving approximate models leads to identifying the dependence of important chemical quantities, such as the electronic-state population at equilibrium, to the parameters and structure of the Hamiltonian. Furthermore, improved understanding may lead to better design of the experiments on an actual molecular system. 

A plethora of classical-digital methods exist to simulate quantum dynamics; the methods can be classified into two categories. First, a classical-digital method is \textit{numerically exact} if it describes the time evolution of a given Hamiltonian up to a controllable error, which can be made arbitrarily small if infinite computing resources are used. Next, a classical-digital method is \textit{approximate} if it relies on a physical or mathematical simplification of the Hamiltonian dynamics, so the method does not converge to the exact solution even given infinite computational resources. For quantum molecular dynamics, typical approximations include the use of classical-dynamical heuristics and the formal expansions of the quantum-mechanical path integral around the classical trajectories of the system. These approximations significantly lessen the requirements for computational resources, such as run time and memory, compared to those of numerically exact methods. See ``The current champion: ...'' subsection for examples of classical-digital methods in both categories. 

The term ``quantum advantage'' is  used broadly to refer to the capability of quantum computers/simulators to solve a problem that cannot be solved using conventional computers. For the challenge of simulating quantum dynamics, we offer a specific working definition. A quantum dynamics problem of finding $O(t)$ up to time $\tau$ and accuracy $\epsilon$ is \textit{classically intractable} if (i) all known numerically exact classical-digital methods require prohibitively long run time or memory given reasonably large computing resources (e.g., computing cluster of a research facility) and (ii) all known approximate classical-digital methods fail to predict $O(t)$ within accuracy $\epsilon$ at some $t \in [0, \tau]$. If a quantum (analog or digital) simulator can solve such a classically intractable problem within run time that is short enough to be practical, then a \textit{quantum advantage} is achieved. This definition can be straightforwardly extended to dissipative dynamics as well. 

For the problem of extracting quantities related to molecular behavior, many valuable predictions may be made even with errors in the estimated values of $O(t)$, as long as general trends are captured accurately. For example, electron transfer rates are estimated by the decay of donor-state population over time, which can be robust to significant errors in the state populations when the errors are unbiased and transient. Also, approximate classical-digital methods (such as semiclassical approximations) may provide acceptable estimates even if some details of the dynamics are not captured. 
\end{mdframed}

%%%%%%%%%%%%%%%%%%%%%%%%%%%%%%%BOX 2%%%%%%%%%%%%%%%%%%%%%%%%%%%%%%%%%%%%%
\begin{mdframed}
\section*{Box 2 | Trapped ions}

A typical trapped-ion system consists of a linear chain of ions confined in free space. Radio frequency (rf) electromagnetic fields are supplied by the electrodes of the trap, creating a  pseudo-potential that constrains the atomic positions. After the ions are sufficiently cooled using lasers, the ions' interactions with the confinement potential and the Coulomb repulsion among the ions reach equilibrium, producing a linear chain of ions.

A typical ion has many accessible electronic states, and two stable states can be exploited to establish an atom-based qubit with excellent coherence properties. Initialization of the qubit to a desired state, and measurement of the qubit state, may be performed through optical excitation with lasers~\cite{Wineland98}. 

Nuclear (bosonic) degrees of freedom exist in the ion chain through its collective vibrating motion. The ion chain's motion can be described in the language of normal modes, where each mode is approximated by a quantum harmonic oscillator. An $n$-ion chain has $n$ motional modes in each of the ion chain's axial direction and two other perpendicular (radial) directions, total $3n$ modes. A subset of these modes can interact with the qubits to mediate the entanglement between qubits or to provide the bosonic ingredient for analog simulations. Since the normal modes are spread through the ion chain, each mode can interact with \textit{all} qubits in the chain (except for qubits located at nodes of the normal mode), allowing flexibility in the structure of the Hamiltonians that can be simulated. 

Panel (c) of the figure below visualizes the $n=4$ radial modes of a four-ion chain. The modes are shown in the descending order of mode frequency from top to bottom, and the relative height (direction) of each arrow represents the magnitude (sign) of the eigenvector component of each normal mode. We denote the mode with equal eigenvector components of all ions [top in (c)] as the center-of-mass (CM) mode and all other modes as non-CM modes. 

To manipulate the qubit states and motional states of the ion chain, electromagnetic fields are applied using lasers or magnetic-field gradients~\cite{Wineland98}. Panels (a) and (b) of the figure below show the example of using lasers to couple the leftmost ion's qubit states to one of the motional modes [third from top in (c)]. The two-photon Raman transition between the qubit states via the excited atomic state $|e\rangle$ is induced using the global (orange) and individually addressed (red and blue) laser beams. The frequency, phase, and amplitude of the laser fields impinging on each ion can be controlled individually, allowing flexibility in defining the Hamiltonian parameters that will be simulated on the ion chain.

\begin{center}
\includegraphics[width=0.95\linewidth]{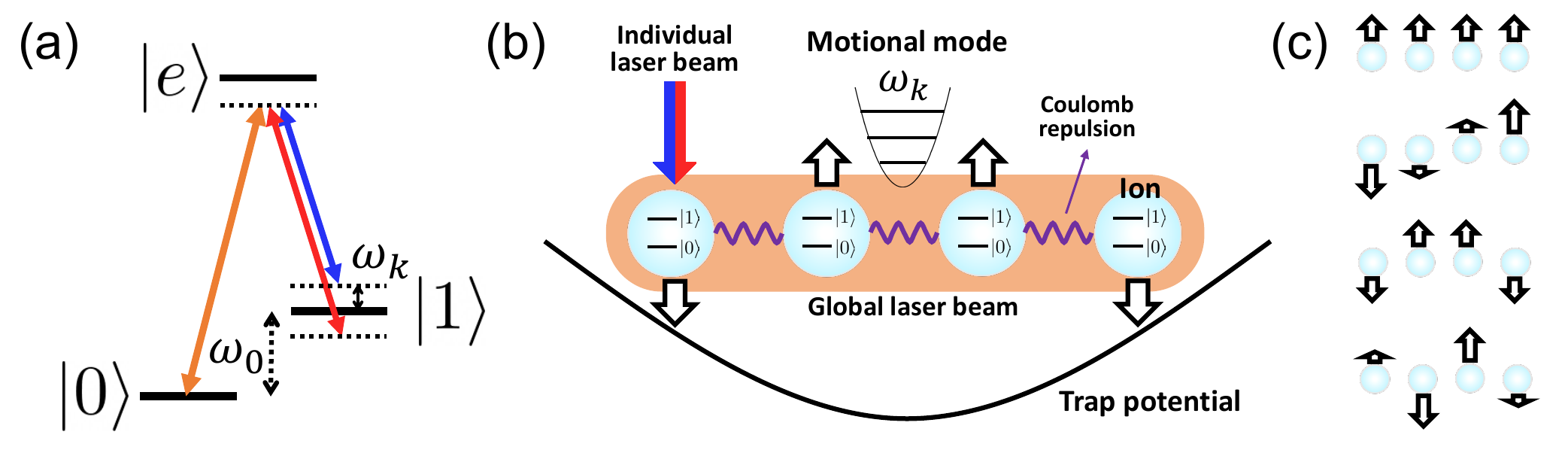}
\end{center}

The architecture of trapped-ion systems considered in this Review is a single chain of many ions, which scales up to 20-30 ions for state-of-art systems~\cite{Pogorelov21, chen2023benchmarking}. Another well-known architecture is the quantum charge-coupled device (QCCD) architecture~\cite{pino2021demonstration}, which utilizes operations such as splitting an ion from a chain, shuttling an ion, and merging an ion to another chain. While the QCCD architecture has advantages such as high-fidelity qubit operations, the motional states are not preserved during the splitting and merging, and thus cannot provide the bosonic ingredient for analog simulations. 
\end{mdframed}

%%%%%%%%%%%%%%%%%%%%%%%%%%%%%%%%%%%%%%%%
%%%%%%%%%%%%%%%%%%%%%%%%%%%%%%%%%%%%%%%%
%% Section
%%%%%%%%%%%%%%%%%%%%%%%%%%%%%%%%%%%%%%%%
%%%%%%%%%%%%%%%%%%%%%%%%%%%%%%%%%%%%%%%%
\section*{The contenders: classical-digital vs trapped-ion quantum-analog}

\subsection*{The current champion: classical-digital simulation methods}

Numerical methods to describe quantum dynamics on classical-digital computers are well-established. The quantum nature of molecular systems, and their coupling to the surrounding environment, place the cost of brute-force simulations on classical-digital computers out of reach for most systems of interest, as the Hilbert space dimension grows exponentially with the number of degrees of freedom in the simulated system. Thus, sophisticated simulation methods provide access to approximate descriptions of a systems' time evolution. There is a trade-off among system size, structure of condensed-phase model, level of approximation, accuracy, and computational cost. 

We will discuss some of the challenge of simulating quantum molecular dynamics in the condensed phase using familiar linear vibronic coupling models (LVCMs)~\cite{Friesner81, Caldeira83, Garg85, MacDonell21}. This class of models is approximate, and real molecules exhibit rich anharmonicities and nonlinear interactions that are beyond the reach of these simple models (see the Outlook section below). A LVCM includes $M$ electronic states and $N$ harmonic bath modes, where the bath modes represent the degrees of freedom in the intra/intermolecular vibrations and the vibrations of solvents. Each electronic state and the linear coupling between electronic states is often coupled linearly to the bath degrees of freedom. The generic Hamiltonian in second-quantized notation is   
\begin{equation}\label{eq:H_mol}
    \hat{H}_{\rm mol}/\hslash = \sum^{M}_{i, j = 1} 
    \hat{\psi}_i^{\dagger} \hat{\psi}_j 
    \Big(\Delta_{i,j} + \sum^{N}_{k=1} \kappa_{i, j, k}(\hat{a}_k + \hat{a}_k^{\dagger})\Big)
    +  \sum^{N}_{k=1} \nu_k \hat{a}_k^\dagger \hat{a}_k,
\end{equation}
where $\hat{\psi}_i^{\dagger}$ and $\hat{\psi}_i$ are the creation and annihilation operators for electronic state $i$, and $\hat{a}_k^\dagger$ and $\hat{a}_k$ are those associated with bath mode $k$. Also, $\Delta_{i,i} := E_i$ is the energy of state $i$, and $\Delta_{i,j} = \Delta^*_{j,i}$ $(i \neq j)$ is the electronic coupling between states $i$ and $j$. Attached to each $\hat{\psi}_i^\dagger \hat{\psi}_j$ term is a linear coupling to the bath mode of frequency $\nu_k$, where the coupling strength is determined by $\kappa_{i,j,k} = \kappa_{j,i,k}^*$. This LVCM is general, in that any pair of states can be coupled and any mode can be coupled to any of the electronic states. 

We often follow the dynamics as one electron moves from a prepared donor state to an acceptor state. The initial vibrational states are often assumed to be at thermal equilibrium, often at room temperature $T_{\rm room}$. Some intramolecular mode frequencies satisfy $\hslash \nu_k \gg k_B T_{\rm room}$ in the condensed phase, where $k_B$ is Boltzmann's constant, so the initial average phonon number of each mode is much less than one. As such, a quantum treatment of these modes is needed. 

Assuming that the electronic and vibrational energy states can be obtained, \textit{numerically exact} methods (see Box 1 for a definition) exist for classical-digital computers to compute quantum molecular dynamics in some regimes. Examples include the multi-configuration time-dependent Hartree (MCTDH) method~\cite{lode2020colloquium,haobinJCP,mctdhBOOK}, the Ab Initio Multiple Spawning (AIMS) method~\cite{ben2000ab,martinez2006insights,curchod2020ssaims}, the hierarchical equations of motion (HEOM) approach~\cite{tanimura1990nonperturbative, jin2008exact, yan2020new}, the quasi-adiabatic propagator path integral (QUAPI) method~\cite{kundu2022intramolecular,makri1992improved,makri2020small,makri2020small2,makri2020small3,makri2021small,TOPALER1993285},  the time-evolving matrix products operators (TEMPO) method~\cite{Strathearn18,cygorek2022simulation,link2023open}, the dissipation-assisted matrix product factorization (DAMPF) method~\cite{Somoza19,somoza2023driving}, and the time-dependent density-matrix renormalization group (tDMRG) method~\cite{White04, Daley04, chin2010exact,Prior10,schollwock2011density,chin2011chain,del2018tensor,dunnett2021efficient,nuomin2022improving,xie2019time,nuomin2022suppressing,xu2022stochastic}. The tDMRG method, for example, can efficiently simulate a composite system of electronic states and bath modes when the composite system may be mapped to a simple configuration, such as one-dimensional chain~\cite{chin2010exact, Prior10} or star geometry~\cite{nuomin2022improving} (see Fig.~1d), through an appropriate basis transformation. For the case of one-dimensional chain, large systems (e.g., $N \sim 10^2$) with moderate interaction strengths (e.g., $|\kappa_{i,j,k}| \lesssim |\Delta_{i,j}|, \: |E_i - E_j|, \: \nu_k$) can be simulated~\cite{Prior10}. However, the tDMRG method is either inaccurate or computationally prohibitive when a transformation to a simple configuration cannot be found, or when the interactions are strong (e.g., $|\kappa_{i,j,k}| \sim$ 10 to $10^2$ times $|\Delta_{i,j}|, \: |E_i - E_j|, \: \nu_k$), which is a regime of interest~\cite{ET_JACS1,ET_JACS2,ET_JPC_LARGE_REORG}, for example, in the study of electron transfer dynamics accompanied by significant nuclear reorganization~\cite{onuchic1987effect,hammes2008proton,hsu2020reorganization,ET_JPC_LARGE_REORG}. Other numerically exact methods also become computationally prohibitive, in terms of computational time or memory demands, when the entanglement among the components becomes significant~\cite{PRB_ComingMode,cygorek2022simulation,makri2020small}. 

It is also worth mentioning that while the AIMS method is fundamentally based on exact principles, its practical application often involves approximations, such as enforcing nuclear wave packets to follow classical trajectories~\cite{mignolet2018walk}. This adaptation positions AIMS as a bridge between numerically exact methods and approximate methods discussed below.

In addition to numerically exact methods, \textit{approximate} methods (see Box 1 for a definition) may be used to describe quantum dynamics. The much larger mass of nuclei compared to electrons motivates approximate classical or semiclassical descriptions of nuclear motion. Approximations include mean-field strategies, localized classical nuclei, and linearization of quantum propagators. Among the battery of methods, mixed quantum-classical (MQC) and semiclassical (SC) methods may model many features of quantum molecular dynamics, including energy and electron transfer. MQC methods include Ehrenfest dynamics~\cite{Eherenfest1927, mclachlan1964variational, agostini2019different}, surface hopping~\cite{tully1990molecular,wang2016recent, barbatti2011nonadiabatic, subotnik2016understanding, wang2015}, and mixed quantum-classical Liouville dynamics~\cite{kapral1999mixed,mac2008trotter,kim2008quantum,hsieh2012nonadiabatic,hsieh2013analysis,kapral2015quantum}. The traditional SC approaches include the semiclassical initial value representation (SC-IVR)~\cite{miller2001semiclassical,miller2009electronically} and linearized SC-IVR (LSC-IVR)~\cite{sun1998semiclassical,shi2003relationship}. Recently, spin-based methods~\cite{bossion2022non, runeson2022spin, runeson2020generalized, mannouch2022partially} were also used to simulate nonadiabatic dynamics. In addition to the MQC and SC methods, other imaginary-time path-integral-based trajectory formalisms~\cite{cao1994formulation,jang1999derivation,habershon2013ring,craig2004quantum,richardson2013communication,ananth2013mapping,chowdhury2017coherent,chowdhury2019state,chowdhury2021non,hele2015boltzmann} can provide approximate methods to describe quantum dynamics.

These approximate methods fail when the bath degrees of freedom cannot be treated classically or semi-classically, as when high-frequency vibrations enter. Specifically, many of these approximations may fail to capture detailed balance \cite{tully-inbook,parandekar2005mixed,kang2019nonadiabatic} and lead to the leakage of zero-point energy (ZPE) \cite{jain2018vibrational,persico2014overview}. One approximate method may be more accurate than another, depending on the details of the model and the dynamical properties of interest. For example, imaginary-time path-integral-based trajectory approaches may provide quantum dynamics that, despite being approximate, preserve quantum detailed balance and avoids ZPE leakage, thus providing better accuracy than traditional SC approaches. However, these trajectory-based methods can be prohibitively expensive computationally~\cite{ananth2022path,liu2022imaginary}.

When the coupling strengths between electronic states and bath modes are weak (e.g., $|\kappa_{i,j,k}| \ll |\Delta_{i,j}|, \: |E_i - E_j|, \: \nu_k$), the accuracy of the approximate methods for describing the electronic-state populations over time can be acceptable for timescales that are much shorter than the inverse of the electronic-coupling energy scale (divided by $\hslash$). Also, when the coupling strengths are extremely strong (e.g., $|\kappa_{i,j,k}| \gg |\Delta_{i,j}|, \: |E_i - E_j|, \: \nu_k$), some approximate methods can estimate quantities such as electron-transfer rates well, as semiclassical treatments of harmonic oscillators are often valid for highly excited states. Yet, for many chemical systems between the regimes of weak and strong coupling to bath modes, where quantum entanglement (between electronic states, between bath modes, and between electronic states and bath modes) plays a significant role in the time evolution, the approximate methods become inaccurate. In particular, it is difficult to derive an approximation for quantum dynamics that performs well at both short and long times, although simulating long-time dynamics may be used to determine important quantities, such as equilibrium electronic-state populations~\cite{bellonzi2016assessment}. Therefore, achieving accuracy for longer timescales and for intermediate coupling $|\kappa_{i,j,k}|$ may serve as a compelling target for analog quantum simulations of chemical dynamics.

\subsection*{The upstart: analog trapped-ion simulations}

A trapped-ion system allows control over its spin and bosonic degrees of freedom, and thus appears to be a natural platform for performing analog quantum simulations of quantum molecular dynamics, as the molecular electronic states can be represented by the spins, and the molecular vibrations can be modeled using the bosonic modes of the trapped-ion system~\cite{MacDonell21, Gorman18}.
First, each ion's atomic states (spin states) can be used to simulate the electronic states of the LVCM. Typically two of the most stable and readily accessible atomic states are used per ion. As such, a \textit{qubit} is encoded in each trapped ion. It is also possible to encode $d > 2$ states per ion, in which case a \textit{qudit} is encoded~\cite{Low20, Ringbauer22}. Next, the normal modes associated with ion motion can be used to simulate the bath modes of the LVCM. 

Lasers or magnetic-field gradients can induce coupling between the ions' atomic states and the motional modes. Each interaction can be mapped to an individual term in the LVCM Hamiltonian, as discussed below. The evolution of the full system to time $\tau$ can be simulated using  \textit{Trotterization}, where the time evolution of  Hamiltonian is simulated for a short time step $\tau/S$, and the process is repeated $S$ times~\cite{lloyd1996universal}. Each evolution with respect to a single Hamiltonian term up to time $\tau/S$ can be considered as a unit operation that comprises the entire evolution. 

This Trotterization-based analog approach to quantum simulation, also suggested for simulating processes in quantum field theory~\cite{Davoudi21}, contrasts with the traditional approach in analog trapped-ion simulations, where interactions for each and every term in the simulated Hamiltonian is simultaneously switched on~\cite{Porras04, Monroe21}. While the traditional approach enabled various quantum-simulation experiments, even with hardware that does not fully support individual control of qubits (see Ref.~\citenum{Maier19} for an example of simulating quantum molecular dynamics), the parameters of the simulated Hamiltonian should meet specific constraints\footnote{For example, when a quantum spin system is simulated using the traditional approach where all ions are simultaneously illuminated by a monochromatic beam, the coupling strength $\Delta_{i,j}$ between spins $i$ and $j$, mapped to ions $i$ and $j$ indexed from the left to right of the ion chain, should satisfy $\Delta_{i,j} \propto |i-j|^{-\alpha}$ ($0<\alpha<3$)~\cite{Porras04, Maier19, Monroe21}.} that do not exist for the Trotterization-based analog approach. 

We now summarize how the LVCM may be simulated using trapped-ion systems. Before we proceed, we note that Ref.~\citenum{MacDonell21} provides an overview of various strategies for simulating the LVCM, which include simultaneous evolution of all terms and its combination with Trotterization. Solely using simultaneous evolution is applicable only to a model with a few electronic states and bath modes, as finding the electromagnetic-field parameters that induce the evolution of a larger LVCM is challenging. Here, we focus on a specific Trotterization strategy that works for any LVCM in its most general form \eqref{eq:H_mol}.

For simplicity, we consider $M$ ion qubits with $N$ motional modes. In this case, each electronic site is mapped to a qubit, and each bath mode is mapped to a motional mode. For example, an electronic state in which the $i^{\rm th}$ electronic state is occupied and all other sites are unoccupied is represented by the $i^{\rm th}$ qubit at $\ket{1}$ and all other qubits at $\ket{0}$, and the $k^{\rm th}$ bath mode is represented by the ion chain's $k^{\rm th}$ motional mode, where the indices of the ions and motional modes are assigned arbitrarily. We note that when the Hamiltonian has symmetries, it can be simulated using fewer ions, as $n$ ion qubits represent $2^n$  states~\cite{Wang18}. The number of ions can be decreased further when qudits are used rather than qubits~\cite{sun2023quantum}. 

At the start of a quantum simulation, the trapped-ion qubits are set to states that represent the initial electronic state of the molecule being simulated. For example, the qubit representing an initial electron donor state is set to $\ket{1}$, and all other qubits are set to $\ket{0}$. The motional modes are also cooled to a low-temperature thermal state~\cite{Lechner16, Feng20} that represents the initial state of the bath modes in the condensed-phase molecular system. Cooling the motional modes to the ground state is a good approximation when the simulated bath-mode frequencies $\nu_k$ satisfy $\hslash \nu_k \gg k_B T_{\rm room}$.

After initializing the simulation, the time evolution with respect to the LVCM is simulated using trapped ions' native operations that are directly executed by applying electromagnetic fields of appropriate frequencies to the ions. Two of the most widely-used native operations are the single-qubit rotation and the spin-dependent force. First, the single-qubit rotation, which is turned on when the electromagnetic field on ion $p$ is resonant with the qubit frequency $\omega_0$, is described by the Hamiltonian
\begin{equation}
    \hat{H}^{(0)}_p/\hslash = \frac{\Omega_p}{2} \hat{\sigma}^\phi_p, \label{eq:H0}
\end{equation}
where $\hat{\sigma}^\phi_p \equiv \hat{\sigma}^+_p e^{i \phi} + \hat{\sigma}^-_p e^{-i \phi}$, $\hat{\sigma}^+_p$ ($\hat{\sigma}^-_p$) is the raising (lowering) operator of qubit $p$, $\Omega_p$ is the Rabi frequency of the transition of qubit $p$, and $\phi$ is the spin phase determined by the phase of the fields. Next, the spin-dependent force is turned on when the field has two frequencies that are nearly resonant with $\omega_0 + \omega_k$ and $\omega_0 - \omega_k$; we denote these as the blue and red sidebands, respectively, where $\omega_k$ is the frequency of motional mode $k$. When applied to ion $p$, the Hamiltonian is  
\begin{equation}
    \hat{H}^{(1)}_{p,k}/\hslash = \frac{\tilde{\Omega}_{p,k}}{2} \hat{\sigma}^\phi_p
    \left(\hat{b}_k e^{i [(\mu - \omega_k) t + \varphi]}  + h.c. \right) 
    \label{eq:H1}
\end{equation}
where $\hat{b}^\dagger_k$ ($\hat{b}_k$) is the creation (annihilation) operator of the resonant mode $k$, $\tilde{\Omega}_{p,k}$ is the Rabi frequency of the sideband transition of qubit $p$ with respect to mode $k$, $\mu$ (which is nearly resonant with $\omega_k$) is the detuning of the electromagnetic field's frequency from $\omega_0$, and $\varphi$ is the motional phase determined by the phase of the fields. We note that the rotating-wave approximation is used to derive both Hamiltonians~\cite{Wineland98}. 

By setting the spin phase $\phi$ of \eqref{eq:H0} to 0 [$-\pi/2$], we can perform single-qubit Pauli-X [Y] rotations $\exp(-i\theta \hat{X}_p/2)$ [$\exp(-i\theta \hat{Y}_p/2)$] by any angle $\theta$, where $\hat{X}_p$ [$\hat{Y}_p$] is the Pauli-X [Y] operator on qubit $p$. An arbitrary single-qubit unitary operation, such as a Pauli-Z rotation, can be decomposed into a sequence of Pauli-X and Y rotations~\cite{NielsenChuang}.  This approach enables full control over each individual qubit. Controlling the motion phase $\varphi$ also allows simulation of the terms $\nu_k \hat{a}_k^\dagger \hat{a}_k$ in (\ref{eq:H_mol}). This is the case because $\hat{H}_{\rm mol}$ can be re-written in the interaction picture with respect to the bath modes~\cite{leibfried2003quantum, lidar2001completely, nuomin2022improving}, removing the terms $\nu_k \hat{a}^\dagger_k \hat{a}_k$ and converting $\hat{a}_k + \hat{a}^\dagger_k$ to $\hat{a}_k e^{-i \nu_k t} + \hat{a}_k^\dagger e^{i \nu_k t}$, which can be mapped to (\ref{eq:H1}) by appropriately setting $\varphi$.

Another important native operation is the M{\o}lmer-S{\o}rensen (MS)~\cite{Molmer99, Sorensen99} interaction, which occurs when a spin-dependent force is applied to two ions $p$ and $p'$ simultaneously. The interaction is described by a Hamiltonian proportional to $\hat{\sigma}^{\phi}_p \hat{\sigma}^{\phi'}_{p'}$, where $\phi$ ($\phi'$) is the spin phase of ion $p$ ($p'$). While the MS interaction is widely used with quantum-computation gates that generate entanglement between two qubits, we use it here to implement electronic transfer, to describe the flow of electronic excitation energy between electronic states. The electronic transfer is simulated by the Hamiltonian
\begin{equation}
    \hat{H}^{(2)}_{p,p'}/\hslash = J(\ket{0}_p \ket{1}_{p'} \bra{1}_p \bra{0}_{p'} + \ket{1}_p \ket{0}_{p'} \bra{0}_p \bra{1}_{p'})
    = \frac{J}{2}(\hat{X}_p \hat{X}_{p'} + \hat{Y}_p \hat{Y}_{p'}) 
    = \frac{J}{2} \left(\hat{\sigma}^{0}_p \hat{\sigma}^{0}_{p'} + \hat{\sigma}^{-\pi/2}_p \hat{\sigma}^{-\pi/2}_{p'} \right). \label{eq:H2}
\end{equation}
Here, $J$ is determined by the field parameters and time duration of the operation, which are carefully chosen so that there is no residual entanglement between the qubits and the motional modes after the operation~\cite{Molmer99, Sorensen99}. This operation allows the simulation of terms proportional to $\hat{\psi}_i^{\dagger} \hat{\psi}_j$ in the LVCM for \textit{arbitrary} pairs of $i$ and $j$. $\hat{X}_p \hat{X}_{p'}$ and $\hat{Y}_p \hat{Y}_{p'}$ commute, so the two corresponding MS interactions can be implemented sequentially. 

Using single-qubit rotation (\ref{eq:H0}), spin-dependent force (\ref{eq:H1}), and MS interaction (\ref{eq:H2}), each term in (\ref{eq:H_mol}) can be simulated. Specifically, the terms $E_i \hat{\psi}_i^\dagger \hat{\psi}_i$ are mapped to (\ref{eq:H0}), the terms $\kappa_{i,i,k} \hat{\psi}_i^\dagger \hat{\psi}_i (\hat{a}_k + \hat{a}_k^\dagger )$ are mapped to (\ref{eq:H1}), and the terms $\Delta_{i,j} \hat{\psi}_i^\dagger \hat{\psi}_j$ ($i \neq j$) are mapped to (\ref{eq:H2}). The terms $\kappa_{i,j,k} \hat{\psi}_i^\dagger \hat{\psi}_j (\hat{a}_k + \hat{a}_k^\dagger)$ ($i \neq j$) are also mapped to a sequence of operations (\ref{eq:H0})-(\ref{eq:H2}), as described in Ref.~\citenum{Zhang20}. The evolution with respect to each term in (\ref{eq:H_mol}) by the Trotterization time step $\tau/S$ is simulated by  choosing the field parameters and time duration of each mapped trapped-ion operation appropriately. By repeating the evolution for all terms and Trotterization steps, the evolution of the entire system described by the LVCM Hamiltonian is simulated.

After simulating the system's time evolution, the state of the qubits are measured simultaneously. Each measurement produces a binary result for each ion, reporting whether each ion $p$ was in the state $\ket{0}_p$ or $\ket{1}_p$. Thus, the sequence of cooling, state initialization, time evolution, and measurement needs to be repeated many times to accurately measure the population of each qubit at a given point in time. Assuming that experimental measurement error is small (the state-of-art measurement error rate for trapped ions~\cite{Christensen20, Ransford21} is roughly $10^{-4}$), the uncertainty of each measured qubit-state population, hereafter referred to as \textit{shot noise}, is given by $\sqrt{P(1-P)/R}$, where $P$ is the measured population and $R$ is the number of repeated simulation runs. To generate the population curve for $S'$ time steps, the state-population measurement needs to be performed separately for each of the $S'$ time steps.

Shot noise can be non-negligibly large for realistic values of $R$ (e.g., $\sqrt{P(1-P)/R} = 0.05$ for $P=0.5$ and $R=100$); thus, analog quantum simulations may not be suitable for estimating the value of each and every point in the population curve with accuracy significantly better than, say, 1\%. However, shot noise has favorable properties. First, it is unbiased and completely independent for each population measurement. Therefore, useful properties of dynamics, such as the overall rate or frequency of population transfer and the equilibrium population, can be  captured reliably by analog quantum simulations. Second, the magnitude of shot noise is straightforwardly given as above. This is unlike the errors in numerically exact classical-digital simulations, whose upper bound is typically obtained by repeating the simulations with varying parameters and checking convergence, and unlike the errors due to noise during the time evolution of the quantum simulator, whose magnitude can only be estimated roughly for simulations of moderate-size models. 

The principal benefit of the trapped-ion simulator described above is the favorable scaling of the simulation time with the coupling strengths in the simulated model. As interactions between the molecular components are directly mapped to interactions between the trapped-ion system's components, under reasonable assumptions on the strength of trapped-ion system's interactions (see the next section), the required time duration for the trapped-ion simulator's physical operations increases only linearly with the coupling strengths in the LVCM, even in the presence of large entanglements created by strong interactions. The magnitude of noise in the trapped-ion simulator determines the upper limit of the evolution time $\tau$ that may be accessed accurately, and this limit may depend on the parameter strengths of the simulated LVCM~\cite{MacDonell21}. 

A disadvantage of this quantum analog method is that only limited information of the system's quantum state is revealed by each measurement. The required number of measurements to obtain the full density matrix of the composite system typically scales exponentially with the number of qubits and motional modes used~\cite{paris2004quantum}. While the number of measurements may scale favorably for a certain class of states~\cite{cramer2010efficient}, in practice, characterization of the motional-mode state is currently limited to a few basis vectors even for 3 modes~\cite{Jia22}. For classical-digital methods, while representing the density matrix of a molecular system can also require memory resources that grow exponentially with the number of electronic states and bath modes, it is often possible to use a compact representation of the composite state. This allows the full density matrix to be obtained for at least a few tens of modes~\cite{MacDonell21}. In this Review, we consider studying the electronic states' population dynamics, as its computational cost is favorable for analog quantum simulations using trapped ions.

%%%%%%%%%%%%%%%%%%%%%%%%%%%%%%%%%%%%%%%%
%%%%%%%%%%%%%%%%%%%%%%%%%%%%%%%%%%%%%%%%
%% Section
%%%%%%%%%%%%%%%%%%%%%%%%%%%%%%%%%%%%%%%%
%%%%%%%%%%%%%%%%%%%%%%%%%%%%%%%%%%%%%%%%
\section*{Sizing each other up: a test match with a simple  model}

We now compare the predicted performance of the analog trapped-ion simulation with two classical-digital algorithms, tDMRG and Ehrenfest, by simulating the dynamics of a simple model Hamiltonian~\cite{Sato18} that is a simple example of a LVCM. The model consists of two electronic states, namely the donor ($\ket{D}$) and acceptor ($\ket{A}$) states, and $N$ bath modes, altogether forming a star geometry as in Fig.~1d. The Hamiltonian is  
\begin{equation}\label{eq:toymodel}
    \hat{H}_{\rm toy}/\hslash =  \frac{\Delta}{2} \big(\ket{D}\bra{A} + \ket{A}\bra{D} \big)
    +  \sum_{k=1}^N \Big[
    \frac{\kappa}{2} \big(\ket{D}\bra{D} - \ket{A}\bra{A} \big)  \big(\hat{a}_k + \hat{a}^\dagger_k\big) 
    + \nu_k \hat{a}_k^\dagger \hat{a}_k 
    \Big],
\end{equation}
where $\Delta$ is the coupling between the two electronic states, $\nu_k$ is the energy (frequency) of the $k$-th bath mode, and $\kappa$ is the state-dependent coupling between the states and each mode. The reorganization energy $\lambda$ of the system is given by $\lambda = \kappa^2 \sum_{k=1}^N \frac{1}{\nu_k}$~\cite{hsu2020reorganization}. Typical reorganization energies range from a few $\Delta$ to hundreds of $\Delta$ \cite{ishizaki2009theoretical,nitzan2006chemical}. We assume that the bath modes have high frequencies compared to the room temperature (times $k_B/\hslash$), such that the average phonon number is $\sim 0.06$ for the initial state of each bath mode in our simulations. 

tDMRG simulations are performed by putting (\ref{eq:toymodel}) into a one-dimensional topology, where the electronic sites and the $N$ bath modes form a linear chain, using the unitary transformation described in Refs.~\citenum{Prior10,nuomin2022improving}. This strategy can be used to describe the dynamics efficiently for a donor-acceptor system coupled to a bath with hundreds of modes that have a moderately large reorganization energy ($\lambda \lesssim 10 \Delta$). However, the computational cost grows as $\kappa \propto \sqrt{\lambda}$ increases, and as the entanglement among the components grows. In contrast, the Ehrenfest simulations can be performed at significantly lower computation cost than the tDMRG simulations, at the expense of accuracy in describing the dynamics only up to a short time compared to $\hslash/\Delta$, as the semiclassical mean-field approximations often break down at longer times. 

We compare the performance of the classical-digital simulations with that of the trapped-ion simulations. Figure~1a shows the population curves for the donor state when $N=2$, calculated using tDMRG, Ehrenfest, and numerical simulations of the trapped-ion quantum simulation, with various reorganization energy values ($\lambda$). To characterize the analog trapped-ion simulations, the evolution of the composite quantum state of the ion qubit and the motional modes is calculated for the specific sequence of pulses, which perform the operations described by \eqref{eq:H1}, that are designed to track the evolution of \eqref{eq:toymodel}.

Specifically, we map $\ket{D}$ and $\ket{A}$ to the states $\ket{0}$ and $\ket{1}$ of a single qubit, and use only the spin-dependent force given by \eqref{eq:H1} that directly simulates the $(\kappa /2) (\ket{D}\bra{D} - \ket{A}\bra{A}) (\hat{a}_k + \hat{a}^\dagger_k)$ terms in an appropriately chosen basis of qubit states. In the interaction picture, the $(\Delta/2) (\ket{D}\bra{A} + \ket{A}\bra{D})$ term can be simulated by tuning the spin phase $\phi$, similar to the way that $\nu_k \hat{a}_k^\dagger \hat{a}_k$ terms are simulated by tuning the motional phase $\varphi$. 

We use $N$ radial~\cite{Zhu06} non-CM (see Box 2 for definition) motional modes to simulate the bath modes, where the CM modes are avoided as they are significantly more susceptible to noise, namely, heating (increase in phonon number over time). We need at least $\lceil N/2 \rceil+1$  trapped ions to generate $N$ radial non-CM modes, although operations are assumed to be performed only on a single, left-most qubit that is mapped to the electronic states. The lasers' propagation direction needs to be aligned to  significantly overlap with the radial directions. We assume that all modes are cooled to zero temperature prior to each simulation run~\cite{Lechner16, Feng20}. For the bath-mode frequencies considered here, assuming zero temperature causes negligible error in the computed electronic-state populations. 

For each of $S$ Trotterization steps, evolution with respect to each $(\kappa /2) (\ket{D}\bra{D} - \ket{A}\bra{A}) (\hat{a}_k + \hat{a}^\dagger_k)$ term up to time $\tau/S$ is simulated using a pulse of constant spin and motional phase that applies the spin-dependent force. With a fixed pulse power\footnote{For the case of MS interactions, the length of each pulse that simulates the evolution of $\Delta (\hat{\psi}^\dagger_i \hat{\psi}_j + h.c.)$ ($i \neq j$) term up to time $\tau/S$ is proportional to $\Delta \tau/S$ only when a larger $\tilde{\Omega}_{p,k}$ is used for a larger value of $S$. This is because inducing the evolution of \eqref{eq:H2} up to a shorter time requires using a pulse that is further detuned from the motional-mode frequencies in order to remove the residual entanglement between the qubits and the modes after the MS interaction. Specifically, when far-detuned pulses are used, the amount of detuning and $\tilde{\Omega}_{p,k}$ should be proportional to $S$ and $S^{1/2}$, respectively.} (quantified by $\tilde{\Omega}_{p,k}$), each pulse length is proportional to $\kappa \tau/S$. Thus, the experimental time for each trapped-ion simulation run, or the sum of all $N \times S$ pulse lengths, is proportional to $\kappa$; however, it does not depend on the value of $S$. This allows us to use sufficiently large $S$, such that the Trotterization error (due to finite $S$) is negligible compared to the error due to the trapped-ion system's decoherence~\cite{sun2023quantum}; for example, we set $S=600$ for generating the $\lambda = 30 \Delta$ curve in Fig.~1a.

%%%%%%%%%%%%%%%%%%%%FIGURE 1%%%%%%%%%%%%%%%%%%%%%%%
\begin{figure}[h!]
\centering
\includegraphics[width=0.9\linewidth]{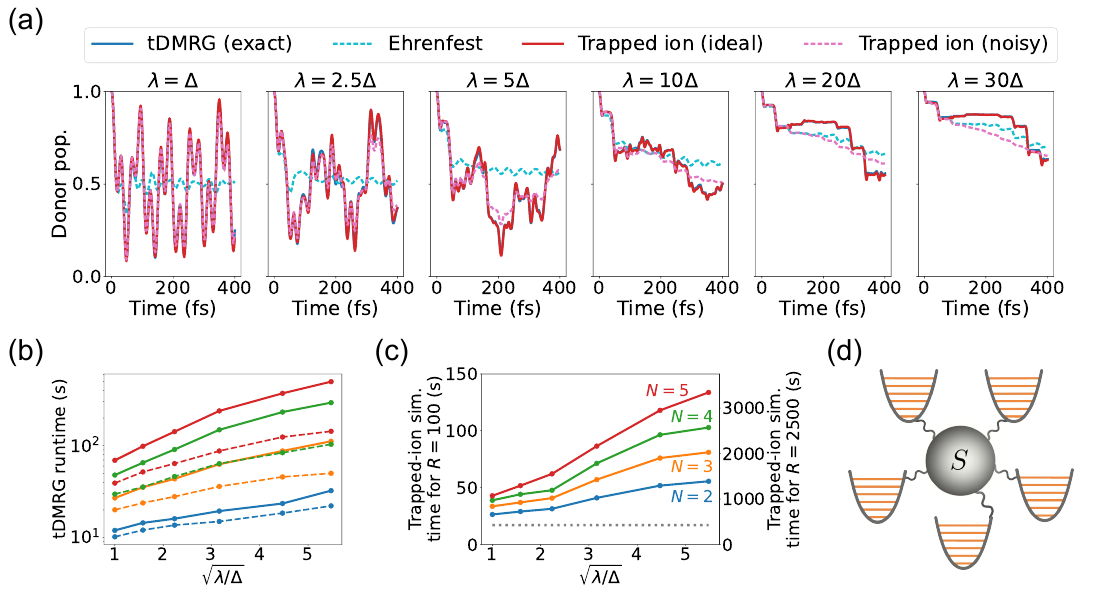}
\caption{Comparison of classical (tDMRG and Ehrenfest) and quantum (trapped ion) methods for simulating the model Hamiltonian (\ref{eq:toymodel}). $\Delta = 0.08679$ eV, $\nu_k = [0.08679 + 0.01240(k-1)/(N-1)$] eV for $k=1,..N$, where $N$ is the number of bath modes. For the trapped-ion method, we assume using $(\lceil N/2 \rceil+1)$-ion chain to generate $N$ radial non-CM motional modes. (a) Evolution of the donor population, for the indicated reorganization energy values ($\lambda$) and $N=2$, simulated using various methods. For the curves labeled ``Trapped ion'', we numerically simulate the evolution of a trapped-ion simulator system. The ideal case assumes no decoherence, and the noisy case assumes state-of-art noise parameters in Table~1. For both cases, shot noise (population uncertainty due to finite number of binary measurements) is not shown. In all panels, the blue and red solid curves overlap. (b) Computational run time for the tDMRG method, with various values of $\lambda/\Delta$ and $N$ [labels shared with (c)]. The additive error for each population value is less than $10^{-3}$ ($10^{-2}$) for the solid (dashed) curves. See Ref.~\citenum{nuomin2022improving} for details of the method (the interaction-picture chain-geometry algorithm). The run time scales nearly exponentially with $\sqrt{\lambda/\Delta}$, where the rate of increase is smaller for the dashed curves. Computational run time for the Ehrenfest method is negligible compared to the tDMRG method. (c) Estimated experimental time for the trapped-ion simulation, with various values of $\lambda/\Delta$ and $N$. We assume 40 equally-spaced time steps and 100 (2500) simulation runs per time step for the left (right) vertical axis, where each run produces a binary measurement result. Dotted line shows the total time spent for cooling, state preparation, and measurement, which is estimated to be \mbox{4.25 ms}~\cite{kang2023efficient} times the total number of simulation runs. The estimated experimental time roughly scales linearly with $\sqrt{\lambda/\Delta}$. (d) A schematic representation of the star-geometry model simulated here, where $N=5$ [see \eqref{eq:toymodel}]. The grey circle, denoted `S', represents the system with two electronic energy levels, and each parabola represents a bath mode.}
\label{fig:toy_model}
\end{figure}
%%%%%%%%%%%%%%%%%%%%%%%%%%%%%%%%%%%%%%%%%%%%%%%%%%%

Even when a pulse is resonant with the sideband frequency of a mode ($\mu = \omega_k$), unwanted off-resonant excitations of other modes, also known as cross-mode coupling, may occur because of the finite spacings between the mode frequencies. Na\"ively, this leads to a constraint that $\tilde{\Omega}_{p,k}$ should be much smaller than the mode-frequency spacing, which imposes a lower bound on the time duration of the corresponding operations. However, cross-mode coupling can be actively removed by modulating the field parameters during the operation. Here, we consider using a frequency($\mu$)-modulated pulse~\cite{Leung18, Kang21} for each spin-dependent-force operation, such that cross-mode coupling is removed using operations with as short a time duration as possible. The required time duration increases as the number of ions increases and mode-frequency spacing decreases. For example, for the case of $\lambda = 30\Delta$ and $S=600$, the length of each pulse that simulates a single Trotterization step averaged over all modes is 15.7 $\mu$s, 17.4 $\mu$s, and 19.0 $\mu$s for a 2-ion ($N=2$), 3-ion ($N=3,4$), and 4-ion ($N=5$) chain, respectively, where the number of ions is set as $\lceil N/2 \rceil +1$. The relevant system parameters for a state-of-art trapped-ion setup~\cite{Wang20FM} are shown in Table~1.   

%%%%%%%%%%%%%%%%%%%%%%%%%%%%%%TABLE 1%%%%%%%%%%%%%%%%%%%%%%%%%%%%%%%%%%%%
\begin{table}[h!]
\centering
\begin{tabular}{|c|c|c|c|c|}
\hline
\begin{tabular}[c]{@{}c@{}}
Motional-mode\\frequencies $\omega_k/2\pi$
\end{tabular} & 
\begin{tabular}[c]{@{}c@{}}
Sideband-transition\\Rabi frequencies $\tilde{\Omega}_{p,k} / 2\pi$
\end{tabular} & 
\begin{tabular}[c]{@{}c@{}}
Coherence time\\of motional modes
\end{tabular} & 
\begin{tabular}[c]{@{}c@{}}
Heating rate of\\non-CM motional modes
\end{tabular} & 
\begin{tabular}[c]{@{}c@{}}
Coherence time\\of lasers
\end{tabular}  \\
\hline
\begin{tabular}[c]{@{}c@{}}1.80 - 1.98 MHz,\\2.45 - 2.58 MHz\end{tabular}                  
& 1.47 - 4.95 kHz & 36 ms & 5 quanta / s & 496 ms \\ \hline    
\end{tabular}
\caption{\textbf{System and noise parameters} for the state-of-art trapped-ion system in Refs.~\citenum{Wang20FM, Kang23} that are used in Fig.~1. Motional modes in two radial directions of ion chain consisting of two to four ions are used. Motional-mode coherence time is taken from Ref.~\citenum{Wang20FM} and motional-mode heating rate and laser coherence time are taken from Ref.~\citenum{Kang23}. The heating rate of CM motional modes is of the order of 100 quanta/s, so we assume that CM modes are not used in our trapped-ion simulations.}
\label{tab:sysnoiseparams}
\end{table}
%%%%%%%%%%%%%%%%%%%%%%%%%%%%%%%%%%%%%%%%%%%%%%%%%%%%%%%%%%%%%%%%%%%%%%%%

In state-of-art trapped-ion systems, the leading sources of errors are motional-mode dephasing, motional-mode heating, and laser dephasing~\cite{Wang20FM, Kang23, Cetina22}. To simulate the evolution of a trapped-ion system in the presence of noise, we use Qutip~\cite{Qutip} to solve the Lindblad master equation~\cite{lindblad1976generators}, following the method in Ref.~\citenum{Wang20FM} which provides reliable predictions of the system performance~\cite{Wang20FM, Kang23}. We use the best recorded noise parameters from a state-of-art trapped-ion system~\cite{Wang20FM, Kang23} as shown in Table~1, which are far from the fundamental limits and can, in principle, be improved.  

For all $\lambda$ values examined, the populations calculated using the tDMRG and the numerical simulation of the ideal trapped-ion simulation without noise match within numerical error. Indeed, both methods capture the quantum dynamics with high accuracy. The Ehrenfest method does not describe the strong population oscillations for smaller values of $\lambda$, as the effects of quantum coherence are ignored. The Ehrenfest simulations perform better when the system-bath coupling is very strong \mbox{($\lambda \gtrsim 20 \Delta$)}, as the bath oscillators become highly excited and behave nearly classically.

For $\lambda \lesssim 5\Delta$, the populations calculated by the numerical simulation of the noisy trapped-ion simulation closely matches the ideal simulation and the tDMRG analysis, capturing many fine details of the population dynamics. However, for larger values of $\lambda$ ($\propto \kappa^2$), the experimental time of each run is closer to the timescale of the noise due to longer operations, so the population curve deviates more from the ideal simulation~\cite{MacDonell21}. Interestingly, for $\lambda = 20\Delta$ and simulated time less than 200 fs, the populations of the noisy trapped-ion simulation approach those of the Ehrenfest calculations. This finding suggests a fascinating possibility that a controlled amount and type of experimental noise in trapped-ion systems can be mapped to certain kinds of semiclassical approximations in chemical dynamics~\cite{MacDonell21}, allowing a noisy device to approximately simulate an intermediate regime between quantum and semiclassical (e.g., Ehrenfest) chemical dynamics. However, for $\lambda = 30 \Delta$, the noisy trapped-ion simulation results deviate from both tDMRG and Ehrenfest calculations.   

Figure~1b shows the run time of the tDMRG simulation, with various $\lambda$ and $N$ values. Here we use two sets of simulation parameters that determine the accuracy and run time, such as the number of states used to represent each bath mode and the singular value decomposition threshold, such that the additive error $\epsilon$ of each electronic-state population is less than a fixed value for all $\lambda$ and $N$ used. For the solid (dashed) curves,  $\epsilon < 10^{-3}$ ($10^{-2}$). Admitting a larger error $\epsilon$ reduces the run time. This trade off of numerically exact classical-digital methods is crucial to the analysis of a potential quantum advantage. 

Figure~1c shows the estimated experimental time for the trapped-ion simulation, with various $\lambda$ and $N$ values. We assume $S'=40$ equally-spaced time steps are used to generate the population curve, and $R=100$ or 2500 simulation runs are repeated for each time step. For $R=100$ (2500), each population $P$ has uncertainty $\sqrt{P(1-P)/R}$ less than 0.05 (0.01), which is not shown in Fig.~1a. These uncertainties are larger than the maximum population error $\epsilon$ values in the tDMRG simulations, which are 0.01 and 0.001; we note that shot noise is unbiased and independent for each measurement, unlike the errors in tDMRG simulations, as explained in the previous section. The time spent for state-of-art cooling, state preparation, and measurement, which are assumed to be, respectively, 4 ms, 100 $\mu$s, and 150 $\mu$s~\cite{kang2023efficient}, times the total number of simulation runs $R \times S'$, is also included to the estimated experimental time. This contribution to the experimental time is dominated by the operations that simulate the time evolution. For each simulation run up to the longest evolution time $\tau$, the estimated total operation time ranges from 5 ms ($\lambda=\Delta$, $N=2$) to 57 ms ($\lambda=30\Delta$, $N=5$). This operation time can potentially be reduced by using motional modes of larger frequency spacings or larger laser power than the trapped-ion system considered in Table 1. 

The run time of the tDMRG simulation increases nearly exponentially with the value of $\lambda$, although the rate of increase is lower when larger error is admitted. The run time may be reduced by using a more highly optimized algorithm, for example, dynamical adaptation of the number of states that represent each bath mode~\cite{Zhang98}. Nonetheless, for a fixed, reasonably small target accuracy, we expect that this near-exponential scaling remains a fundamental limitation for numerically exact classical-digital methods, as the entanglement between the bath modes (mediated by the electronic states) grows as $\lambda$ increases. Ref.~\citenum{MacDonell21} shows another case of exponential scaling of required computational resource (classical memory with respect to number of modes) for the numerically exact MCTDH method.

The estimated experimental time required of the trapped-ion simulations is only $\mathcal{O}(\sqrt{\lambda})$, or $\mathcal{O}(\kappa)$, as the operation time of the spin-dependent force is directly proportional to $\kappa$ in \eqref{eq:toymodel}. Combining (i) the comparison between Fig.~1b and c and (ii) the observation that Ehrenfest calculation can be more accurate than noisy trapped-ion simulation for very large $\lambda$ values, we expect that trapped-ion simulations may have advantages over classical-digital methods, in terms of both computational cost and accuracy, for an \textit{intermediate} regime of reorganization energy compared to the electronic coupling and energy gap. Box 3 provides a general qualitative analysis of the potential advantages of noisy trapped-ion simulation over classical-digital methods for simulating LVCMs. 

%%%%%%%%%%%%%%%%%%%%%%%%%%%%%%%BOX 3%%%%%%%%%%%%%%%%%%%%%%%%%%%%%%%%%%%%%
\begin{mdframed}
\section*{Box 3 | LVCM: Where might analog trapped-ion simulations win?}
Color gradients in the figure illustrate the qualitative performance (accuracy and computational cost) of three classes of quantum dynamics simulations of LVCMs: numerically exact classical-digital methods, approximate classical-digital methods, and analog trapped-ion simulations on a noisy device. The qualitative performance of these methods is illustrated over several regimes of coupling strength between the electronic states and bath modes, quantified by the ratio of the reorganization energy $\lambda$ to the coupling strength between electronic states $\Delta$. 

If $\lambda \ll \Delta$, the numerically exact methods are generally efficient and accurate, since the entanglement among the electronic states and bath modes is weak. In this regime, the approximate methods are also accurate, at least for a few dynamical cycles of the system. This is because the influence of the bath on the electronic states is weak, and errors associated with neglecting non-classical trajectories of the bath modes have little influence on the evolution of the electronic states.

In the regime  where $\lambda \gg \Delta$, the numerically exact methods converge slowly and generally perform poorly because of strong entanglement among the electronic states and bath modes. However, the approximate methods generally perform well, providing reliable predictions for the donor-state population decay rate, for example. This acceptable performance arises when the coupling is strong and the bath modes are highly-excited harmonic oscillators that often behave almost classically. 

In the intermediate regime where $\lambda \gtrsim \Delta$, the coupling can be sufficiently strong that large entanglement renders the numerically exact methods computationally intractable, yet weak enough that the bath modes are not sufficiently highly excited to behave classically. As such, this regime appears to be a compelling target for testing whether an analog trapped-ion simulation may outperform classical-digital algorithms for simulating quantum molecular dynamics economically, at least in the space of LVCMs. The upper bound of $\lambda/\Delta$ that may allow a trapped-ion simulation to produce sufficiently accurate predictions is determined by the magnitude of the noise relative to the strength of native operations of the trapped-ion system. 

\begin{center}
\includegraphics[width=0.7\linewidth]{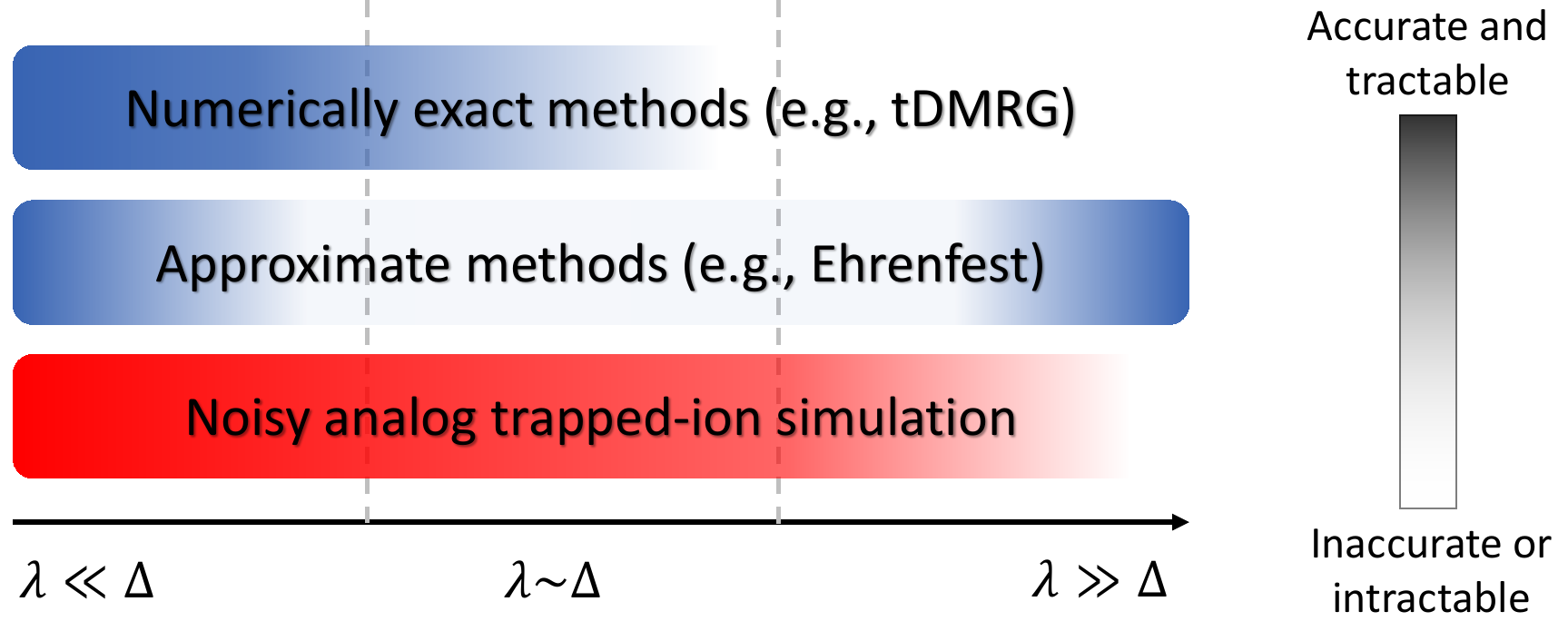}
\end{center}
\end{mdframed}
%%%%%%%%%%%%%%%%%%%%%%%%%%%%%%%%%%%%%%%%%%%%%%%%%%%%%%%%%%%%%%%%%%%%%%%%

\subsection*{LVCM beyond the simple model: seeking a quantum advantage}

While the simplicity of the  model in \eqref{eq:toymodel} allows a head-to-head comparison between classical-digital methods and noisy analog trapped-ion simulations in Fig.~1, achieving a \textit{quantum advantage} (defined in Box 1) requires simulating a more complicated model that is \textit{classically intractable} (also defined in Box 1). In the space of LVCMs described by (\ref{eq:H_mol}), a model in which multiple electronic states are strongly coupled to each other - and to many bath modes - may be considered. The couplings may have a complex connectivity (when characterized as a graph), which makes classical-digital simulations more challenging, as each bath mode can be quickly entangled to many other bath modes and electronic states. For example, if the Hamiltonian cannot be transformed into one-dimensional chain or star geometry, the tDMRG method is inefficient even for systems with moderately large reorganization energy compared to the electronic coupling strength. 

Providing a concrete model on which a quantum advantage can be achieved requires a rigorous study of the cost of classical-digital computation, which is beyond the scope of this Review. Instead, we guide the readers to the state-of-art numerically exact classical-digital simulation results in Fig. 4(a) of Ref.~\citenum{Somoza19} and Sec. V B of Ref.~\citenum{Mascherpa20}. While the models simulated here are not in the form of LVCM in \eqref{eq:H_mol}, as the bath modes are dissipative (see the Outlook section below for trapped-ion simulations of dissipative bath modes), we expect these results to give a rough sense of the computational cost of numerically exact simulations of LVCMs and similar models that are more complicated than the simple model in \eqref{eq:toymodel}. Ref.~\citenum{Somoza19} simulated a homogeneous polymer of 20 electronic states with nearest-neighbor electronic coupling $\Delta$. Each electronic state is independently coupled to 2 bath modes (total 40 modes) with reorganization energy 1.25 times larger than the coupling strength between neighboring electronic states ($\lambda/\Delta = 1.25$). Simulations that achieve numerical convergence required 60 hours of computation on 2 processors with 8 cores each, although allowing larger errors led to significantly reduced run time. Ref.~\citenum{Mascherpa20} simulated a similar model with 10 electronic states each coupled to 4 bath modes (total 40 modes) and $\lambda /\Delta = 1.5$, which required several days of computation on a 16-core node of a computing cluster. Based on the observation in Fig.~1b that the run time of a numerically exact classical-digital simulation method increases nearly exponentially with $\lambda/\Delta$, it seems possible that a similar model with a few tens of electronic states and bath modes and larger $\lambda/\Delta$ (e.g., $\lambda/\Delta \approx 10$) is challenging for numerically exact simulations, especially when a more complicated connectivity of couplings between the electronic states and bath modes is considered. 

Analog quantum simulations of LVCMs with a few tens of electronic states and bath modes can be performed on a chain of a few tens of ions, which is often used in state-of-art quantum computation and simulation experiments~\cite{Pogorelov21, chen2023benchmarking}. A model beyond the LVCM, for example, with dissipative bath modes, can also be simulated using trapped ions, as discussed in the Outlook section below. To achieve a quantum advantage, the experimental time for trapped-ion simulations should be short enough to be practical. The experimental time depends on the specific hardware, simulated model, and simulated evolution time. Nonetheless, compared to the longest experimental time plotted in Fig.~1, we estimate that the experimental time for simulating a LVCM with a few tens of electronic states and bath modes and $\lambda/\Delta \approx 10$ is roughly two orders of magnitude longer, where one order of magnitude comes from the larger number of Hamiltonian terms, and another comes from requiring longer qubit-mode interactions in longer ion chains in order to remove the cross-mode coupling, as the spacing between neighboring radial motional-mode frequencies is roughly inversely proportional to the number of ions~\cite{Zhu06}. Thus, the experimental time for trapped-ion simulations in this regime is expected to be in the order of 1s for each run (followed by binary qubit-state measurement) and several hours to days in total. 

An important difference between the trapped-ion simulation of \eqref{eq:toymodel} and that of more complicated LVCMs is that the former only requires spin-dependent forces, while the latter additionally requires MS interactions. When a moderately large value of $\lambda/\Delta$ is considered, as in the previous paragraph, the time spent for MS interactions is expected to be comparable to or shorter than that for spin-dependent forces. Also, as MS interactions are mediated by the motional modes, the leading sources of errors are the same as those simulated in Fig.~1a, namely, motional-mode dephasing and heating~\cite{Kang23, Cetina22}. Thus, for a fixed time, we expect that the contribution of noise to errors in the qubit population is similar for MS interactions and spin-dependent forces.

Performing accurate analog-quantum simulations using longer evolution time requires less noise. Thus, in order to achieve a quantum advantage, a number of technological innovations should be made for trapped ions. First, the coherence time of the motional modes can be extended by improving the stability of the rf voltage supply that determines the trapping potential. Indeed, motional coherence time of a single ion in the order of 1s has already been demonstrated~\cite{jarlaud2020coherence}. Second, the heating of the motional modes can be suppressed by using a cryogenic system~\cite{pagano2018cryogenic, Spivey21} rather than a commonly-used room-temperature system. Ref.~\citenum{spivey_dissert} reports the heating rate of the non-CM mode of a two-ion chain that is less than 0.5 quanta/s. Better material for the surface of the ion-trap chip needs to be explored as well~\cite{brown2021materials}. Third, the coherence time of the laser phase can be improved by better protecting the optical setup from mechanical vibrations~\cite{ZhangDissert}. Finally, remaining sources of errors, such as optical crosstalk, miscalibration, and spontaneous decay of the excited atomic level ($|e\rangle$ in the figure of Box 2) can be mitigated by using control techniques~\cite{parrado2021crosstalk, Fang22}, efficient system parameter characterization~\cite{kang2023efficient}, and increased detuning from the excited level~\cite{Moore23}, respectively. With all these technological innovations combined, accurate simulations of LVCMs and similar models in the classically intractable regime may be within reach, potentially leading to a quantum advantage.

%%%%%%%%%%%%%%%%%%%%%%%%%%%%%%%%%%%%%%%%
%%%%%%%%%%%%%%%%%%%%%%%%%%%%%%%%%%%%%%%%
%% Section
%%%%%%%%%%%%%%%%%%%%%%%%%%%%%%%%%%%%%%%%
%%%%%%%%%%%%%%%%%%%%%%%%%%%%%%%%%%%%%%%%
\section*{Candidate models for near-term trapped-ion simulations}

As described in the previous section, performing accurate quantum simulations of classically intractable LVCMs may require trapped-ion simulators that are significantly improved with respect to noise level over state-of-art systems. In this section, we describe three candidate models that can be simulated using current or expected near-term trapped-ion hardware. While each candidate model has a small number of components, each model's dynamics may exhibit signatures of quantum interference and entanglement of interest. Extending these models to larger systems may provide a route toward a quantum advantage in developing simulation tools to understand the complex dynamics of potentially valuable molecular systems.

\subsection*{Conical intersections}

The efficient conversion of light into chemical energy plays a central role in the biology of vision~\cite{polli2010conical}, DNA damage and repair~\cite{barbatti2010relaxation, matsika2005three}, and photosynthesis\cite{ishizaki2009theoretical}. The underpinning non-adiabatic dynamics typically occur near a conical intersection (CI), or crossing of potential-energy surfaces \cite{larson2020conical, yarkony1996diabolical, baer2006beyond}. Near a CI, the Born-Oppenheimer approximation fails. Thus, a Hilbert space of entangled electronic and nuclear states needs to be considered when considering nonadiabatic dynamics, which increases the memory and computational time required for dynamical simulations. 

%%%%%%%%%%%%%%%%%%%%FIGURE 2%%%%%%%%%%%%%%%%%%%%%%%
\begin{figure}[h!]
\centering
\includegraphics[width=0.8\linewidth]{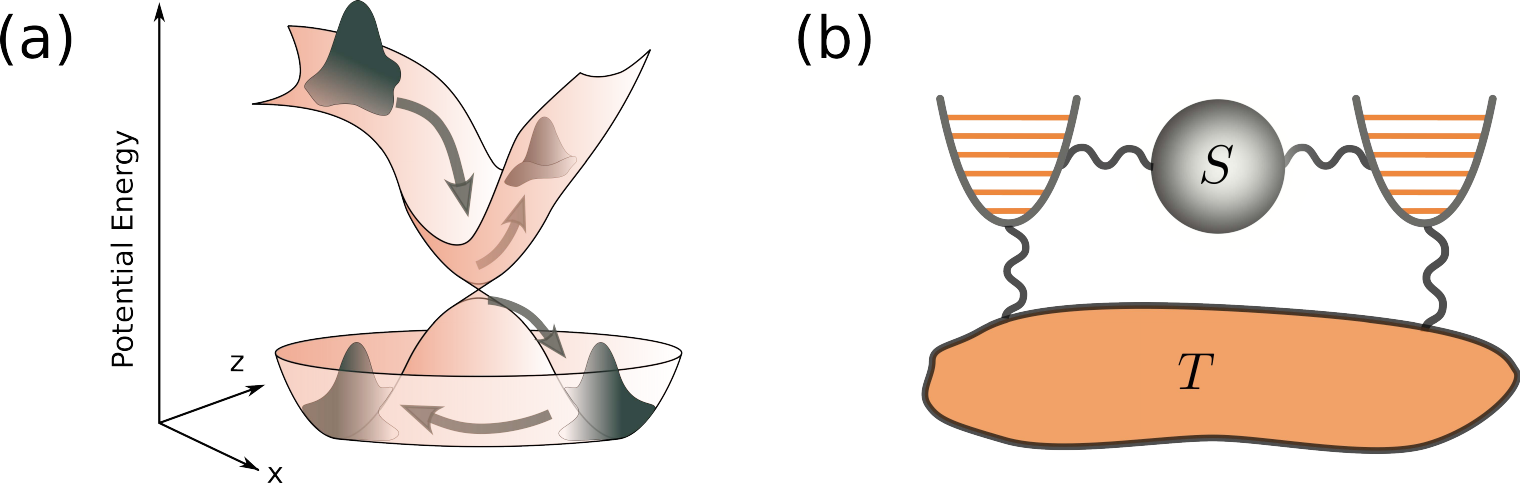}
\caption{(a) Potential-energy surfaces defined as functions of two reaction coordinates $x$ and $z$. A conical intersection (CI) occurs when the two potential-energy surfaces cross. A wave packet traveling through the CI displays highly nonadiabatic behavior, including quantum interference between pathways. For the case of symmetric mode frequencies and couplings, a degenerate ground-state space surrounds the CI, as depicted by the lower potential-energy surface. (b) A schematic representation of the model [see \eqref{eq:H_CI}]. The grey circle, denoted `S', represents the system with two electronic energy levels. The parabola on the left represents a mode that modulates the energy difference of the two levels (i.e., an accepting mode)~\cite{lin1968effect}. The parabola on the right represents a mode that modulates the coupling strength between the two levels (i.e., a promoting mode)~\cite{lin1968effect}. The two modes can be coupled to a thermal reservoir of oscillators, denoted `T' [not included in \eqref{eq:H_CI}]. }
\label{fig:conicalIntersections}
\end{figure}
%%%%%%%%%%%%%%%%%%%%%%%%%%%%%%%%%%%%%%%%%%%%%%%%

A quantum-interference effect, known as \textit{geometric phase}, enters the dynamics near CIs~\cite{ryabinkin2017geometric, joubert2013geometric, li2017geometric}. When a wave packet evolves, a time-independent phase accumulates. The phase is path-independent unless a CI is present. When a CI is present, the difference in phase between two pathways depends on the solid angle enclosed by the paths around the CI. This phase can have important effects on the chemical reactions of molecules, even when the nuclei involved have energies well below the energy of the intersection itself \cite{kendrick2015geometric, yuan2018observation}.

Consider a model consisting of two electronic states $\ket{D}$ and $\ket{A}$ and two bath modes, labeled $x$ and $z$:
%Such a model is associated with the Jahn-Teller effect \cite{larson2020conical}. 
\begin{equation}\label{eq:H_CI}
    \hat{H}_{\rm CI}/\hslash = \kappa_{x} (\ket{D}\bra{A} + \ket{A}\bra{D}) 
    (\hat{a}_x + \hat{a}_x^{\dagger}) + \kappa_{z}(\ket{D}\bra{D} - \ket{A}\bra{A})
    (\hat{a}_z + \hat{a}_z^{\dagger})
    + \nu_x \hat{a}_x^\dagger \hat{a}_x
    + \nu_z \hat{a}_z^\dagger \hat{a}_z.
\end{equation}
The $x$ mode influences the coupling between the electronic states, and the $z$ mode influences the energy splitting between the electronic states~\cite{lin1968effect}. This Hamiltonian is a simple example of a LVCM in \eqref{eq:H_mol}.

Using the Born-Oppenheimer approximation and parameterizing the eigenvector of the electronic subspace in terms of classical nuclear coordinates $x$, $z$, $p_x$, and $p_z$, where $\xi=\frac{1}{\sqrt{2}}(\hat{a}_\xi + \hat{a}_\xi^{\dagger})$ and $p_\xi=\frac{-i}{\sqrt{2}}(\hat{a}_\xi - \hat{a}_\xi^{\dagger})$ for $\xi \in \{x, z\}$, the two eigenenergies are $E_{\pm} = \frac{\nu_x}{2}(x^2 + p_x^2) + \frac{\nu_z}{2}(z^2 + p_z^2) \pm \sqrt{2\kappa_x^2 x^2 + 2\kappa_z^2 z^2}$. This choice of basis is referred to as an adiabatic representation, and the associated nuclear potential energies produce a surface for the system to move on adiabatically. For the case of symmetric mode frequencies and couplings ($\nu_x = \nu_z$ and $\kappa_x = \kappa_z$), the two potential-energy surfaces for the donor and acceptor states form a CI at the center, which is surrounded by a space of degenerate ground states, as shown in Fig.~2a. The existence of a CI invalidates the Born-Oppenheimer approximation, necessitating a quantum treatment of the full (electronic plus nuclear) dynamics. 

Observing the geometric phase, a signature of quantum interference in molecular dynamics, by simulations of \eqref{eq:H_CI} with trapped ions was recently proposed~\cite{gambetta2021exploring, MacDonell21} and demonstrated experimentally~\cite{whitlow2023quantum, valahu2023direct}. We note that Ref.~\citenum{whitlow2023quantum} uses Trotterization (similar to the scheme described in the previous section), while Ref.~\citenum{valahu2023direct} uses an approach where the interactions corresponding to every term in \eqref{eq:H_CI} are driven simultaneously. Another signature, namely, the wavepacket branching due to a CI, was demonstrated experimentally  using a superconducting qubit coupled to electromagnetic cavities~\cite{wang2023observation}. 

Systems with larger numbers of electronic and vibrational states, as well as multiple CIs, represent a next-level challenge. Adding dissipation, or coupling to a thermal reservoir, mimics more realistic chemical systems of interest, as described in Fig.~2b. While such condensed-phase modeling is beyond the LVCM that has a finite number of bath modes, it provides a potentially more realistic model than the LVCM for describing quantum molecular dynamics. Possible methods of simulating dissipative bath modes are introduced in the Outlook section below. Experimentally isolating and observing the influence of CIs on molecular systems is extremely challenging, so employing analog quantum simulators for simulating models beyond the reach of classical-digital methods may address this challenge.

\subsection*{Vibrationally-assisted energy transfer}

Energy transfer in molecular systems, such as biological  light-harvesting complexes, occurs in the presence of coupling among chromophores and a bath, defined by protein, solvent, counter-ions, and other species that may be present. The spectral properties of the bath modes, and their coupling strengths, influence the  energy-transfer dynamics \cite{Huelga13, Adolphs06, Chin13, Irish14, NalBach15, Fujihashi15}. Extensive research has addressed how quantum coherence  and vibronic dynamics may influence energy-transfer efficiencies in photosynthetic systems \cite{Engel07, Christensson12, Duan17, Womick11, fuller2014vibronic, plenio2013origin, Wang19, Cao20, Tiwari13, killoran2015enhancing}. Understanding the influence of vibrations on quantum coherence and energy transfer may require the study of dynamics in systems where both the electronic and vibrational states are quantized, as in the LVCM.

%%%%%%%%%%%%%%%%%%%%FIGURE 3%%%%%%%%%%%%%%%%%%%%%%%
\begin{figure}[h!]
\centering
\includegraphics[width=0.9\linewidth]{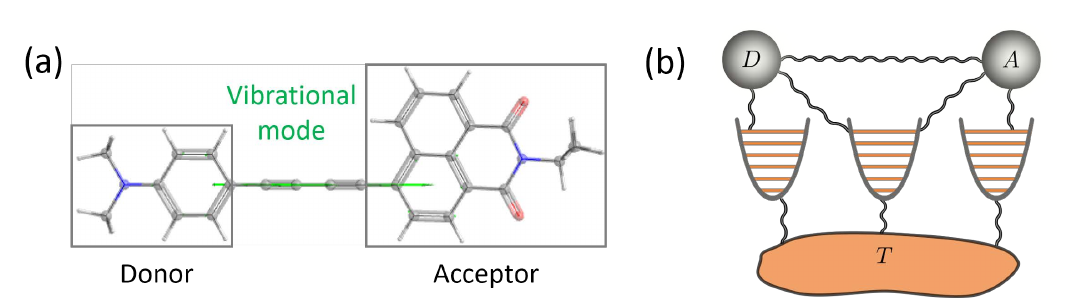}
\caption{(a) Illustration of a molecular donor-acceptor energy transfer system with intramolecular vibrations. (b) A schematic representation of the vibrationally assisted energy-transfer model [see (\ref{eq:H_VAET})]. Each grey circle represents an electronic state and each parabola represents a vibrational mode. The vibrational modes can be coupled to a thermal reservoir of oscillators, denoted `T' [not included in (\ref{eq:H_VAET})]. }
\label{fig:vibrationalET}
\end{figure}
%%%%%%%%%%%%%%%%%%%%%%%%%%%%%%%%%%%%%%%%%%%%%%%%%%

We now provide an example of a simplified version of the LVCM, which can be readily simulated  with trapped-ion systems. The model consists of two electronic states and three bath modes. The Hamiltonian (see Fig.~3) is
\begin{equation}
    \hat{H}_{\rm VAET}/\hslash = (E_A - E_D) |A \rangle \langle A|
    + \frac{\Delta}{2}(|D\rangle \langle A| + |A\rangle \langle D|)
    + \sum_{k=1}^2 \kappa_{D,k} |D \rangle \langle D| (\hat{a}_k + \hat{a}^\dagger_k)
    + \sum_{k=2}^3 \kappa_{A,k} |A \rangle \langle A| (\hat{a}_k + \hat{a}^\dagger_k)
    + \sum_{k=1}^3 \nu_k \hat{a}^\dagger_k \hat{a}_k,
\label{eq:H_VAET}
\end{equation}
 where $E_i$ is the energy of state $i$, $\Delta$ is the coupling strength between the two states, $\nu_{k}$ is the frequency of mode $k$, and $\kappa_{i,k}$ is the coupling strength between occupied state $i$ and mode $k$. Mode 2 is coupled to \textit{both} states, and is thus a correlated or anti-correlated mode that represents intramolecular vibrations, depending on the relative signs of $\kappa_{D,2}$ and $\kappa_{A,2}$.

One reason to simulate (\ref{eq:H_VAET}) is to determine the relationship between the rate of electronic-energy transfer and the bath-mode frequencies. Certain combinations of $\nu_k$'s may produce constructive (destructive) interference, leading to an increase (decrease) in the transfer rate~\cite{Gorman18, Li21}. Also, the correlated or anti-correlated nature of vibrational modes can profoundly influence the chemical dynamics and reactivity~\cite{Fassioli10, Sarovar11, Tiwari13, Uchiyama18, Li21}. 

Experimental progress has already been made towards analog quantum simulations of models similar to \eqref{eq:H_VAET}, but in limited regimes. A single-mode version of (\ref{eq:H_VAET}) (with only mode 3) has been simulated on a two-ion chain, where the relationship between the energy-transfer probability and the vibrational mode frequency is shown~\cite{Gorman18}. A similar model is simulated on a single ion to generate the molecular vibronic spectra~\cite{macdonell2023predicting}. We note that this approach of using time-dynamics simulation for the task of generating the vibronic spectra can be compared to the previous approach of using boson sampling for the same task, which is also experimentally demonstrated on trapped ions~\cite{shen2018quantum}. 

A similar energy-transfer model where electronic states are coupled to a structured semiclassical bath~\cite{rebentrost2009environment} was also simulated using many qubits in various platforms, including trapped ions \cite{Maier19}, superconducting qubits \cite{Potovcnik18}, and NMR qubits \cite{Wang18}. In these experiments, interactions between the electronic states and the bath are simulated by noise injected to the qubits, rather than by interactions between the qubits and the quantum harmonic oscillators. This resembles classical-digital simulation methods that treat the effects of \textit{low-frequency} bath modes (e.g., $\hslash \nu_k \lesssim k_B T_{\rm room}$) as time-dependent fluctuations~\cite{berkelbach2012reduced} or random offsets~\cite{montoya2015extending} to the electronic energies. The method of adding noise to the qubits can be combined with the trapped-ion simulations of LVCMs to capture the interplay between the effects of low-frequency (semiclassical) and high-frequency (quantum) bath modes~\cite{Li22}.  

To move beyond the simulations of VAET in the limited regimes mentioned above, the primary challenge is to achieve full quantum simulations of models with larger numbers of electronic states and quantum vibrational oscillators, so that the model begins to describe some of the complexities of molecular energy-harvesting complexes, such as the Fenna–Matthews–Olson complex~\cite{Cao20}. For a more realistic description of such molecular systems, the vibrational oscillators can be considered to be dissipative and coupled to a thermal reservoir, so that the energy can flow into the reservoir instead of remaining in the electronic states or a few reaction coordinates.

\subsection*{Polarized-light-induced electron transfer}

We now consider a molecular system where an electron may tunnel from one part (atom or molecule) of the system to another. Such electron-transfer phenomenon is crucial to understanding catalysis in biological and chemical systems~\cite{etmarcus, etdb}. Quantum effects may play a central role in the dynamics of the electron transfer~\cite{prytkova2007coupling}. The underlying mechanism can be approximately described as linear couplings between the electronic states~\cite{skourtis2004inelastic, goldsmith2006electron, zarea2013decoherence, spiroscpl, xiao2009turning}, with potential additional coupling to the bath modes~\cite{skourtis2004inelastic, xiao2009turning, lin2009modulating}, similarly to the energy-transfer model above. 

Specifically, we consider as an example a donor-acceptor system, where an electron may hop between the donor and the acceptor when the donor is in one of its excited states. Multiple excited states of the donor lead to multiple pathways of electron transfer, and the possibility of quantum interference between these pathways. When the excited states of the donor are near degenerate, specific superpositions of the excited states may be prepared for electron transfer by controlling the polarization of light that excites the donor. These controllable superposition states produce interferences that can influence the overall electron-transfer rate from the donor to the acceptor~\cite{spiroscpl}.

%%%%%%%%%%%%%%%%%%%%FIGURE 4%%%%%%%%%%%%%%%%%%%%%%%
\begin{figure}[h!]
\centering
\includegraphics[width=0.95\linewidth]{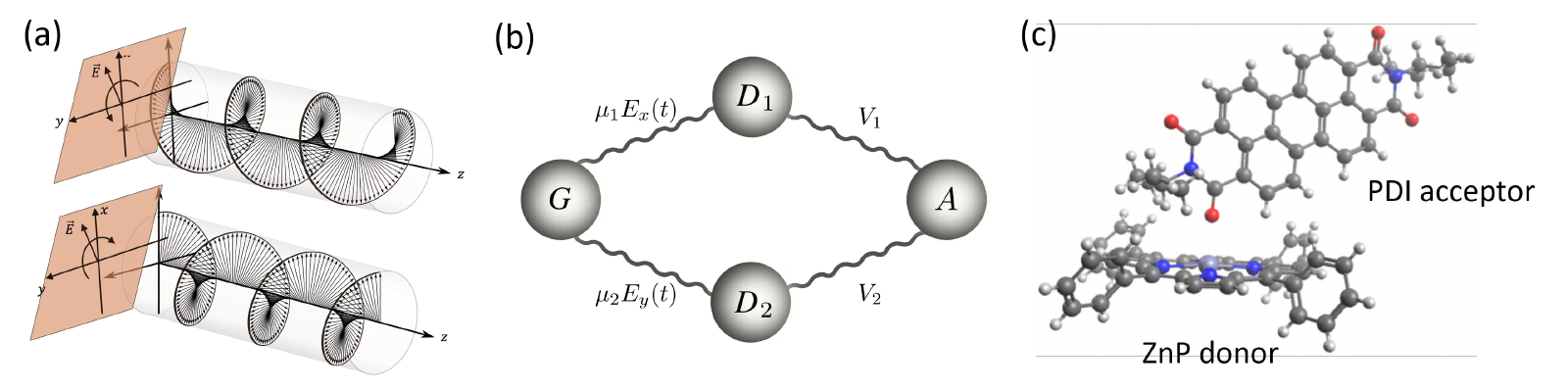}
\caption{(a) An illustration of left- (top) and right-polarized (bottom) light, to be used to excite a molecular electron donor-acceptor system. (b) A schematic representation of the polarized-light-induced electron-transfer process [see (\ref{eq:PLET})]. In this example  $\ket{D_1}$ ($\ket{D_2}$) is polarized in the $x$ ($y$) direction, such that the $x$ ($y$) component of the electric field excites $\ket{G}$ to $\ket{D_1}$ ($\ket{D_2}$). Electron transfer occurs from $\ket{D_1}$ and $\ket{D_2}$ to $\ket{A}$. Bath modes are not included in this particular model. (c) An example molecular system (ZnP-PDI), where the polarization of the driving light may influence the electron-transfer rate. }
\label{fig:polarizedET}
\end{figure}
%%%%%%%%%%%%%%%%%%%%%%%%%%%%%%%%%%%%%%%%%%%%%%%%%%%%

A simple model that shows the influence of light polarization on the electronic dynamics consists of a ground state $\ket{G}$, two excited states $\ket{D_1}$ and $\ket{D_2}$ in which the electron is localized at the donor, and another excited state $\ket{A}$ in which the electron is localized at the acceptor. The couplings between $\ket{G}$ and $\ket{D_i}$ ($i=1,2$), which represent the photo-excitation processes of the donor molecule, have orthogonal electric-dipole moments $\vec{\mu}_i$ and thus couple differently to the electric field of the driving light. The couplings between $\ket{D_i}$ and $\ket{A}$ represents the electron-transfer processes. The Hamiltonian (see Fig.~4) is given by
\begin{align}\label{eq:PLET}
\hat{H}_{\rm PLET}(t)/\hslash = \sum_j \omega_j \ket{j}\bra{j}
 + \sum_{i=1}^2 \big[ \vec{\mu}_i \cdot \vec{E}(t) \ket{G}\bra{D_i} + h.c.]
+ \sum_{i=1}^2 \big[V_i \ket{D_i}\bra{A} +  h.c. \big],
\end{align}
where $j = \{G, D_1, D_2, A\}$, $\omega_j$'s are the corresponding energy levels, $\vec{E}(t)$ is the electric field, and $V_i$ is the coupling strength between the state with the donor excited ($\ket{D_i}$) and the state with the electron localized on the acceptor ($\ket{A}$). Bath modes can be straightforwardly added to the Hamiltonian. This model is also a simplified version of the LVCM that can be efficiently simulated using trapped ions. In a recent experiment, the photo-excitation and electron-transfer processes are separately simulated on a trapped-ion qutrit (three-level) system~\cite{sun2023quantum}.

The polarization of the driving light affects the population of the two excited donor states and the relative phase between them. Thus, the rate of electron transfer can be controlled by the light polarization. This quantum-interference effect is likely to be diminished when the electronic states' coherence is lost by coupling to a thermal bath~\cite{skourtis2004inelastic, goldsmith2006electron, zarea2013decoherence}, and may be rich for further exploration. The donor-acceptor model in \eqref{eq:PLET} can also be extended to models of more complex molecular systems with a larger number of excited donor or acceptor states as well as to systems with excitable bridges between the donor and acceptor~\cite{evenson1992effective, skourtis1993two, troisi2003rate, xiao2009turning, lin2009modulating}. Exploring how the driving light and bath modes affect the quantum interference, and eventually the electron-transfer rate in a complex molecular system, is within the LVCM framework and may be a promising avenue for analog trapped-ion simulations. 

%%%%%%%%%%%%%%%%%%%%%%%%%%%%%%%%%%%%%%%%
%%%%%%%%%%%%%%%%%%%%%%%%%%%%%%%%%%%%%%%%
%% Section
%%%%%%%%%%%%%%%%%%%%%%%%%%%%%%%%%%%%%%%%
%%%%%%%%%%%%%%%%%%%%%%%%%%%%%%%%%%%%%%%%

\section*{Outlook: beyond the LVCM}

The LVCM offers a simplified framework to investigate molecular quantum dynamics. It is important to note that the LVCM class of Hamiltonians is simple, as (i) it contains only a finite number of vibrational modes that are (ii) harmonic and (iii) only linearly coupled to the electronic states. As a consequence of these simplifications, the LVCM may fail to capture several key aspects of real molecular systems. 

For example, the primary vibrational modes that play a crucial role in intermolecular energy transfer are themselves subject to energy dissipation processes, which may arise through interactions with solvent or other vibrational modes. One approach to incorporate dissipation of this kind into simulations is to use a model where the modes that represent nuclear motion are dampened by an external reservoir of modes. This external reservoir could represent the solvent and other molecular modes that are orthogonal to the reaction coordinate~\cite{Garg85}. 

In some approximations~\cite{Caldeira83,Garg85,sun2016exact,Lemmer18} or analytical regimes~\cite{Tamascelli18, Mascherpa20}, a bath consisting of a finite number of damped modes may have the same effects on electronic energy transfer with a bath of infinitely many undamped modes with a continuous (e.g., sum of Lorentzian) spectral density. A bath of continuously distributed modes is different from that described in the LVCM, but is also used widely to study molecule-bath interactions~\cite{Caldeira83,Garg85,onuchic1986some,onuchic1987effect,caldeira1981influence,leggett1984quantum,leggett1987dynamics}. For trapped ions, directly simulating a finite number of damped modes is more feasible \cite{MacDonell21, Lemmer18, Schlawin21} than simulating infinitely many undamped modes, as discussed in the next subsection. Learning appropriate finite-bath approximations for chemical dynamics~\cite{Lemmer18, Tamascelli18, Mascherpa20} (or developing approaches to include effectively infinite baths) will be required for the broad applicability of trapped-ion simulations for condensed-phase chemical dynamics. 

Aside from the limitations of the LVCM in describing some aspects of chemical dynamics, models beyond the LVCM might provide a compelling target to achieve a quantum advantage in simulating molecular dynamics, as the LVCM is a favorable framework for classical-digital simulation. The conceptual and mathematical simplicity of \eqref{eq:H_mol}, and decades of research into the quantum dynamics of molecules, have led to sophisticated approximation schemes that walk a tightrope between accuracy and computational effort~\cite{kundu2022intramolecular,tanimura2020numerically,zhao2022hierarchy,ye2016heom,yan2020new,gelin2022equation,ren2022time,weimer2021simulation,de2017dynamics}. There are fundamental reasons to suggest that dynamics generated by the LVCM are particularly well suited for quantum-classical~\cite{kapral2015quantum} or semiclassical~\cite{thoss2004semiclassical,lee2016semiclassical} analysis using classical computers, as discussed in the next paragraph. Thus, building quantum simulators for models beyond the LVCM may lead to a more immediate quantum advantage, as cutting-edge classical-digital algorithms often rely on the effectiveness of approximations that seem especially suited for the LVCM. Simulating nonlinear system-bath couplings and anharmonic bath modes using trapped ions is promising, although this may require operations that are more challenging to implement than the ones described by \eqref{eq:H0}-\eqref{eq:H2} (see the next subsection).

Two results illustrate the relative effectiveness of quantum-classical and semiclassical approximations for simulating the LVCM. First, the quantum-classical Liouville equations \cite{kapral1999mixed} are used to approximate open-system quantum dynamics. This approach becomes exact for the linearly-coupled vibronic dynamics of \eqref{eq:H_mol} \cite{mac2002surface}.
Second, more generally in semiclassical theory, the Feynman-Vernon influence functional plays a vital role \cite{makri1991feynman,feynman1963,shi2004derivation,may2011charge}. This functional encodes the influence of a bath on the dynamics of the system of interest as a function of the trajectory of the system of interest only~\cite{feynman1963}, including any influence that encodes memory of previous states of the system of interest. Remarkably, the form of this functional is analytic in the limit of a linearly-coupled harmonic bath in \eqref{eq:H_mol} \cite{feynman1948space, may2011charge}. This analytic form may then be used instead of numerically calculating the functional. While the quantum-classical Liouville equations and semiclassical approximations based on the Feynman-Vernon influence functional~\cite{TOPALER1993285} remain difficult to solve, they are computationally less expensive than direct propagation of the time-dependent Schrodinger equation for simulating the LVCM. This suggests that the LVCMs may not be the best target for attempts to achieve a quantum advantage using analog quantum simulations, as the LVCMs constitute a regime of exceptional classical-digital performance.

\subsection*{Analog trapped-ion simulation}

There are at least three ways to go beyond the LVCM to describe molecular systems more realistically, and these strategies can make classical-digital simulation methods less tractable (as described in the previous section). First, dissipation of the bath modes may be added. Second, the coupling between electronic states and bath modes may be of higher order, involving terms proportional to $\hat{\psi}^\dagger_i \hat{\psi}_j (\hat{a}_k + \hat{a}_k^\dagger)^n$ where $n \geq 2$ ($n=2$ corresponds to the quadratic vibronic-coupling model \cite{MacDonell21, zobel2021surface}). Third, the bath modes may be  anharmonic (fermionic bath modes may exhibit extreme anharmonicity, accommodating only a finite number of states, besides their different statistics from bosons), involving terms proportional to $(\hat{a}_k^\dagger)^n (\hat{a}_k)^n$ where $n \geq 2$. 

Trapped ions can be used to simulate all three cases above. First, damping (phonon loss) or heating (phonon gain) of the bath modes can be simulated. Phonon loss can be simulated using laser cooling of the trapped ions' motional modes, such as sideband cooling and electromagnetically induced transparency cooling \cite{Lemmer18, MacDonell21, Schlawin21}. Laser cooling can be understood as engineered dissipation~\cite{Harrington22} of the motional modes. The rate of cooling and the final temperature after cooling can be controlled~\cite{Monroe95}. By interleaving cooling with the usual operations that, by themselves, simulate coherent dynamics of the LVCM, damping of the bath modes during molecular dynamics may be simulated. This may require trapping two kinds of ions, one dedicated to representing electronic states and another dedicated to cooling the motional modes~\cite{Larson86, Blinov02, Barrett03, negnevitsky2018repeated}. 

Phonon gain also can be similarly simulated by intentionally heating the motional modes, which also requires two kinds of ions~\cite{Clark10}. Alternatively, phonon gain can be simulated by averaging over many instances of random stochastic operations interleaved with the usual operations for simulating the LVCM. Here, the random stochastic operations are described by (\ref{eq:H1}) with varying amplitude ($\tilde{\Omega}$), phase ($\varphi$), and/or detuning ($\mu$) of the laser, which is applied to an auxiliary ion that shares the motional mode. The rate of phonon gain determines the range over which the laser parameters are randomly drawn. This method relies on the fact that evolution of an open quantum system with respect to the Lindblad master equation can be realized by an average of many evolutions of closed systems, each consisting of stochastic Hamiltonian terms~\cite{Cai13, Chenu17}. Trapping two kinds of ions is not required with this alternative method. Instead, averaging over a large number of analog-simulation runs may be required in order to achieve a sufficiently low uncertainty in simulating the electronic-state population evolution~\cite{Wang18}.

Second, the second-order coupling between electronic states and bath modes can be simulated~\cite{MacDonell21}. Intuitively, as a spin-dependent force (resonant with the first-order sideband $\omega_0 \pm \omega_k$) can be used in simulating a linear coupling that is proportional to $\hat{\psi}^\dagger_i \hat{\psi}_j (\hat{a}_k + \hat{a}_k^\dagger)$, a force resonant with the second-order sideband $\omega_0 \pm \omega_k \pm [\mp] \omega_l$ can be used to simulate a second-order coupling that is proportional to $\hat{\psi}^\dagger_i \hat{\psi}_j (\hat{a}_k \hat{a}_l + \hat{a}_k^\dagger \hat{a}_l^\dagger)$ [$\hat{\psi}^\dagger_i \hat{\psi}_j (\hat{a}_k \hat{a}^\dagger_l + \hat{a}_k^\dagger \hat{a}_l)$]. This was demonstrated in several experiments~\cite{marshall2017linear, shen2018quantum, gan2020hybrid, chen2021quantum, nguyen2021experimental, chen2023scalable}. A protocol for simultaneously simulating multiple second-order coupling terms is shown in Ref.~\citenum{Katz23}. The challenge in experimental implementations is that the second-order sideband interactions are typically an order of magnitude weaker than the first-order interactions. Thus, longer coherence times of the experimental system may be required to simulate the dynamics with second-order vibronic couplings accurately. Similar arguments apply to simulating higher-order vibronic couplings. 

Third, the anharmonicity of the bath modes can be simulated. Consider, for example, a model where the anharmonicity  $\sum_{k=1}^N (\chi_k / 2) \hat{a}^\dagger_k \hat{a}^\dagger_k \hat{a}_k \hat{a}_k$ is added to \eqref{eq:H_mol}, where $\chi_k$ is the anharmonicity of mode $k$. When $\chi_k$ is small, one can simulate the Hamiltonian in the interaction picture where the anharmonic term is rotated out and removed, leaving a transformation $\hat{a}_k \rightarrow \hat{a}_k - i \chi_k t  \hat{a}^\dagger_k \hat{a}_k \hat{a}_k$ that is valid up to first order in $\chi_k$. This leads to adding weak third-order couplings, which can be simulated as described in the previous paragraph. A general method of simulating anharmonic modes, including the case where the anharmonicity is large, using the motional modes of trapped ions, remains open to further investigation.

%%%%%%%%%%%%%%%%%%%%%%%%%%%%%%%%%%%%%%%%
%%%%%%%%%%%%%%%%%%%%%%%%%%%%%%%%%%%%%%%%
%% Section
%%%%%%%%%%%%%%%%%%%%%%%%%%%%%%%%%%%%%%%%
%%%%%%%%%%%%%%%%%%%%%%%%%%%%%%%%%%%%%%%%

\section*{Conclusions}

In order to identify quantum advantages in simulating molecular quantum dynamics, it is essential to understand the capabilities of both analog quantum simulations on noisy devices and classical-digital algorithms; in particular, the trade off between the accuracy and computational cost of classical-digital algorithms needs to be analyzed. Using a simplified model Hamiltonian based on linear vibronic couplings, we suggest that analog trapped-ion simulations may have an advantage over classical-digital algorithms, in terms of accuracy and computational cost, in an intermediate regime of coupling strength between the electronic states and bath modes. We present three candidate models relevant to chemical phenomena that can be simulated using current or near-term trapped-ion hardware. In order to achieve a quantum advantage, these models will likely need to be extended to larger sizes, where classical-digital simulations are prohibitively expensive. LVCMs with complex connectivity between the electronic states and bath modes, or models where the bath modes are themselves dissipative, are of particular interest. Quantum advantages may also be achieved in models with nonlinear system-bath couplings and anharmonic bath modes, features that semiclassical or quantum-classical approximations struggle to treat accurately. 

This Review is intended to inspire collaboration between the communities of quantum-chemical theory and analog quantum simulation. Analog quantum simulation may serve as a catalyst for advancing our understanding of complex chemical dynamics, and may enable  access to elements of molecular realism that be inaccessible in current classical-digital simulations.

%%%%%%%%%%%%%%%%%%%%%%%%%%%%%%%%%%%%%%%%
%%%%%%%%%%%%%%%%%%%%%%%%%%%%%%%%%%%%%%%%
%% End of text
%%%%%%%%%%%%%%%%%%%%%%%%%%%%%%%%%%%%%%%%
%%%%%%%%%%%%%%%%%%%%%%%%%%%%%%%%%%%%%%%%

\bibliography{sample}  

\begin{thebibliography}{100}
\urlstyle{rm}
\expandafter\ifx\csname url\endcsname\relax
  \def\url#1{\texttt{#1}}\fi
\expandafter\ifx\csname urlprefix\endcsname\relax\def\urlprefix{URL }\fi
\expandafter\ifx\csname doiprefix\endcsname\relax\def\doiprefix{DOI: }\fi
\providecommand{\bibinfo}[2]{#2}
\providecommand{\eprint}[2][]{\url{#2}}

\bibitem{park2020multireference}
\bibinfo{author}{Park, J.~W.}, \bibinfo{author}{Al-Saadon, R.}, \bibinfo{author}{MacLeod, M.~K.}, \bibinfo{author}{Shiozaki, T.} \& \bibinfo{author}{Vlaisavljevich, B.}
\newblock \bibinfo{journal}{\bibinfo{title}{Multireference electron correlation methods: Journeys along potential energy surfaces}}.
\newblock {\emph{\JournalTitle{Chem. Rev.}}} \textbf{\bibinfo{volume}{120}}, \bibinfo{pages}{5878--5909} (\bibinfo{year}{2020}).

\bibitem{park2017fly}
\bibinfo{author}{Park, J.~W.} \& \bibinfo{author}{Shiozaki, T.}
\newblock \bibinfo{journal}{\bibinfo{title}{On-the-fly {CASPT2} surface-hopping dynamics}}.
\newblock {\emph{\JournalTitle{J. Chem. Theory Comput.}}} \textbf{\bibinfo{volume}{13}}, \bibinfo{pages}{3676--3683} (\bibinfo{year}{2017}).

\bibitem{larsson2022chromium}
\bibinfo{author}{Larsson, H.~R.}, \bibinfo{author}{Zhai, H.}, \bibinfo{author}{Umrigar, C.~J.} \& \bibinfo{author}{Chan, G. K.-L.}
\newblock \bibinfo{journal}{\bibinfo{title}{The chromium dimer: closing a chapter of quantum chemistry}}.
\newblock {\emph{\JournalTitle{J. Am. Chem. Soc.}}} \textbf{\bibinfo{volume}{144}}, \bibinfo{pages}{15932--15937} (\bibinfo{year}{2022}).

\bibitem{Cirac12}
\bibinfo{author}{Cirac, J.~I.} \& \bibinfo{author}{Zoller, P.}
\newblock \bibinfo{journal}{\bibinfo{title}{Goals and opportunities in quantum simulation}}.
\newblock {\emph{\JournalTitle{Nat. Phys.}}} \textbf{\bibinfo{volume}{8}}, \bibinfo{pages}{264--266} (\bibinfo{year}{2012}).

\bibitem{Georgescu14}
\bibinfo{author}{Georgescu, I.~M.}, \bibinfo{author}{Ashhab, S.} \& \bibinfo{author}{Nori, F.}
\newblock \bibinfo{journal}{\bibinfo{title}{Quantum simulation}}.
\newblock {\emph{\JournalTitle{Rev. Mod. Phys.}}} \textbf{\bibinfo{volume}{86}}, \bibinfo{pages}{153} (\bibinfo{year}{2014}).

\bibitem{Daley22}
\bibinfo{author}{Daley, A.~J.} \emph{et~al.}
\newblock \bibinfo{journal}{\bibinfo{title}{Practical quantum advantage in quantum simulation}}.
\newblock {\emph{\JournalTitle{Nature}}} \textbf{\bibinfo{volume}{607}}, \bibinfo{pages}{667--676} (\bibinfo{year}{2022}).

\bibitem{Alexeev21}
\bibinfo{author}{Alexeev, Y.} \emph{et~al.}
\newblock \bibinfo{journal}{\bibinfo{title}{Quantum computer systems for scientific discovery}}.
\newblock {\emph{\JournalTitle{PRX Quantum}}} \textbf{\bibinfo{volume}{2}}, \bibinfo{pages}{017001} (\bibinfo{year}{2021}).

\bibitem{lee2023evaluating}
\bibinfo{author}{Lee, S.} \emph{et~al.}
\newblock \bibinfo{journal}{\bibinfo{title}{Evaluating the evidence for exponential quantum advantage in ground-state quantum chemistry}}.
\newblock {\emph{\JournalTitle{Nat. Commun.}}} \textbf{\bibinfo{volume}{14}}, \bibinfo{pages}{1952} (\bibinfo{year}{2023}).

\bibitem{Kassal08}
\bibinfo{author}{Kassal, I.}, \bibinfo{author}{Jordan, S.~P.}, \bibinfo{author}{Love, P.~J.}, \bibinfo{author}{Mohseni, M.} \& \bibinfo{author}{Aspuru-Guzik, A.}
\newblock \bibinfo{journal}{\bibinfo{title}{Polynomial-time quantum algorithm for the simulation of chemical dynamics}}.
\newblock {\emph{\JournalTitle{Proc. Natl. Acad. Sci.}}} \textbf{\bibinfo{volume}{105}}, \bibinfo{pages}{18681--18686} (\bibinfo{year}{2008}).

\bibitem{Sawaya20}
\bibinfo{author}{Sawaya, N.~P.} \emph{et~al.}
\newblock \bibinfo{journal}{\bibinfo{title}{Resource-efficient digital quantum simulation of $d$-level systems for photonic, vibrational, and spin-$s$ hamiltonians}}.
\newblock {\emph{\JournalTitle{npj Quantum Inf.}}} \textbf{\bibinfo{volume}{6}}, \bibinfo{pages}{1--13} (\bibinfo{year}{2020}).

\bibitem{Jahangiri20}
\bibinfo{author}{Jahangiri, S.}, \bibinfo{author}{Arrazola, J.~M.}, \bibinfo{author}{Quesada, N.} \& \bibinfo{author}{Delgado, A.}
\newblock \bibinfo{journal}{\bibinfo{title}{Quantum algorithm for simulating molecular vibrational excitations}}.
\newblock {\emph{\JournalTitle{Phys. Chem. Chem. Phys.}}} \textbf{\bibinfo{volume}{22}}, \bibinfo{pages}{25528--25537} (\bibinfo{year}{2020}).

\bibitem{MacDonell21}
\bibinfo{author}{MacDonell, R.~J.} \emph{et~al.}
\newblock \bibinfo{journal}{\bibinfo{title}{Analog quantum simulation of chemical dynamics}}.
\newblock {\emph{\JournalTitle{Chem. Sci.}}} \textbf{\bibinfo{volume}{12}}, \bibinfo{pages}{9794--9805} (\bibinfo{year}{2021}).

\bibitem{Saha21}
\bibinfo{author}{Saha, D.}, \bibinfo{author}{Iyengar, S.~S.}, \bibinfo{author}{Richerme, P.}, \bibinfo{author}{Smith, J.~M.} \& \bibinfo{author}{Sabry, A.}
\newblock \bibinfo{journal}{\bibinfo{title}{Mapping quantum chemical dynamics problems to spin-lattice simulators}}.
\newblock {\emph{\JournalTitle{J. Chem. Theory Comput.}}} \textbf{\bibinfo{volume}{17}}, \bibinfo{pages}{6713--6732} (\bibinfo{year}{2021}).

\bibitem{Kitaev96}
\bibinfo{author}{Kitaev, A.~Y.}
\newblock \bibinfo{journal}{\bibinfo{title}{Quantum measurements and the {Abelian Stabilizer Problem}}}.
\newblock {\emph{\JournalTitle{Electron. Colloquium Comput. Complex.}}} \textbf{\bibinfo{volume}{3}} (\bibinfo{year}{1996}).

\bibitem{Aspuru05}
\bibinfo{author}{Aspuru-Guzik, A.}, \bibinfo{author}{Dutoi, A.~D.}, \bibinfo{author}{Love, P.~J.} \& \bibinfo{author}{Head-Gordon, M.}
\newblock \bibinfo{journal}{\bibinfo{title}{Simulated quantum computation of molecular energies}}.
\newblock {\emph{\JournalTitle{Science}}} \textbf{\bibinfo{volume}{309}}, \bibinfo{pages}{1704--1707} (\bibinfo{year}{2005}).

\bibitem{Lanyon10}
\bibinfo{author}{Lanyon, B.~P.} \emph{et~al.}
\newblock \bibinfo{journal}{\bibinfo{title}{Towards quantum chemistry on a quantum computer}}.
\newblock {\emph{\JournalTitle{Nat. Chem.}}} \textbf{\bibinfo{volume}{2}}, \bibinfo{pages}{106--111} (\bibinfo{year}{2010}).

\bibitem{Whitfield11}
\bibinfo{author}{Whitfield, J.~D.}, \bibinfo{author}{Biamonte, J.} \& \bibinfo{author}{Aspuru-Guzik, A.}
\newblock \bibinfo{journal}{\bibinfo{title}{Simulation of electronic structure hamiltonians using quantum computers}}.
\newblock {\emph{\JournalTitle{Mol. Phys.}}} \textbf{\bibinfo{volume}{109}}, \bibinfo{pages}{735--750} (\bibinfo{year}{2011}).

\bibitem{Peruzzo14}
\bibinfo{author}{Peruzzo, A.} \emph{et~al.}
\newblock \bibinfo{journal}{\bibinfo{title}{A variational eigenvalue solver on a photonic quantum processor}}.
\newblock {\emph{\JournalTitle{Nat. Commun.}}} \textbf{\bibinfo{volume}{5}}, \bibinfo{pages}{1--7} (\bibinfo{year}{2014}).

\bibitem{OMalley16}
\bibinfo{author}{O’Malley, P.~J.} \emph{et~al.}
\newblock \bibinfo{journal}{\bibinfo{title}{Scalable quantum simulation of molecular energies}}.
\newblock {\emph{\JournalTitle{Phys. Rev. X}}} \textbf{\bibinfo{volume}{6}}, \bibinfo{pages}{031007} (\bibinfo{year}{2016}).

\bibitem{Kandala17}
\bibinfo{author}{Kandala, A.} \emph{et~al.}
\newblock \bibinfo{journal}{\bibinfo{title}{Hardware-efficient variational quantum eigensolver for small molecules and quantum magnets}}.
\newblock {\emph{\JournalTitle{Nature}}} \textbf{\bibinfo{volume}{549}}, \bibinfo{pages}{242--246} (\bibinfo{year}{2017}).

\bibitem{Nam20}
\bibinfo{author}{Nam, Y.} \emph{et~al.}
\newblock \bibinfo{journal}{\bibinfo{title}{Ground-state energy estimation of the water molecule on a trapped-ion quantum computer}}.
\newblock {\emph{\JournalTitle{npj Quantum Inf.}}} \textbf{\bibinfo{volume}{6}}, \bibinfo{pages}{1--6} (\bibinfo{year}{2020}).

\bibitem{Wang19}
\bibinfo{author}{Wang, L.}, \bibinfo{author}{Allodi, M.~A.} \& \bibinfo{author}{Engel, G.~S.}
\newblock \bibinfo{journal}{\bibinfo{title}{Quantum coherences reveal excited-state dynamics in biophysical systems}}.
\newblock {\emph{\JournalTitle{Nat. Rev. Chem.}}} \textbf{\bibinfo{volume}{3}}, \bibinfo{pages}{477--490} (\bibinfo{year}{2019}).

\bibitem{Cao20}
\bibinfo{author}{Cao, J.} \emph{et~al.}
\newblock \bibinfo{journal}{\bibinfo{title}{Quantum biology revisited}}.
\newblock {\emph{\JournalTitle{Sci. Adv.}}} \textbf{\bibinfo{volume}{6}}, \bibinfo{pages}{eaaz4888} (\bibinfo{year}{2020}).

\bibitem{hammes2008proton}
\bibinfo{author}{Hammes-Schiffer, S.} \& \bibinfo{author}{Soudackov, A.~V.}
\newblock \bibinfo{journal}{\bibinfo{title}{Proton-coupled electron transfer in solution, proteins, and electrochemistry}}.
\newblock {\emph{\JournalTitle{J. Phys. Chem. B}}} \textbf{\bibinfo{volume}{112}}, \bibinfo{pages}{14108--14123} (\bibinfo{year}{2008}).

\bibitem{hammes2015proton}
\bibinfo{author}{Hammes-Schiffer, S.}
\newblock \bibinfo{journal}{\bibinfo{title}{Proton-coupled electron transfer: Moving together and charging forward}}.
\newblock {\emph{\JournalTitle{J. Am. Chem. Soc.}}} \textbf{\bibinfo{volume}{137}}, \bibinfo{pages}{8860--8871} (\bibinfo{year}{2015}).

\bibitem{Reiher17}
\bibinfo{author}{Reiher, M.}, \bibinfo{author}{Wiebe, N.}, \bibinfo{author}{Svore, K.~M.}, \bibinfo{author}{Wecker, D.} \& \bibinfo{author}{Troyer, M.}
\newblock \bibinfo{journal}{\bibinfo{title}{Elucidating reaction mechanisms on quantum computers}}.
\newblock {\emph{\JournalTitle{Proc. Natl. Acad. Sci.}}} \textbf{\bibinfo{volume}{114}}, \bibinfo{pages}{7555--7560} (\bibinfo{year}{2017}).

\bibitem{Babbush18}
\bibinfo{author}{Babbush, R.} \emph{et~al.}
\newblock \bibinfo{journal}{\bibinfo{title}{Encoding electronic spectra in quantum circuits with linear {T} complexity}}.
\newblock {\emph{\JournalTitle{Phys. Rev. X}}} \textbf{\bibinfo{volume}{8}}, \bibinfo{pages}{041015} (\bibinfo{year}{2018}).

\bibitem{Su21}
\bibinfo{author}{Su, Y.}, \bibinfo{author}{Berry, D.~W.}, \bibinfo{author}{Wiebe, N.}, \bibinfo{author}{Rubin, N.} \& \bibinfo{author}{Babbush, R.}
\newblock \bibinfo{journal}{\bibinfo{title}{Fault-tolerant quantum simulations of chemistry in first quantization}}.
\newblock {\emph{\JournalTitle{PRX Quantum}}} \textbf{\bibinfo{volume}{2}}, \bibinfo{pages}{040332} (\bibinfo{year}{2021}).

\bibitem{Kim22}
\bibinfo{author}{Kim, I.~H.} \emph{et~al.}
\newblock \bibinfo{journal}{\bibinfo{title}{Fault-tolerant resource estimate for quantum chemical simulations: Case study on {Li}-ion battery electrolyte molecules}}.
\newblock {\emph{\JournalTitle{Phys. Rev. Res.}}} \textbf{\bibinfo{volume}{4}}, \bibinfo{pages}{023019} (\bibinfo{year}{2022}).

\bibitem{Houck12}
\bibinfo{author}{Houck, A.~A.}, \bibinfo{author}{T{\"u}reci, H.~E.} \& \bibinfo{author}{Koch, J.}
\newblock \bibinfo{journal}{\bibinfo{title}{On-chip quantum simulation with superconducting circuits}}.
\newblock {\emph{\JournalTitle{Nat. Phys.}}} \textbf{\bibinfo{volume}{8}}, \bibinfo{pages}{292--299} (\bibinfo{year}{2012}).

\bibitem{Hartmann16}
\bibinfo{author}{Hartmann, M.~J.}
\newblock \bibinfo{journal}{\bibinfo{title}{Quantum simulation with interacting photons}}.
\newblock {\emph{\JournalTitle{J. Opt.}}} \textbf{\bibinfo{volume}{18}}, \bibinfo{pages}{104005} (\bibinfo{year}{2016}).

\bibitem{Greiner02}
\bibinfo{author}{Greiner, M.}, \bibinfo{author}{Mandel, O.}, \bibinfo{author}{Esslinger, T.}, \bibinfo{author}{H{\"a}nsch, T.~W.} \& \bibinfo{author}{Bloch, I.}
\newblock \bibinfo{journal}{\bibinfo{title}{Quantum phase transition from a superfluid to a mott insulator in a gas of ultracold atoms}}.
\newblock {\emph{\JournalTitle{Nature}}} \textbf{\bibinfo{volume}{415}}, \bibinfo{pages}{39--44} (\bibinfo{year}{2002}).

\bibitem{Gross17}
\bibinfo{author}{Gross, C.} \& \bibinfo{author}{Bloch, I.}
\newblock \bibinfo{journal}{\bibinfo{title}{Quantum simulations with ultracold atoms in optical lattices}}.
\newblock {\emph{\JournalTitle{Science}}} \textbf{\bibinfo{volume}{357}}, \bibinfo{pages}{995--1001} (\bibinfo{year}{2017}).

\bibitem{Aspuru12}
\bibinfo{author}{Aspuru-Guzik, A.} \& \bibinfo{author}{Walther, P.}
\newblock \bibinfo{journal}{\bibinfo{title}{Photonic quantum simulators}}.
\newblock {\emph{\JournalTitle{Nat. Phys.}}} \textbf{\bibinfo{volume}{8}}, \bibinfo{pages}{285--291} (\bibinfo{year}{2012}).

\bibitem{Blatt12}
\bibinfo{author}{Blatt, R.} \& \bibinfo{author}{Roos, C.~F.}
\newblock \bibinfo{journal}{\bibinfo{title}{Quantum simulations with trapped ions}}.
\newblock {\emph{\JournalTitle{Nat. Phys.}}} \textbf{\bibinfo{volume}{8}}, \bibinfo{pages}{277--284} (\bibinfo{year}{2012}).

\bibitem{Monroe21}
\bibinfo{author}{Monroe, C.} \emph{et~al.}
\newblock \bibinfo{journal}{\bibinfo{title}{Programmable quantum simulations of spin systems with trapped ions}}.
\newblock {\emph{\JournalTitle{Rev. Mod. Phys.}}} \textbf{\bibinfo{volume}{93}}, \bibinfo{pages}{025001} (\bibinfo{year}{2021}).

\bibitem{Brown16}
\bibinfo{author}{Brown, K.~R.}, \bibinfo{author}{Kim, J.} \& \bibinfo{author}{Monroe, C.}
\newblock \bibinfo{journal}{\bibinfo{title}{Co-designing a scalable quantum computer with trapped atomic ions}}.
\newblock {\emph{\JournalTitle{npj Quantum Inf.}}} \textbf{\bibinfo{volume}{2}}, \bibinfo{pages}{1--10} (\bibinfo{year}{2016}).

\bibitem{Bruzewicz19}
\bibinfo{author}{Bruzewicz, C.~D.}, \bibinfo{author}{Chiaverini, J.}, \bibinfo{author}{McConnell, R.} \& \bibinfo{author}{Sage, J.~M.}
\newblock \bibinfo{journal}{\bibinfo{title}{Trapped-ion quantum computing: Progress and challenges}}.
\newblock {\emph{\JournalTitle{Appl. Phys. Rev.}}} \textbf{\bibinfo{volume}{6}}, \bibinfo{pages}{021314} (\bibinfo{year}{2019}).

\bibitem{Kienzler16}
\bibinfo{author}{Kienzler, D.} \emph{et~al.}
\newblock \bibinfo{journal}{\bibinfo{title}{Observation of quantum interference between separated mechanical oscillator wave packets}}.
\newblock {\emph{\JournalTitle{Phys. Rev. Lett.}}} \textbf{\bibinfo{volume}{116}}, \bibinfo{pages}{140402} (\bibinfo{year}{2016}).

\bibitem{Um16}
\bibinfo{author}{Um, M.} \emph{et~al.}
\newblock \bibinfo{journal}{\bibinfo{title}{Phonon arithmetic in a trapped ion system}}.
\newblock {\emph{\JournalTitle{Nat. Commun.}}} \textbf{\bibinfo{volume}{7}}, \bibinfo{pages}{1--7} (\bibinfo{year}{2016}).

\bibitem{Zhang18}
\bibinfo{author}{Zhang, J.} \emph{et~al.}
\newblock \bibinfo{journal}{\bibinfo{title}{{NOON} states of nine quantized vibrations in two radial modes of a trapped ion}}.
\newblock {\emph{\JournalTitle{Phys. Rev. Lett.}}} \textbf{\bibinfo{volume}{121}}, \bibinfo{pages}{160502} (\bibinfo{year}{2018}).

\bibitem{Fluhmann19}
\bibinfo{author}{Fl{\"u}hmann, C.} \emph{et~al.}
\newblock \bibinfo{journal}{\bibinfo{title}{Encoding a qubit in a trapped-ion mechanical oscillator}}.
\newblock {\emph{\JournalTitle{Nature}}} \textbf{\bibinfo{volume}{566}}, \bibinfo{pages}{513--517} (\bibinfo{year}{2019}).

\bibitem{Neeve22}
\bibinfo{author}{de~Neeve, B.}, \bibinfo{author}{Nguyen, T.-L.}, \bibinfo{author}{Behrle, T.} \& \bibinfo{author}{Home, J.~P.}
\newblock \bibinfo{journal}{\bibinfo{title}{Error correction of a logical grid state qubit by dissipative pumping}}.
\newblock {\emph{\JournalTitle{Nat. Phys.}}} \textbf{\bibinfo{volume}{18}}, \bibinfo{pages}{296--300} (\bibinfo{year}{2022}).

\bibitem{Jia22}
\bibinfo{author}{Jia, Z.} \emph{et~al.}
\newblock \bibinfo{journal}{\bibinfo{title}{Determination of multimode motional quantum states in a trapped ion system}}.
\newblock {\emph{\JournalTitle{Phys. Rev. Lett.}}} \textbf{\bibinfo{volume}{129}}, \bibinfo{pages}{103602} (\bibinfo{year}{2022}).

\bibitem{Gorman18}
\bibinfo{author}{Gorman, D.~J.} \emph{et~al.}
\newblock \bibinfo{journal}{\bibinfo{title}{Engineering vibrationally assisted energy transfer in a trapped-ion quantum simulator}}.
\newblock {\emph{\JournalTitle{Phys. Rev. X}}} \textbf{\bibinfo{volume}{8}}, \bibinfo{pages}{011038} (\bibinfo{year}{2018}).

\bibitem{karplus1959analog}
\bibinfo{author}{Karplus, W.} \& \bibinfo{author}{Soroka, W.}
\newblock \emph{\bibinfo{title}{Analog Methods: Computation and Simulation}}.
\newblock McGraw-Hill series in engineering sciences (\bibinfo{publisher}{McGraw-Hill}, \bibinfo{year}{1959}).

\bibitem{jackson1960analog}
\bibinfo{author}{Jackson, A.}
\newblock \emph{\bibinfo{title}{Analog Computation}} (\bibinfo{publisher}{McGraw-Hill}, \bibinfo{year}{1960}).

\bibitem{lindblad1976generators}
\bibinfo{author}{Lindblad, G.}
\newblock \bibinfo{journal}{\bibinfo{title}{On the generators of quantum dynamical semigroups}}.
\newblock {\emph{\JournalTitle{Commun. Math. Phys.}}} \textbf{\bibinfo{volume}{48}}, \bibinfo{pages}{119--130} (\bibinfo{year}{1976}).

\bibitem{Wineland98}
\bibinfo{author}{Wineland, D.~J.} \emph{et~al.}
\newblock \bibinfo{journal}{\bibinfo{title}{Experimental issues in coherent quantum-state manipulation of trapped atomic ions}}.
\newblock {\emph{\JournalTitle{J. Res. Natl. Inst. Stan.}}} \textbf{\bibinfo{volume}{103}}, \bibinfo{pages}{259} (\bibinfo{year}{1998}).

\bibitem{Pogorelov21}
\bibinfo{author}{Pogorelov, I.} \emph{et~al.}
\newblock \bibinfo{journal}{\bibinfo{title}{Compact ion-trap quantum computing demonstrator}}.
\newblock {\emph{\JournalTitle{PRX Quantum}}} \textbf{\bibinfo{volume}{2}}, \bibinfo{pages}{020343} (\bibinfo{year}{2021}).

\bibitem{chen2023benchmarking}
\bibinfo{author}{Chen, J.-S.} \emph{et~al.}
\newblock \bibinfo{journal}{\bibinfo{title}{Benchmarking a trapped-ion quantum computer with 29 algorithmic qubits}}.
\newblock {\emph{\JournalTitle{arXiv preprint arXiv:2308.05071}}}  (\bibinfo{year}{2023}).

\bibitem{pino2021demonstration}
\bibinfo{author}{Pino, J.~M.} \emph{et~al.}
\newblock \bibinfo{journal}{\bibinfo{title}{Demonstration of the trapped-ion quantum ccd computer architecture}}.
\newblock {\emph{\JournalTitle{Nature}}} \textbf{\bibinfo{volume}{592}}, \bibinfo{pages}{209--213} (\bibinfo{year}{2021}).

\bibitem{Friesner81}
\bibinfo{author}{Friesner, R.~A.} \& \bibinfo{author}{Silbey, R.}
\newblock \bibinfo{journal}{\bibinfo{title}{Linear vibronic coupling in a general two level system}}.
\newblock {\emph{\JournalTitle{J. Chem. Phys.}}} \textbf{\bibinfo{volume}{75}}, \bibinfo{pages}{3925--3936} (\bibinfo{year}{1981}).

\bibitem{Caldeira83}
\bibinfo{author}{Caldeira, A.~O.} \& \bibinfo{author}{Leggett, A.~J.}
\newblock \bibinfo{journal}{\bibinfo{title}{Quantum tunnelling in a dissipative system}}.
\newblock {\emph{\JournalTitle{Ann. Phys.}}} \textbf{\bibinfo{volume}{149}}, \bibinfo{pages}{374--456} (\bibinfo{year}{1983}).

\bibitem{Garg85}
\bibinfo{author}{Garg, A.}, \bibinfo{author}{Onuchic, J.~N.} \& \bibinfo{author}{Ambegaokar, V.}
\newblock \bibinfo{journal}{\bibinfo{title}{Effect of friction on electron transfer in biomolecules}}.
\newblock {\emph{\JournalTitle{J. Chem. Phys.}}} \textbf{\bibinfo{volume}{83}}, \bibinfo{pages}{4491--4503} (\bibinfo{year}{1985}).

\bibitem{lode2020colloquium}
\bibinfo{author}{Lode, A.~U.}, \bibinfo{author}{L{\'e}v{\^e}que, C.}, \bibinfo{author}{Madsen, L.~B.}, \bibinfo{author}{Streltsov, A.~I.} \& \bibinfo{author}{Alon, O.~E.}
\newblock \bibinfo{journal}{\bibinfo{title}{Colloquium: {Multiconfigurational} time-dependent {Hartree} approaches for indistinguishable particles}}.
\newblock {\emph{\JournalTitle{Rev. Mod. Phys.}}} \textbf{\bibinfo{volume}{92}}, \bibinfo{pages}{011001} (\bibinfo{year}{2020}).

\bibitem{haobinJCP}
\bibinfo{author}{Wang, H.} \& \bibinfo{author}{Thoss, M.}
\newblock \bibinfo{journal}{\bibinfo{title}{Multilayer formulation of the multiconfiguration time-dependent {Hartree} theory}}.
\newblock {\emph{\JournalTitle{J. Chem. Phys.}}} \textbf{\bibinfo{volume}{119}}, \bibinfo{pages}{1289--1299} (\bibinfo{year}{2003}).

\bibitem{mctdhBOOK}
\bibinfo{author}{Meyer, H.-D.}, \bibinfo{author}{Gatti, F.} \& \bibinfo{author}{Worth, G.~A.}
\newblock \emph{\bibinfo{title}{Multidimensional quantum dynamics: {MCTDH} theory and applications}} (\bibinfo{publisher}{John Wiley \& Sons}, \bibinfo{year}{2009}).

\bibitem{ben2000ab}
\bibinfo{author}{Ben-Nun, M.}, \bibinfo{author}{Quenneville, J.} \& \bibinfo{author}{Mart{\'\i}nez, T.~J.}
\newblock \bibinfo{journal}{\bibinfo{title}{Ab initio multiple spawning: Photochemistry from first principles quantum molecular dynamics}}.
\newblock {\emph{\JournalTitle{J. Phys. Chem. A}}} \textbf{\bibinfo{volume}{104}}, \bibinfo{pages}{5161--5175} (\bibinfo{year}{2000}).

\bibitem{martinez2006insights}
\bibinfo{author}{Martinez, T.~J.}
\newblock \bibinfo{journal}{\bibinfo{title}{Insights for light-driven molecular devices from ab initio multiple spawning excited-state dynamics of organic and biological chromophores}}.
\newblock {\emph{\JournalTitle{Acc. Chem. Res.}}} \textbf{\bibinfo{volume}{39}}, \bibinfo{pages}{119--126} (\bibinfo{year}{2006}).

\bibitem{curchod2020ssaims}
\bibinfo{author}{Curchod, B.~F.}, \bibinfo{author}{Glover, W.~J.} \& \bibinfo{author}{Mart{\'\i}nez, T.~J.}
\newblock \bibinfo{journal}{\bibinfo{title}{Ssaims—stochastic-selection ab initio multiple spawning for efficient nonadiabatic molecular dynamics}}.
\newblock {\emph{\JournalTitle{J. Phys. Chem. A}}} \textbf{\bibinfo{volume}{124}}, \bibinfo{pages}{6133--6143} (\bibinfo{year}{2020}).

\bibitem{tanimura1990nonperturbative}
\bibinfo{author}{Tanimura, Y.}
\newblock \bibinfo{journal}{\bibinfo{title}{Nonperturbative expansion method for a quantum system coupled to a harmonic-oscillator bath}}.
\newblock {\emph{\JournalTitle{Phys. Rev. A}}} \textbf{\bibinfo{volume}{41}}, \bibinfo{pages}{6676} (\bibinfo{year}{1990}).

\bibitem{jin2008exact}
\bibinfo{author}{Jin, J.}, \bibinfo{author}{Zheng, X.} \& \bibinfo{author}{Yan, Y.}
\newblock \bibinfo{journal}{\bibinfo{title}{Exact dynamics of dissipative electronic systems and quantum transport: Hierarchical equations of motion approach}}.
\newblock {\emph{\JournalTitle{J. Chem. Phys.}}} \textbf{\bibinfo{volume}{128}}, \bibinfo{pages}{234703} (\bibinfo{year}{2008}).

\bibitem{yan2020new}
\bibinfo{author}{Yan, Y.}, \bibinfo{author}{Xing, T.} \& \bibinfo{author}{Shi, Q.}
\newblock \bibinfo{journal}{\bibinfo{title}{A new method to improve the numerical stability of the hierarchical equations of motion for discrete harmonic oscillator modes}}.
\newblock {\emph{\JournalTitle{J. Chem. Phys.}}} \textbf{\bibinfo{volume}{153}}, \bibinfo{pages}{204109} (\bibinfo{year}{2020}).

\bibitem{kundu2022intramolecular}
\bibinfo{author}{Kundu, S.} \& \bibinfo{author}{Makri, N.}
\newblock \bibinfo{journal}{\bibinfo{title}{Intramolecular vibrations in excitation energy transfer: Insights from real-time path integral calculations}}.
\newblock {\emph{\JournalTitle{Annu. Rev. Phys. Chem.}}} \textbf{\bibinfo{volume}{73}}, \bibinfo{pages}{349--375} (\bibinfo{year}{2022}).

\bibitem{makri1992improved}
\bibinfo{author}{Makri, N.}
\newblock \bibinfo{journal}{\bibinfo{title}{Improved {Feynman} propagators on a grid and non-adiabatic corrections within the path integral framework}}.
\newblock {\emph{\JournalTitle{Chem. Phys. Lett.}}} \textbf{\bibinfo{volume}{193}}, \bibinfo{pages}{435--445} (\bibinfo{year}{1992}).

\bibitem{makri2020small}
\bibinfo{author}{Makri, N.}
\newblock \bibinfo{journal}{\bibinfo{title}{Small matrix disentanglement of the path integral: overcoming the exponential tensor scaling with memory length}}.
\newblock {\emph{\JournalTitle{J. Chem. Phys.}}} \textbf{\bibinfo{volume}{152}}, \bibinfo{pages}{041104} (\bibinfo{year}{2020}).

\bibitem{makri2020small2}
\bibinfo{author}{Makri, N.}
\newblock \bibinfo{journal}{\bibinfo{title}{Small matrix path integral for system-bath dynamics}}.
\newblock {\emph{\JournalTitle{J. Chem. Theory Comput.}}} \textbf{\bibinfo{volume}{16}}, \bibinfo{pages}{4038--4049} (\bibinfo{year}{2020}).

\bibitem{makri2020small3}
\bibinfo{author}{Makri, N.}
\newblock \bibinfo{journal}{\bibinfo{title}{Small matrix path integral with extended memory}}.
\newblock {\emph{\JournalTitle{J. Chem. Theory Comput.}}} \textbf{\bibinfo{volume}{17}}, \bibinfo{pages}{1--6} (\bibinfo{year}{2020}).

\bibitem{makri2021small}
\bibinfo{author}{Makri, N.}
\newblock \bibinfo{journal}{\bibinfo{title}{Small matrix path integral for driven dissipative dynamics}}.
\newblock {\emph{\JournalTitle{J. Phys. Chem. A}}} \textbf{\bibinfo{volume}{125}}, \bibinfo{pages}{10500--10506} (\bibinfo{year}{2021}).

\bibitem{TOPALER1993285}
\bibinfo{author}{Topaler, M.} \& \bibinfo{author}{Makri, N.}
\newblock \bibinfo{journal}{\bibinfo{title}{Quasi-adiabatic propagator path integral methods. {Exact} quantum rate constants for condensed phase reactions}}.
\newblock {\emph{\JournalTitle{Chem. Phys. Lett.}}} \textbf{\bibinfo{volume}{210}}, \bibinfo{pages}{285--293} (\bibinfo{year}{1993}).

\bibitem{Strathearn18}
\bibinfo{author}{Strathearn, A.}, \bibinfo{author}{Kirton, P.}, \bibinfo{author}{Kilda, D.}, \bibinfo{author}{Keeling, J.} \& \bibinfo{author}{Lovett, B.~W.}
\newblock \bibinfo{journal}{\bibinfo{title}{Efficient non-markovian quantum dynamics using time-evolving matrix product operators}}.
\newblock {\emph{\JournalTitle{Nat. Commun.}}} \textbf{\bibinfo{volume}{9}}, \bibinfo{pages}{3322} (\bibinfo{year}{2018}).

\bibitem{cygorek2022simulation}
\bibinfo{author}{Cygorek, M.} \emph{et~al.}
\newblock \bibinfo{journal}{\bibinfo{title}{Simulation of open quantum systems by automated compression of arbitrary environments}}.
\newblock {\emph{\JournalTitle{Nat. Phys.}}} \bibinfo{pages}{1--7} (\bibinfo{year}{2022}).

\bibitem{link2023open}
\bibinfo{author}{Link, V.}, \bibinfo{author}{Tu, H.-H.} \& \bibinfo{author}{Strunz, W.~T.}
\newblock \bibinfo{journal}{\bibinfo{title}{Open quantum system dynamics from infinite tensor network contraction}}.
\newblock {\emph{\JournalTitle{arXiv preprint arXiv:2307.01802}}}  (\bibinfo{year}{2023}).

\bibitem{Somoza19}
\bibinfo{author}{Somoza, A.~D.}, \bibinfo{author}{Marty, O.}, \bibinfo{author}{Lim, J.}, \bibinfo{author}{Huelga, S.~F.} \& \bibinfo{author}{Plenio, M.~B.}
\newblock \bibinfo{journal}{\bibinfo{title}{Dissipation-assisted matrix product factorization}}.
\newblock {\emph{\JournalTitle{Phys. Rev. Lett.}}} \textbf{\bibinfo{volume}{123}}, \bibinfo{pages}{100502} (\bibinfo{year}{2019}).

\bibitem{somoza2023driving}
\bibinfo{author}{Somoza, A.~D.}, \bibinfo{author}{Lorenzoni, N.}, \bibinfo{author}{Lim, J.}, \bibinfo{author}{Huelga, S.~F.} \& \bibinfo{author}{Plenio, M.~B.}
\newblock \bibinfo{journal}{\bibinfo{title}{Driving force and nonequilibrium vibronic dynamics in charge separation of strongly bound electron--hole pairs}}.
\newblock {\emph{\JournalTitle{Commun. Phys.}}} \textbf{\bibinfo{volume}{6}}, \bibinfo{pages}{65} (\bibinfo{year}{2023}).

\bibitem{White04}
\bibinfo{author}{White, S.~R.} \& \bibinfo{author}{Feiguin, A.~E.}
\newblock \bibinfo{journal}{\bibinfo{title}{Real-time evolution using the density matrix renormalization group}}.
\newblock {\emph{\JournalTitle{Phys. Rev. Lett.}}} \textbf{\bibinfo{volume}{93}}, \bibinfo{pages}{076401} (\bibinfo{year}{2004}).

\bibitem{Daley04}
\bibinfo{author}{Daley, A.~J.}, \bibinfo{author}{Kollath, C.}, \bibinfo{author}{Schollw{\"o}ck, U.} \& \bibinfo{author}{Vidal, G.}
\newblock \bibinfo{journal}{\bibinfo{title}{Time-dependent density-matrix renormalization-group using adaptive effective hilbert spaces}}.
\newblock {\emph{\JournalTitle{J. Stat. Mech.: Theory Exp.}}} \textbf{\bibinfo{volume}{2004}}, \bibinfo{pages}{P04005} (\bibinfo{year}{2004}).

\bibitem{chin2010exact}
\bibinfo{author}{Chin, A.~W.}, \bibinfo{author}{Rivas, {\'A}.}, \bibinfo{author}{Huelga, S.~F.} \& \bibinfo{author}{Plenio, M.~B.}
\newblock \bibinfo{journal}{\bibinfo{title}{Exact mapping between system-reservoir quantum models and semi-infinite discrete chains using orthogonal polynomials}}.
\newblock {\emph{\JournalTitle{J. Math. Phys.}}} \textbf{\bibinfo{volume}{51}}, \bibinfo{pages}{092109} (\bibinfo{year}{2010}).

\bibitem{Prior10}
\bibinfo{author}{Prior, J.}, \bibinfo{author}{Chin, A.~W.}, \bibinfo{author}{Huelga, S.~F.} \& \bibinfo{author}{Plenio, M.~B.}
\newblock \bibinfo{journal}{\bibinfo{title}{Efficient simulation of strong system-environment interactions}}.
\newblock {\emph{\JournalTitle{Phys. Rev. Lett.}}} \textbf{\bibinfo{volume}{105}}, \bibinfo{pages}{050404} (\bibinfo{year}{2010}).

\bibitem{schollwock2011density}
\bibinfo{author}{Schollw{\"o}ck, U.}
\newblock \bibinfo{journal}{\bibinfo{title}{The density-matrix renormalization group in the age of matrix product states}}.
\newblock {\emph{\JournalTitle{Ann. Phys.}}} \textbf{\bibinfo{volume}{326}}, \bibinfo{pages}{96--192} (\bibinfo{year}{2011}).

\bibitem{chin2011chain}
\bibinfo{author}{Chin, A.~W.}, \bibinfo{author}{Huelga, S.~F.} \& \bibinfo{author}{Plenio, M.~B.}
\newblock \bibinfo{title}{Chain representations of open quantum systems and their numerical simulation with time-adaptive density matrix renormalisation group methods}.
\newblock In \emph{\bibinfo{booktitle}{Semiconductors and Semimetals}}, vol.~\bibinfo{volume}{85}, \bibinfo{pages}{115--143} (\bibinfo{publisher}{Elsevier}, \bibinfo{year}{2011}).

\bibitem{del2018tensor}
\bibinfo{author}{Del~Pino, J.}, \bibinfo{author}{Schr{\"o}der, F.~A.}, \bibinfo{author}{Chin, A.~W.}, \bibinfo{author}{Feist, J.} \& \bibinfo{author}{Garcia-Vidal, F.~J.}
\newblock \bibinfo{journal}{\bibinfo{title}{Tensor network simulation of polaron-polaritons in organic microcavities}}.
\newblock {\emph{\JournalTitle{Phys. Rev. B}}} \textbf{\bibinfo{volume}{98}}, \bibinfo{pages}{165416} (\bibinfo{year}{2018}).

\bibitem{dunnett2021efficient}
\bibinfo{author}{Dunnett, A.~J.} \& \bibinfo{author}{Chin, A.~W.}
\newblock \bibinfo{journal}{\bibinfo{title}{Efficient bond-adaptive approach for finite-temperature open quantum dynamics using the one-site time-dependent variational principle for matrix product states}}.
\newblock {\emph{\JournalTitle{Phys. Rev. B}}} \textbf{\bibinfo{volume}{104}}, \bibinfo{pages}{214302} (\bibinfo{year}{2021}).

\bibitem{nuomin2022improving}
\bibinfo{author}{Nuomin, H.}, \bibinfo{author}{Beratan, D.~N.} \& \bibinfo{author}{Zhang, P.}
\newblock \bibinfo{journal}{\bibinfo{title}{Improving the efficiency of open-quantum-system simulations using matrix product states in the interaction picture}}.
\newblock {\emph{\JournalTitle{Phys. Rev. A}}} \textbf{\bibinfo{volume}{105}}, \bibinfo{pages}{032406} (\bibinfo{year}{2022}).

\bibitem{xie2019time}
\bibinfo{author}{Xie, X.} \emph{et~al.}
\newblock \bibinfo{journal}{\bibinfo{title}{Time-dependent density matrix renormalization group quantum dynamics for realistic chemical systems}}.
\newblock {\emph{\JournalTitle{J. Chem. Phys.}}} \textbf{\bibinfo{volume}{151}}, \bibinfo{pages}{224101} (\bibinfo{year}{2019}).

\bibitem{nuomin2022suppressing}
\bibinfo{author}{Nuomin, H.}, \bibinfo{author}{Song, F.-F.}, \bibinfo{author}{Beratan, D.~N.} \& \bibinfo{author}{Zhang, P.}
\newblock \bibinfo{journal}{\bibinfo{title}{Suppressing the entanglement growth in matrix product state evolution of quantum systems through nonunitary similarity transformations}}.
\newblock {\emph{\JournalTitle{Phys. Rev. B}}} \textbf{\bibinfo{volume}{106}}, \bibinfo{pages}{104306} (\bibinfo{year}{2022}).

\bibitem{xu2022stochastic}
\bibinfo{author}{Xu, Y.}, \bibinfo{author}{Xie, Z.}, \bibinfo{author}{Xie, X.}, \bibinfo{author}{Schollw\"ock, U.} \& \bibinfo{author}{Ma, H.}
\newblock \bibinfo{journal}{\bibinfo{title}{Stochastic adaptive single-site time-dependent variational principle}}.
\newblock {\emph{\JournalTitle{JACS Au}}} \textbf{\bibinfo{volume}{2}}, \bibinfo{pages}{335--340} (\bibinfo{year}{2022}).

\bibitem{ET_JACS1}
\bibinfo{author}{Miller, J.~R.}, \bibinfo{author}{Calcaterra, L.} \& \bibinfo{author}{Closs, G.}
\newblock \bibinfo{journal}{\bibinfo{title}{Intramolecular long-distance electron transfer in radical anions. the effects of free energy and solvent on the reaction rates}}.
\newblock {\emph{\JournalTitle{J. Am. Chem. Soc.}}} \textbf{\bibinfo{volume}{106}}, \bibinfo{pages}{3047--3049} (\bibinfo{year}{1984}).

\bibitem{ET_JACS2}
\bibinfo{author}{Miller, J.~R.}, \bibinfo{author}{Beitz, J.~V.} \& \bibinfo{author}{Huddleston, R.~K.}
\newblock \bibinfo{journal}{\bibinfo{title}{Effect of free energy on rates of electron transfer between molecules}}.
\newblock {\emph{\JournalTitle{J. Am. Chem. Soc.}}} \textbf{\bibinfo{volume}{106}}, \bibinfo{pages}{5057--5068} (\bibinfo{year}{1984}).

\bibitem{ET_JPC_LARGE_REORG}
\bibinfo{author}{Shibano, Y.} \emph{et~al.}
\newblock \bibinfo{journal}{\bibinfo{title}{Large reorganization energy of pyrrolidine-substituted perylenediimide in electron transfer}}.
\newblock {\emph{\JournalTitle{J. Phys. Chem. C}}} \textbf{\bibinfo{volume}{111}}, \bibinfo{pages}{6133--6142} (\bibinfo{year}{2007}).

\bibitem{onuchic1987effect}
\bibinfo{author}{Onuchic, J.~N.}
\newblock \bibinfo{journal}{\bibinfo{title}{Effect of friction on electron transfer: {The} two reaction coordinate case}}.
\newblock {\emph{\JournalTitle{J. Chem. Phys.}}} \textbf{\bibinfo{volume}{86}}, \bibinfo{pages}{3925--3943} (\bibinfo{year}{1987}).

\bibitem{hsu2020reorganization}
\bibinfo{author}{Hsu, C.-P.}
\newblock \bibinfo{journal}{\bibinfo{title}{Reorganization energies and spectral densities for electron transfer problems in charge transport materials}}.
\newblock {\emph{\JournalTitle{Phys. Chem. Chem. Phys.}}} \textbf{\bibinfo{volume}{22}}, \bibinfo{pages}{21630--21641} (\bibinfo{year}{2020}).

\bibitem{PRB_ComingMode}
\bibinfo{author}{Polyakov, E.~A.}
\newblock \bibinfo{journal}{\bibinfo{title}{Real-time motion of open quantum systems: Structure of entanglement, renormalization group, and trajectories}}.
\newblock {\emph{\JournalTitle{Phys. Rev. B}}} \textbf{\bibinfo{volume}{105}}, \bibinfo{pages}{054306} (\bibinfo{year}{2022}).

\bibitem{mignolet2018walk}
\bibinfo{author}{Mignolet, B.} \& \bibinfo{author}{Curchod, B.~F.}
\newblock \bibinfo{journal}{\bibinfo{title}{A walk through the approximations of ab initio multiple spawning}}.
\newblock {\emph{\JournalTitle{J. Chem. Phys.}}} \textbf{\bibinfo{volume}{148}} (\bibinfo{year}{2018}).

\bibitem{Eherenfest1927}
\bibinfo{author}{Ehrenfest, P.}
\newblock \bibinfo{journal}{\bibinfo{title}{Bemerkung über die angenäherte gültigkeit der klassischen mechanik innerhalb der quantenmechanik}}.
\newblock {\emph{\JournalTitle{Zeitschrift fuer Physik}}} \textbf{\bibinfo{volume}{45}}, \bibinfo{pages}{455–457} (\bibinfo{year}{1927}).

\bibitem{mclachlan1964variational}
\bibinfo{author}{McLachlan, A.}
\newblock \bibinfo{journal}{\bibinfo{title}{A variational solution of the time-dependent schrodinger equation}}.
\newblock {\emph{\JournalTitle{Mol. Phys.}}} \textbf{\bibinfo{volume}{8}}, \bibinfo{pages}{39--44} (\bibinfo{year}{1964}).

\bibitem{agostini2019different}
\bibinfo{author}{Agostini, F.} \& \bibinfo{author}{Curchod, B.~F.}
\newblock \bibinfo{journal}{\bibinfo{title}{Different flavors of nonadiabatic molecular dynamics}}.
\newblock {\emph{\JournalTitle{Wiley Interdiscip. Rev. Comput. Mol. Sci.}}} \textbf{\bibinfo{volume}{9}}, \bibinfo{pages}{e1417} (\bibinfo{year}{2019}).

\bibitem{tully1990molecular}
\bibinfo{author}{Tully, J.~C.}
\newblock \bibinfo{journal}{\bibinfo{title}{Molecular dynamics with electronic transitions}}.
\newblock {\emph{\JournalTitle{J. Chem. Phys.}}} \textbf{\bibinfo{volume}{93}}, \bibinfo{pages}{1061--1071} (\bibinfo{year}{1990}).

\bibitem{wang2016recent}
\bibinfo{author}{Wang, L.}, \bibinfo{author}{Akimov, A.} \& \bibinfo{author}{Prezhdo, O.~V.}
\newblock \bibinfo{journal}{\bibinfo{title}{Recent progress in surface hopping: 2011--2015}}.
\newblock {\emph{\JournalTitle{J. Phys. Chem. Lett.}}} \textbf{\bibinfo{volume}{7}}, \bibinfo{pages}{2100--2112} (\bibinfo{year}{2016}).

\bibitem{barbatti2011nonadiabatic}
\bibinfo{author}{Barbatti, M.}
\newblock \bibinfo{journal}{\bibinfo{title}{Nonadiabatic dynamics with trajectory surface hopping method}}.
\newblock {\emph{\JournalTitle{Wiley Interdiscip. Rev. Comput. Mol. Sci.}}} \textbf{\bibinfo{volume}{1}}, \bibinfo{pages}{620--633} (\bibinfo{year}{2011}).

\bibitem{subotnik2016understanding}
\bibinfo{author}{Subotnik, J.~E.} \emph{et~al.}
\newblock \bibinfo{journal}{\bibinfo{title}{Understanding the surface hopping view of electronic transitions and decoherence}}.
\newblock {\emph{\JournalTitle{Annu. Rev. Phys. Chem.}}} \textbf{\bibinfo{volume}{67}}, \bibinfo{pages}{387--417} (\bibinfo{year}{2016}).

\bibitem{wang2015}
\bibinfo{author}{Wang, L.}, \bibinfo{author}{Sifain, A.~E.} \& \bibinfo{author}{Prezhdo, O.~V.}
\newblock \bibinfo{journal}{\bibinfo{title}{Fewest switches surface hopping in {Liouville} space}}.
\newblock {\emph{\JournalTitle{J. Phys. Chem. Lett.}}} \textbf{\bibinfo{volume}{6}}, \bibinfo{pages}{3827--3833} (\bibinfo{year}{2015}).

\bibitem{kapral1999mixed}
\bibinfo{author}{Kapral, R.} \& \bibinfo{author}{Ciccotti, G.}
\newblock \bibinfo{journal}{\bibinfo{title}{Mixed quantum-classical dynamics}}.
\newblock {\emph{\JournalTitle{J. Chem. Phys.}}} \textbf{\bibinfo{volume}{110}}, \bibinfo{pages}{8919--8929} (\bibinfo{year}{1999}).

\bibitem{mac2008trotter}
\bibinfo{author}{Mac~Kernan, D.}, \bibinfo{author}{Ciccotti, G.} \& \bibinfo{author}{Kapral, R.}
\newblock \bibinfo{journal}{\bibinfo{title}{Trotter-based simulation of quantum-classical dynamics}}.
\newblock {\emph{\JournalTitle{J. Phys. Chem. B}}} \textbf{\bibinfo{volume}{112}}, \bibinfo{pages}{424--432} (\bibinfo{year}{2008}).

\bibitem{kim2008quantum}
\bibinfo{author}{Kim, H.}, \bibinfo{author}{Nassimi, A.} \& \bibinfo{author}{Kapral, R.}
\newblock \bibinfo{journal}{\bibinfo{title}{Quantum-classical {Liouville} dynamics in the mapping basis}}.
\newblock {\emph{\JournalTitle{J. Chem. Phys.}}} \textbf{\bibinfo{volume}{129}}, \bibinfo{pages}{084102} (\bibinfo{year}{2008}).

\bibitem{hsieh2012nonadiabatic}
\bibinfo{author}{Hsieh, C.-Y.} \& \bibinfo{author}{Kapral, R.}
\newblock \bibinfo{journal}{\bibinfo{title}{Nonadiabatic dynamics in open quantum-classical systems: {Forward-backward} trajectory solution}}.
\newblock {\emph{\JournalTitle{J. Chem. Phys.}}} \textbf{\bibinfo{volume}{137}}, \bibinfo{pages}{22A507} (\bibinfo{year}{2012}).

\bibitem{hsieh2013analysis}
\bibinfo{author}{Hsieh, C.-Y.} \& \bibinfo{author}{Kapral, R.}
\newblock \bibinfo{journal}{\bibinfo{title}{Analysis of the forward-backward trajectory solution for the mixed quantum-classical {Liouville} equation}}.
\newblock {\emph{\JournalTitle{J. Chem. Phys.}}} \textbf{\bibinfo{volume}{138}}, \bibinfo{pages}{134110} (\bibinfo{year}{2013}).

\bibitem{kapral2015quantum}
\bibinfo{author}{Kapral, R.}
\newblock \bibinfo{journal}{\bibinfo{title}{Quantum dynamics in open quantum-classical systems}}.
\newblock {\emph{\JournalTitle{J. Condens. Matter Phys.}}} \textbf{\bibinfo{volume}{27}}, \bibinfo{pages}{073201} (\bibinfo{year}{2015}).

\bibitem{miller2001semiclassical}
\bibinfo{author}{Miller, W.~H.}
\newblock \bibinfo{journal}{\bibinfo{title}{The semiclassical initial value representation: {A} potentially practical way for adding quantum effects to classical molecular dynamics simulations}}.
\newblock {\emph{\JournalTitle{J. Phys. Chem. A}}} \textbf{\bibinfo{volume}{105}}, \bibinfo{pages}{2942--2955} (\bibinfo{year}{2001}).

\bibitem{miller2009electronically}
\bibinfo{author}{Miller, W.~H.}
\newblock \bibinfo{journal}{\bibinfo{title}{Electronically nonadiabatic dynamics via semiclassical initial value methods}}.
\newblock {\emph{\JournalTitle{J. Phys. Chem. A}}} \textbf{\bibinfo{volume}{113}}, \bibinfo{pages}{1405--1415} (\bibinfo{year}{2009}).

\bibitem{sun1998semiclassical}
\bibinfo{author}{Sun, X.}, \bibinfo{author}{Wang, H.} \& \bibinfo{author}{Miller, W.~H.}
\newblock \bibinfo{journal}{\bibinfo{title}{Semiclassical theory of electronically nonadiabatic dynamics: {Results} of a linearized approximation to the initial value representation}}.
\newblock {\emph{\JournalTitle{J. Chem. Phys.}}} \textbf{\bibinfo{volume}{109}}, \bibinfo{pages}{7064--7074} (\bibinfo{year}{1998}).

\bibitem{shi2003relationship}
\bibinfo{author}{Shi, Q.} \& \bibinfo{author}{Geva, E.}
\newblock \bibinfo{journal}{\bibinfo{title}{A relationship between semiclassical and centroid correlation functions}}.
\newblock {\emph{\JournalTitle{J. Chem. Phys.}}} \textbf{\bibinfo{volume}{118}}, \bibinfo{pages}{8173--8184} (\bibinfo{year}{2003}).

\bibitem{bossion2022non}
\bibinfo{author}{Bossion, D.}, \bibinfo{author}{Ying, W.}, \bibinfo{author}{Chowdhury, S.~N.} \& \bibinfo{author}{Huo, P.}
\newblock \bibinfo{journal}{\bibinfo{title}{Non-adiabatic mapping dynamics in the phase space of the $su(n)$ {Lie} group}}.
\newblock {\emph{\JournalTitle{J. Chem. Phys.}}} \textbf{\bibinfo{volume}{157}}, \bibinfo{pages}{084105} (\bibinfo{year}{2022}).

\bibitem{runeson2022spin}
\bibinfo{author}{Runeson, J.~E.}, \bibinfo{author}{Mannouch, J.~R.}, \bibinfo{author}{Amati, G.}, \bibinfo{author}{Fiechter, M.~R.} \& \bibinfo{author}{Richardson, J.}
\newblock \bibinfo{journal}{\bibinfo{title}{Spin-mapping methods for simulating ultrafast nonadiabatic dynamics}}.
\newblock {\emph{\JournalTitle{Chimia}}} \textbf{\bibinfo{volume}{76}}, \bibinfo{pages}{582--588} (\bibinfo{year}{2022}).

\bibitem{runeson2020generalized}
\bibinfo{author}{Runeson, J.~E.} \& \bibinfo{author}{Richardson, J.~O.}
\newblock \bibinfo{journal}{\bibinfo{title}{Generalized spin mapping for quantum-classical dynamics}}.
\newblock {\emph{\JournalTitle{J. Chem. Phys.}}} \textbf{\bibinfo{volume}{152}}, \bibinfo{pages}{084110} (\bibinfo{year}{2020}).

\bibitem{mannouch2022partially}
\bibinfo{author}{Mannouch, J.~R.} \& \bibinfo{author}{Richardson, J.~O.}
\newblock \bibinfo{journal}{\bibinfo{title}{A partially linearized spin-mapping approach for simulating nonlinear optical spectra}}.
\newblock {\emph{\JournalTitle{J. Chem. Phys.}}} \textbf{\bibinfo{volume}{156}}, \bibinfo{pages}{024108} (\bibinfo{year}{2022}).

\bibitem{cao1994formulation}
\bibinfo{author}{Cao, J.} \& \bibinfo{author}{Voth, G.~A.}
\newblock \bibinfo{journal}{\bibinfo{title}{The formulation of quantum statistical mechanics based on the {Feynman} path centroid density. ii. dynamical properties}}.
\newblock {\emph{\JournalTitle{J. Chem. Phys.}}} \textbf{\bibinfo{volume}{100}}, \bibinfo{pages}{5106--5117} (\bibinfo{year}{1994}).

\bibitem{jang1999derivation}
\bibinfo{author}{Jang, S.} \& \bibinfo{author}{Voth, G.~A.}
\newblock \bibinfo{journal}{\bibinfo{title}{A derivation of centroid molecular dynamics and other approximate time evolution methods for path integral centroid variables}}.
\newblock {\emph{\JournalTitle{J. Chem. Phys.}}} \textbf{\bibinfo{volume}{111}}, \bibinfo{pages}{2371--2384} (\bibinfo{year}{1999}).

\bibitem{habershon2013ring}
\bibinfo{author}{Habershon, S.}, \bibinfo{author}{Manolopoulos, D.~E.}, \bibinfo{author}{Markland, T.~E.} \& \bibinfo{author}{Miller~III, T.~F.}
\newblock \bibinfo{journal}{\bibinfo{title}{Ring-polymer molecular dynamics: {Quantum} effects in chemical dynamics from classical trajectories in an extended phase space}}.
\newblock {\emph{\JournalTitle{Annu. Rev. Phys. Chem.}}} \textbf{\bibinfo{volume}{64}}, \bibinfo{pages}{387--413} (\bibinfo{year}{2013}).

\bibitem{craig2004quantum}
\bibinfo{author}{Craig, I.~R.} \& \bibinfo{author}{Manolopoulos, D.~E.}
\newblock \bibinfo{journal}{\bibinfo{title}{Quantum statistics and classical mechanics: {Real} time correlation functions from ring polymer molecular dynamics}}.
\newblock {\emph{\JournalTitle{J. Chem. Phys.}}} \textbf{\bibinfo{volume}{121}}, \bibinfo{pages}{3368--3373} (\bibinfo{year}{2004}).

\bibitem{richardson2013communication}
\bibinfo{author}{Richardson, J.~O.} \& \bibinfo{author}{Thoss, M.}
\newblock \bibinfo{journal}{\bibinfo{title}{Communication: {Nonadiabatic} ring-polymer molecular dynamics}}.
\newblock {\emph{\JournalTitle{J. Chem. Phys.}}} \textbf{\bibinfo{volume}{139}}, \bibinfo{pages}{031102} (\bibinfo{year}{2013}).

\bibitem{ananth2013mapping}
\bibinfo{author}{Ananth, N.}
\newblock \bibinfo{journal}{\bibinfo{title}{Mapping variable ring polymer molecular dynamics: {A} path-integral based method for nonadiabatic processes}}.
\newblock {\emph{\JournalTitle{J. Chem. Phys.}}} \textbf{\bibinfo{volume}{139}}, \bibinfo{pages}{124102} (\bibinfo{year}{2013}).

\bibitem{chowdhury2017coherent}
\bibinfo{author}{Chowdhury, S.~N.} \& \bibinfo{author}{Huo, P.}
\newblock \bibinfo{journal}{\bibinfo{title}{Coherent state mapping ring polymer molecular dynamics for non-adiabatic quantum propagations}}.
\newblock {\emph{\JournalTitle{J. Chem. Phys.}}} \textbf{\bibinfo{volume}{147}}, \bibinfo{pages}{214109} (\bibinfo{year}{2017}).

\bibitem{chowdhury2019state}
\bibinfo{author}{Chowdhury, S.~N.} \& \bibinfo{author}{Huo, P.}
\newblock \bibinfo{journal}{\bibinfo{title}{State dependent ring polymer molecular dynamics for investigating excited nonadiabatic dynamics}}.
\newblock {\emph{\JournalTitle{J. Chem. Phys.}}} \textbf{\bibinfo{volume}{150}}, \bibinfo{pages}{244102} (\bibinfo{year}{2019}).

\bibitem{chowdhury2021non}
\bibinfo{author}{Chowdhury, S.~N.} \& \bibinfo{author}{Huo, P.}
\newblock \bibinfo{journal}{\bibinfo{title}{Non-adiabatic {Matsubara} dynamics and non-adiabatic ring-polymer molecular dynamics}}.
\newblock {\emph{\JournalTitle{J. Chem. Phys.}}} \textbf{\bibinfo{volume}{154}}, \bibinfo{pages}{124124} (\bibinfo{year}{2021}).

\bibitem{hele2015boltzmann}
\bibinfo{author}{Hele, T.~J.}, \bibinfo{author}{Willatt, M.~J.}, \bibinfo{author}{Muolo, A.} \& \bibinfo{author}{Althorpe, S.~C.}
\newblock \bibinfo{journal}{\bibinfo{title}{Boltzmann-conserving classical dynamics in quantum time-correlation functions: “{Matsubara} dynamics”}}.
\newblock {\emph{\JournalTitle{J. Chem. Phys.}}} \textbf{\bibinfo{volume}{142}}, \bibinfo{pages}{134103} (\bibinfo{year}{2015}).

\bibitem{tully-inbook}
\bibinfo{author}{Tully, J.~C.}
\newblock \emph{\bibinfo{title}{Modern methods for multidimensional dynamics computations in chemistry}}, chap. \bibinfo{chapter}{Nonadiabatic Dynamics}, \bibinfo{pages}{34--72} (\bibinfo{publisher}{World Scientific}, \bibinfo{year}{1998}).

\bibitem{parandekar2005mixed}
\bibinfo{author}{Parandekar, P.~V.} \& \bibinfo{author}{Tully, J.~C.}
\newblock \bibinfo{journal}{\bibinfo{title}{Mixed quantum-classical equilibrium}}.
\newblock {\emph{\JournalTitle{J. Chem. Phys.}}} \textbf{\bibinfo{volume}{122}}, \bibinfo{pages}{094102} (\bibinfo{year}{2005}).

\bibitem{kang2019nonadiabatic}
\bibinfo{author}{Kang, J.} \& \bibinfo{author}{Wang, L.-W.}
\newblock \bibinfo{journal}{\bibinfo{title}{Nonadiabatic molecular dynamics with decoherence and detailed balance under a density matrix ensemble formalism}}.
\newblock {\emph{\JournalTitle{Phys. Rev. B}}} \textbf{\bibinfo{volume}{99}}, \bibinfo{pages}{224303} (\bibinfo{year}{2019}).

\bibitem{jain2018vibrational}
\bibinfo{author}{Jain, A.} \& \bibinfo{author}{Subotnik, J.~E.}
\newblock \bibinfo{journal}{\bibinfo{title}{Vibrational energy relaxation: {A} benchmark for mixed quantum--classical methods}}.
\newblock {\emph{\JournalTitle{J. Phys. Chem. A}}} \textbf{\bibinfo{volume}{122}}, \bibinfo{pages}{16--27} (\bibinfo{year}{2018}).

\bibitem{persico2014overview}
\bibinfo{author}{Persico, M.} \& \bibinfo{author}{Granucci, G.}
\newblock \bibinfo{journal}{\bibinfo{title}{An overview of nonadiabatic dynamics simulations methods, with focus on the direct approach versus the fitting of potential energy surfaces}}.
\newblock {\emph{\JournalTitle{Theor. Chem. Acc.}}} \textbf{\bibinfo{volume}{133}}, \bibinfo{pages}{1--28} (\bibinfo{year}{2014}).

\bibitem{ananth2022path}
\bibinfo{author}{Ananth, N.}
\newblock \bibinfo{journal}{\bibinfo{title}{Path integrals for nonadiabatic dynamics: Multistate ring polymer molecular dynamics}}.
\newblock {\emph{\JournalTitle{Ann. Rev. Phys. Chem.}}} \textbf{\bibinfo{volume}{73}}, \bibinfo{pages}{299--322} (\bibinfo{year}{2022}).

\bibitem{liu2022imaginary}
\bibinfo{author}{Liu, Z.}, \bibinfo{author}{Xu, W.}, \bibinfo{author}{Tuckerman, M.~E.} \& \bibinfo{author}{Sun, X.}
\newblock \bibinfo{journal}{\bibinfo{title}{Imaginary-time open-chain path-integral approach for two-state time correlation functions and applications in charge transfer}}.
\newblock {\emph{\JournalTitle{J. Chem. Phys.}}} \textbf{\bibinfo{volume}{157}}, \bibinfo{pages}{114111} (\bibinfo{year}{2022}).

\bibitem{bellonzi2016assessment}
\bibinfo{author}{Bellonzi, N.}, \bibinfo{author}{Jain, A.} \& \bibinfo{author}{Subotnik, J.~E.}
\newblock \bibinfo{journal}{\bibinfo{title}{An assessment of mean-field mixed semiclassical approaches: Equilibrium populations and algorithm stability}}.
\newblock {\emph{\JournalTitle{J. Chem. Phys.}}} \textbf{\bibinfo{volume}{144}}, \bibinfo{pages}{154110} (\bibinfo{year}{2016}).

\bibitem{Low20}
\bibinfo{author}{Low, P.~J.}, \bibinfo{author}{White, B.~M.}, \bibinfo{author}{Cox, A.~A.}, \bibinfo{author}{Day, M.~L.} \& \bibinfo{author}{Senko, C.}
\newblock \bibinfo{journal}{\bibinfo{title}{Practical trapped-ion protocols for universal qudit-based quantum computing}}.
\newblock {\emph{\JournalTitle{Phys. Rev. Res.}}} \textbf{\bibinfo{volume}{2}}, \bibinfo{pages}{033128} (\bibinfo{year}{2020}).

\bibitem{Ringbauer22}
\bibinfo{author}{Ringbauer, M.} \emph{et~al.}
\newblock \bibinfo{journal}{\bibinfo{title}{A universal qudit quantum processor with trapped ions}}.
\newblock {\emph{\JournalTitle{Nat. Phys.}}} \bibinfo{pages}{1--5} (\bibinfo{year}{2022}).

\bibitem{lloyd1996universal}
\bibinfo{author}{Lloyd, S.}
\newblock \bibinfo{journal}{\bibinfo{title}{Universal quantum simulators}}.
\newblock {\emph{\JournalTitle{Science}}} \textbf{\bibinfo{volume}{273}}, \bibinfo{pages}{1073--1078} (\bibinfo{year}{1996}).

\bibitem{Davoudi21}
\bibinfo{author}{Davoudi, Z.}, \bibinfo{author}{Linke, N.~M.} \& \bibinfo{author}{Pagano, G.}
\newblock \bibinfo{journal}{\bibinfo{title}{Toward simulating quantum field theories with controlled phonon-ion dynamics: A hybrid analog-digital approach}}.
\newblock {\emph{\JournalTitle{Phys. Rev. Res.}}} \textbf{\bibinfo{volume}{3}}, \bibinfo{pages}{043072} (\bibinfo{year}{2021}).

\bibitem{Porras04}
\bibinfo{author}{Porras, D.} \& \bibinfo{author}{Cirac, J.~I.}
\newblock \bibinfo{journal}{\bibinfo{title}{Effective quantum spin systems with trapped ions}}.
\newblock {\emph{\JournalTitle{Phys. Rev. Lett.}}} \textbf{\bibinfo{volume}{92}}, \bibinfo{pages}{207901} (\bibinfo{year}{2004}).

\bibitem{Maier19}
\bibinfo{author}{Maier, C.} \emph{et~al.}
\newblock \bibinfo{journal}{\bibinfo{title}{Environment-assisted quantum transport in a 10-qubit network}}.
\newblock {\emph{\JournalTitle{Phys. Rev. Lett.}}} \textbf{\bibinfo{volume}{122}}, \bibinfo{pages}{050501} (\bibinfo{year}{2019}).

\bibitem{Wang18}
\bibinfo{author}{Wang, B.-X.} \emph{et~al.}
\newblock \bibinfo{journal}{\bibinfo{title}{Efficient quantum simulation of photosynthetic light harvesting}}.
\newblock {\emph{\JournalTitle{npj Quantum Inf.}}} \textbf{\bibinfo{volume}{4}}, \bibinfo{pages}{52} (\bibinfo{year}{2018}).

\bibitem{sun2023quantum}
\bibinfo{author}{Sun, K.} \emph{et~al.}
\newblock \bibinfo{journal}{\bibinfo{title}{Quantum simulation of polarized light-induced electron transfer with a trapped-ion qutrit system}}.
\newblock {\emph{\JournalTitle{J. Phys. Chem. Lett.}}} \textbf{\bibinfo{volume}{14}}, \bibinfo{pages}{6071--6077} (\bibinfo{year}{2023}).

\bibitem{Lechner16}
\bibinfo{author}{Lechner, R.} \emph{et~al.}
\newblock \bibinfo{journal}{\bibinfo{title}{Electromagnetically-induced-transparency ground-state cooling of long ion strings}}.
\newblock {\emph{\JournalTitle{Phys. Rev. A}}} \textbf{\bibinfo{volume}{93}}, \bibinfo{pages}{053401} (\bibinfo{year}{2016}).

\bibitem{Feng20}
\bibinfo{author}{Feng, L.} \emph{et~al.}
\newblock \bibinfo{journal}{\bibinfo{title}{Efficient ground-state cooling of large trapped-ion chains with an electromagnetically-induced-transparency tripod scheme}}.
\newblock {\emph{\JournalTitle{Phys. Rev. Lett.}}} \textbf{\bibinfo{volume}{125}}, \bibinfo{pages}{053001} (\bibinfo{year}{2020}).

\bibitem{NielsenChuang}
\bibinfo{author}{Nielsen, M.~A.} \& \bibinfo{author}{Chuang, I.~L.}
\newblock \emph{\bibinfo{title}{Quantum Computation and Quantum Information: 10th Anniversary Edition}} (\bibinfo{publisher}{Cambridge University Press}, \bibinfo{address}{USA}, \bibinfo{year}{2011}), \bibinfo{edition}{10th} edn.

\bibitem{leibfried2003quantum}
\bibinfo{author}{Leibfried, D.}, \bibinfo{author}{Blatt, R.}, \bibinfo{author}{Monroe, C.} \& \bibinfo{author}{Wineland, D.}
\newblock \bibinfo{journal}{\bibinfo{title}{Quantum dynamics of single trapped ions}}.
\newblock {\emph{\JournalTitle{Rev. Mod. Phys.}}} \textbf{\bibinfo{volume}{75}}, \bibinfo{pages}{281} (\bibinfo{year}{2003}).

\bibitem{lidar2001completely}
\bibinfo{author}{Lidar, D.~A.}, \bibinfo{author}{Bihary, Z.} \& \bibinfo{author}{Whaley, K.~B.}
\newblock \bibinfo{journal}{\bibinfo{title}{From completely positive maps to the quantum {Markovian} semigroup master equation}}.
\newblock {\emph{\JournalTitle{Chem. Phys.}}} \textbf{\bibinfo{volume}{268}}, \bibinfo{pages}{35--53} (\bibinfo{year}{2001}).

\bibitem{Molmer99}
\bibinfo{author}{M{\o}lmer, K.} \& \bibinfo{author}{S{\o}rensen, A.}
\newblock \bibinfo{journal}{\bibinfo{title}{Multiparticle entanglement of hot trapped ions}}.
\newblock {\emph{\JournalTitle{Phys. Rev. Lett.}}} \textbf{\bibinfo{volume}{82}}, \bibinfo{pages}{1835} (\bibinfo{year}{1999}).

\bibitem{Sorensen99}
\bibinfo{author}{S{\o}rensen, A.} \& \bibinfo{author}{M{\o}lmer, K.}
\newblock \bibinfo{journal}{\bibinfo{title}{Quantum computation with ions in thermal motion}}.
\newblock {\emph{\JournalTitle{Phys. Rev. Lett.}}} \textbf{\bibinfo{volume}{82}}, \bibinfo{pages}{1971} (\bibinfo{year}{1999}).

\bibitem{Zhang20}
\bibinfo{author}{Zhang, J.-N.} \emph{et~al.}
\newblock \bibinfo{journal}{\bibinfo{title}{Probabilistic eigensolver with a trapped-ion quantum processor}}.
\newblock {\emph{\JournalTitle{Phys. Rev. A}}} \textbf{\bibinfo{volume}{101}}, \bibinfo{pages}{052333} (\bibinfo{year}{2020}).

\bibitem{Christensen20}
\bibinfo{author}{Christensen, J.~E.}, \bibinfo{author}{Hucul, D.}, \bibinfo{author}{Campbell, W.~C.} \& \bibinfo{author}{Hudson, E.~R.}
\newblock \bibinfo{journal}{\bibinfo{title}{High-fidelity manipulation of a qubit enabled by a manufactured nucleus}}.
\newblock {\emph{\JournalTitle{npj Quantum Inf.}}} \textbf{\bibinfo{volume}{6}}, \bibinfo{pages}{1--5} (\bibinfo{year}{2020}).

\bibitem{Ransford21}
\bibinfo{author}{Ransford, A.}, \bibinfo{author}{Roman, C.}, \bibinfo{author}{Dellaert, T.}, \bibinfo{author}{McMillin, P.} \& \bibinfo{author}{Campbell, W.~C.}
\newblock \bibinfo{journal}{\bibinfo{title}{Weak dissipation for high-fidelity qubit-state preparation and measurement}}.
\newblock {\emph{\JournalTitle{Phys. Rev. A}}} \textbf{\bibinfo{volume}{104}}, \bibinfo{pages}{L060402} (\bibinfo{year}{2021}).

\bibitem{paris2004quantum}
\bibinfo{author}{Paris, M.} \& \bibinfo{author}{Rehacek, J.}
\newblock \emph{\bibinfo{title}{Quantum state estimation}}, vol. \bibinfo{volume}{649} (\bibinfo{publisher}{Springer Science \& Business Media}, \bibinfo{year}{2004}).

\bibitem{cramer2010efficient}
\bibinfo{author}{Cramer, M.} \emph{et~al.}
\newblock \bibinfo{journal}{\bibinfo{title}{Efficient quantum state tomography}}.
\newblock {\emph{\JournalTitle{Nat. Commun.}}} \textbf{\bibinfo{volume}{1}}, \bibinfo{pages}{149} (\bibinfo{year}{2010}).

\bibitem{Sato18}
\bibinfo{author}{Sato, S.~A.}, \bibinfo{author}{Kelly, A.} \& \bibinfo{author}{Rubio, A.}
\newblock \bibinfo{journal}{\bibinfo{title}{Coupled forward-backward trajectory approach for nonequilibrium electron-ion dynamics}}.
\newblock {\emph{\JournalTitle{Phys. Rev. B}}} \textbf{\bibinfo{volume}{97}}, \bibinfo{pages}{134308} (\bibinfo{year}{2018}).

\bibitem{ishizaki2009theoretical}
\bibinfo{author}{Ishizaki, A.} \& \bibinfo{author}{Fleming, G.~R.}
\newblock \bibinfo{journal}{\bibinfo{title}{Theoretical examination of quantum coherence in a photosynthetic system at physiological temperature}}.
\newblock {\emph{\JournalTitle{Proc. Natl. Acad. Sci. U.S.A.}}} \textbf{\bibinfo{volume}{106}}, \bibinfo{pages}{17255--17260} (\bibinfo{year}{2009}).

\bibitem{nitzan2006chemical}
\bibinfo{author}{Nitzan, A.}
\newblock \emph{\bibinfo{title}{Chemical dynamics in condensed phases: relaxation, transfer and reactions in condensed molecular systems}} (\bibinfo{publisher}{Oxford university press}, \bibinfo{year}{2006}).

\bibitem{Zhu06}
\bibinfo{author}{Zhu, S.-L.}, \bibinfo{author}{Monroe, C.} \& \bibinfo{author}{Duan, L.-M.}
\newblock \bibinfo{journal}{\bibinfo{title}{Trapped ion quantum computation with transverse phonon modes}}.
\newblock {\emph{\JournalTitle{Phys. Rev. Lett.}}} \textbf{\bibinfo{volume}{97}}, \bibinfo{pages}{050505} (\bibinfo{year}{2006}).

\bibitem{kang2023efficient}
\bibinfo{author}{Kang, M.}, \bibinfo{author}{Liang, Q.}, \bibinfo{author}{Li, M.} \& \bibinfo{author}{Nam, Y.}
\newblock \bibinfo{journal}{\bibinfo{title}{Efficient motional-mode characterization for high-fidelity trapped-ion quantum computing}}.
\newblock {\emph{\JournalTitle{Quantum Sci. Technol.}}} \textbf{\bibinfo{volume}{8}}, \bibinfo{pages}{024002} (\bibinfo{year}{2023}).

\bibitem{Leung18}
\bibinfo{author}{Leung, P.~H.} \emph{et~al.}
\newblock \bibinfo{journal}{\bibinfo{title}{Robust 2-qubit gates in a linear ion crystal using a frequency-modulated driving force}}.
\newblock {\emph{\JournalTitle{Phys. Rev. Lett.}}} \textbf{\bibinfo{volume}{120}}, \bibinfo{pages}{020501} (\bibinfo{year}{2018}).

\bibitem{Kang21}
\bibinfo{author}{Kang, M.} \emph{et~al.}
\newblock \bibinfo{journal}{\bibinfo{title}{Batch optimization of frequency-modulated pulses for robust two-qubit gates in ion chains}}.
\newblock {\emph{\JournalTitle{Phys. Rev. Appl.}}} \textbf{\bibinfo{volume}{16}}, \bibinfo{pages}{024039} (\bibinfo{year}{2021}).

\bibitem{Wang20FM}
\bibinfo{author}{Wang, Y.} \emph{et~al.}
\newblock \bibinfo{journal}{\bibinfo{title}{High-fidelity two-qubit gates using a microelectromechanical-system-based beam steering system for individual qubit addressing}}.
\newblock {\emph{\JournalTitle{Phys. Rev. Lett.}}} \textbf{\bibinfo{volume}{125}}, \bibinfo{pages}{150505} (\bibinfo{year}{2020}).

\bibitem{Kang23}
\bibinfo{author}{Kang, M.} \emph{et~al.}
\newblock \bibinfo{journal}{\bibinfo{title}{Designing filter functions of frequency-modulated pulses for high-fidelity two-qubit gates in ion chains}}.
\newblock {\emph{\JournalTitle{Phys. Rev. Appl.}}} \textbf{\bibinfo{volume}{19}}, \bibinfo{pages}{014014} (\bibinfo{year}{2023}).

\bibitem{Cetina22}
\bibinfo{author}{Cetina, M.} \emph{et~al.}
\newblock \bibinfo{journal}{\bibinfo{title}{Control of transverse motion for quantum gates on individually addressed atomic qubits}}.
\newblock {\emph{\JournalTitle{PRX Quantum}}} \textbf{\bibinfo{volume}{3}}, \bibinfo{pages}{010334} (\bibinfo{year}{2022}).

\bibitem{Qutip}
\bibinfo{author}{Johansson, J.~R.}, \bibinfo{author}{Nation, P.~D.} \& \bibinfo{author}{Nori, F.}
\newblock \bibinfo{journal}{\bibinfo{title}{Qutip: An open-source python framework for the dynamics of open quantum systems}}.
\newblock {\emph{\JournalTitle{Comput. Phys. Commun.}}} \textbf{\bibinfo{volume}{183}}, \bibinfo{pages}{1760--1772} (\bibinfo{year}{2012}).

\bibitem{Zhang98}
\bibinfo{author}{Zhang, C.}, \bibinfo{author}{Jeckelmann, E.} \& \bibinfo{author}{White, S.~R.}
\newblock \bibinfo{journal}{\bibinfo{title}{Density matrix approach to local hilbert space reduction}}.
\newblock {\emph{\JournalTitle{Phys. Rev. Lett.}}} \textbf{\bibinfo{volume}{80}}, \bibinfo{pages}{2661--2664} (\bibinfo{year}{1998}).

\bibitem{Mascherpa20}
\bibinfo{author}{Mascherpa, F.} \emph{et~al.}
\newblock \bibinfo{journal}{\bibinfo{title}{Optimized auxiliary oscillators for the simulation of general open quantum systems}}.
\newblock {\emph{\JournalTitle{Phys. Rev. A}}} \textbf{\bibinfo{volume}{101}}, \bibinfo{pages}{052108} (\bibinfo{year}{2020}).

\bibitem{jarlaud2020coherence}
\bibinfo{author}{Jarlaud, V.}, \bibinfo{author}{Hrmo, P.}, \bibinfo{author}{Joshi, M.~K.} \& \bibinfo{author}{Thompson, R.~C.}
\newblock \bibinfo{journal}{\bibinfo{title}{Coherence properties of highly-excited motional states of a trapped ion}}.
\newblock {\emph{\JournalTitle{Journal of Physics B: Atomic, Molecular and Optical Physics}}} \textbf{\bibinfo{volume}{54}}, \bibinfo{pages}{015501} (\bibinfo{year}{2020}).

\bibitem{pagano2018cryogenic}
\bibinfo{author}{Pagano, G.} \emph{et~al.}
\newblock \bibinfo{journal}{\bibinfo{title}{Cryogenic trapped-ion system for large scale quantum simulation}}.
\newblock {\emph{\JournalTitle{Quantum Sci. Technol.}}} \textbf{\bibinfo{volume}{4}}, \bibinfo{pages}{014004} (\bibinfo{year}{2018}).

\bibitem{Spivey21}
\bibinfo{author}{Spivey, R.~F.} \emph{et~al.}
\newblock \bibinfo{journal}{\bibinfo{title}{High-stability cryogenic system for quantum computing with compact packaged ion traps}}.
\newblock {\emph{\JournalTitle{IEEE Trans. Quantum Eng.}}} \textbf{\bibinfo{volume}{3}}, \bibinfo{pages}{1--11} (\bibinfo{year}{2021}).

\bibitem{spivey_dissert}
\bibinfo{author}{Spivey~III, R.~F.}
\newblock \emph{\bibinfo{title}{A Compact Cryogenic Package Approach to Ion Trap Quantum Computing}}.
\newblock Ph.D. thesis, \bibinfo{school}{Duke University} (\bibinfo{year}{2022}).

\bibitem{brown2021materials}
\bibinfo{author}{Brown, K.~R.}, \bibinfo{author}{Chiaverini, J.}, \bibinfo{author}{Sage, J.~M.} \& \bibinfo{author}{H{\"a}ffner, H.}
\newblock \bibinfo{journal}{\bibinfo{title}{Materials challenges for trapped-ion quantum computers}}.
\newblock {\emph{\JournalTitle{Nat. Rev. Mater.}}} \textbf{\bibinfo{volume}{6}}, \bibinfo{pages}{892--905} (\bibinfo{year}{2021}).

\bibitem{ZhangDissert}
\bibinfo{author}{Zhang, B.}
\newblock \emph{\bibinfo{title}{Improving Circuit Performance in a Trapped-Ion Quantum Computer}}.
\newblock Ph.D. thesis, \bibinfo{school}{Duke University} (\bibinfo{year}{2021}).

\bibitem{parrado2021crosstalk}
\bibinfo{author}{Parrado-Rodr{\'\i}guez, P.}, \bibinfo{author}{Ryan-Anderson, C.}, \bibinfo{author}{Bermudez, A.} \& \bibinfo{author}{M{\"u}ller, M.}
\newblock \bibinfo{journal}{\bibinfo{title}{Crosstalk suppression for fault-tolerant quantum error correction with trapped ions}}.
\newblock {\emph{\JournalTitle{Quantum}}} \textbf{\bibinfo{volume}{5}}, \bibinfo{pages}{487} (\bibinfo{year}{2021}).

\bibitem{Fang22}
\bibinfo{author}{Fang, C.}, \bibinfo{author}{Wang, Y.}, \bibinfo{author}{Huang, S.}, \bibinfo{author}{Brown, K.~R.} \& \bibinfo{author}{Kim, J.}
\newblock \bibinfo{journal}{\bibinfo{title}{Crosstalk suppression in individually addressed two-qubit gates in a trapped-ion quantum computer}}.
\newblock {\emph{\JournalTitle{Phys. Rev. Lett.}}} \textbf{\bibinfo{volume}{129}}, \bibinfo{pages}{240504} (\bibinfo{year}{2022}).

\bibitem{Moore23}
\bibinfo{author}{Moore, I.~D.} \emph{et~al.}
\newblock \bibinfo{journal}{\bibinfo{title}{Photon scattering errors during stimulated raman transitions in trapped-ion qubits}}.
\newblock {\emph{\JournalTitle{Phys. Rev. A}}} \textbf{\bibinfo{volume}{107}}, \bibinfo{pages}{032413} (\bibinfo{year}{2023}).

\bibitem{polli2010conical}
\bibinfo{author}{Polli, D.} \emph{et~al.}
\newblock \bibinfo{journal}{\bibinfo{title}{Conical intersection dynamics of the primary photoisomerization event in vision}}.
\newblock {\emph{\JournalTitle{Nature}}} \textbf{\bibinfo{volume}{467}}, \bibinfo{pages}{440--443} (\bibinfo{year}{2010}).

\bibitem{barbatti2010relaxation}
\bibinfo{author}{Barbatti, M.} \emph{et~al.}
\newblock \bibinfo{journal}{\bibinfo{title}{Relaxation mechanisms of {UV}-photoexcited {DNA} and {RNA} nucleobases}}.
\newblock {\emph{\JournalTitle{Proc. Natl. Acad. Sci.}}} \textbf{\bibinfo{volume}{107}}, \bibinfo{pages}{21453--21458} (\bibinfo{year}{2010}).

\bibitem{matsika2005three}
\bibinfo{author}{Matsika, S.}
\newblock \bibinfo{journal}{\bibinfo{title}{Three-state conical intersections in nucleic acid bases}}.
\newblock {\emph{\JournalTitle{J. Phys. Chem. A}}} \textbf{\bibinfo{volume}{109}}, \bibinfo{pages}{7538--7545} (\bibinfo{year}{2005}).

\bibitem{larson2020conical}
\bibinfo{author}{Larson, J.}, \bibinfo{author}{Sj{\"o}qvist, E.} \& \bibinfo{author}{{\"O}hberg, P.}
\newblock \emph{\bibinfo{title}{Conical Intersections in Physics}} (\bibinfo{publisher}{Springer}, \bibinfo{year}{2020}).

\bibitem{yarkony1996diabolical}
\bibinfo{author}{Yarkony, D.~R.}
\newblock \bibinfo{journal}{\bibinfo{title}{Diabolical conical intersections}}.
\newblock {\emph{\JournalTitle{Rev. Mod. Phys.}}} \textbf{\bibinfo{volume}{68}}, \bibinfo{pages}{985} (\bibinfo{year}{1996}).

\bibitem{baer2006beyond}
\bibinfo{author}{Baer, M.}
\newblock \emph{\bibinfo{title}{Beyond Born-Oppenheimer: electronic nonadiabatic coupling terms and conical intersections}} (\bibinfo{publisher}{John Wiley \& Sons}, \bibinfo{year}{2006}).

\bibitem{lin1968effect}
\bibinfo{author}{Lin, S.-H.} \& \bibinfo{author}{Bersohn, R.}
\newblock \bibinfo{journal}{\bibinfo{title}{Effect of partial deuteration and temperature on triplet-state lifetimes}}.
\newblock {\emph{\JournalTitle{J. Chem. Phys.}}} \textbf{\bibinfo{volume}{48}}, \bibinfo{pages}{2732--2736} (\bibinfo{year}{1968}).

\bibitem{ryabinkin2017geometric}
\bibinfo{author}{Ryabinkin, I.~G.}, \bibinfo{author}{Joubert-Doriol, L.} \& \bibinfo{author}{Izmaylov, A.~F.}
\newblock \bibinfo{journal}{\bibinfo{title}{Geometric phase effects in nonadiabatic dynamics near conical intersections}}.
\newblock {\emph{\JournalTitle{Acc. Chem. Res.}}} \textbf{\bibinfo{volume}{50}}, \bibinfo{pages}{1785--1793} (\bibinfo{year}{2017}).

\bibitem{joubert2013geometric}
\bibinfo{author}{Joubert-Doriol, L.}, \bibinfo{author}{Ryabinkin, I.~G.} \& \bibinfo{author}{Izmaylov, A.~F.}
\newblock \bibinfo{journal}{\bibinfo{title}{Geometric phase effects in low-energy dynamics near conical intersections: A study of the multidimensional linear vibronic coupling model}}.
\newblock {\emph{\JournalTitle{J. Chem. Phys.}}} \textbf{\bibinfo{volume}{139}}, \bibinfo{pages}{234103} (\bibinfo{year}{2013}).

\bibitem{li2017geometric}
\bibinfo{author}{Li, J.}, \bibinfo{author}{Joubert-Doriol, L.} \& \bibinfo{author}{Izmaylov, A.~F.}
\newblock \bibinfo{journal}{\bibinfo{title}{Geometric phase effects in excited state dynamics through a conical intersection in large molecules: {N}-dimensional linear vibronic coupling model study}}.
\newblock {\emph{\JournalTitle{J. Chem. Phys.}}} \textbf{\bibinfo{volume}{147}}, \bibinfo{pages}{064106} (\bibinfo{year}{2017}).

\bibitem{kendrick2015geometric}
\bibinfo{author}{Kendrick, B.}, \bibinfo{author}{Hazra, J.} \& \bibinfo{author}{Balakrishnan, N.}
\newblock \bibinfo{journal}{\bibinfo{title}{The geometric phase controls ultracold chemistry}}.
\newblock {\emph{\JournalTitle{Nat. Commun.}}} \textbf{\bibinfo{volume}{6}}, \bibinfo{pages}{1--7} (\bibinfo{year}{2015}).

\bibitem{yuan2018observation}
\bibinfo{author}{Yuan, D.} \emph{et~al.}
\newblock \bibinfo{journal}{\bibinfo{title}{Observation of the geometric phase effect in the {H}+{HD}→{H}$_2$+{D} reaction}}.
\newblock {\emph{\JournalTitle{Science}}} \textbf{\bibinfo{volume}{362}}, \bibinfo{pages}{1289--1293} (\bibinfo{year}{2018}).

\bibitem{gambetta2021exploring}
\bibinfo{author}{Gambetta, F.~M.}, \bibinfo{author}{Zhang, C.}, \bibinfo{author}{Hennrich, M.}, \bibinfo{author}{Lesanovsky, I.} \& \bibinfo{author}{Li, W.}
\newblock \bibinfo{journal}{\bibinfo{title}{Exploring the many-body dynamics near a conical intersection with trapped {Rydberg} ions}}.
\newblock {\emph{\JournalTitle{Phys. Rev. Lett.}}} \textbf{\bibinfo{volume}{126}}, \bibinfo{pages}{233404} (\bibinfo{year}{2021}).

\bibitem{whitlow2023quantum}
\bibinfo{author}{Whitlow, J.} \emph{et~al.}
\newblock \bibinfo{journal}{\bibinfo{title}{Quantum simulation of conical intersections using trapped ions}}.
\newblock {\emph{\JournalTitle{Nat. Chem.}}} \textbf{\bibinfo{volume}{15}}, \bibinfo{pages}{1509--1514} (\bibinfo{year}{2023}).

\bibitem{valahu2023direct}
\bibinfo{author}{Valahu, C.~H.} \emph{et~al.}
\newblock \bibinfo{journal}{\bibinfo{title}{Direct observation of geometric-phase interference in dynamics around a conical intersection}}.
\newblock {\emph{\JournalTitle{Nat. Chem.}}} \textbf{\bibinfo{volume}{15}}, \bibinfo{pages}{1503--1508} (\bibinfo{year}{2023}).

\bibitem{wang2023observation}
\bibinfo{author}{Wang, C.~S.} \emph{et~al.}
\newblock \bibinfo{journal}{\bibinfo{title}{Observation of wave-packet branching through an engineered conical intersection}}.
\newblock {\emph{\JournalTitle{Phys. Rev. X}}} \textbf{\bibinfo{volume}{13}}, \bibinfo{pages}{011008} (\bibinfo{year}{2023}).

\bibitem{Huelga13}
\bibinfo{author}{Huelga, S.~F.} \& \bibinfo{author}{Plenio, M.~B.}
\newblock \bibinfo{journal}{\bibinfo{title}{Vibrations, quanta and biology}}.
\newblock {\emph{\JournalTitle{Contemp. Phys.}}} \textbf{\bibinfo{volume}{54}}, \bibinfo{pages}{181--207} (\bibinfo{year}{2013}).

\bibitem{Adolphs06}
\bibinfo{author}{Adolphs, J.} \& \bibinfo{author}{Renger, T.}
\newblock \bibinfo{journal}{\bibinfo{title}{How proteins trigger excitation energy transfer in the fmo complex of green sulfur bacteria}}.
\newblock {\emph{\JournalTitle{Biophys. J.}}} \textbf{\bibinfo{volume}{91}}, \bibinfo{pages}{2778--97} (\bibinfo{year}{2006}).

\bibitem{Chin13}
\bibinfo{author}{Chin, A.~W.} \emph{et~al.}
\newblock \bibinfo{journal}{\bibinfo{title}{The role of non-equilibrium vibrational structures in electronic coherence and recoherence in pigment--protein complexes}}.
\newblock {\emph{\JournalTitle{Nat. Phys.}}} \textbf{\bibinfo{volume}{9}}, \bibinfo{pages}{113--118} (\bibinfo{year}{2013}).

\bibitem{Irish14}
\bibinfo{author}{Irish, E.}, \bibinfo{author}{G{\'o}mez-Bombarelli, R.} \& \bibinfo{author}{Lovett, B.}
\newblock \bibinfo{journal}{\bibinfo{title}{Vibration-assisted resonance in photosynthetic excitation-energy transfer}}.
\newblock {\emph{\JournalTitle{Phys. Rev. A}}} \textbf{\bibinfo{volume}{90}}, \bibinfo{pages}{012510} (\bibinfo{year}{2014}).

\bibitem{NalBach15}
\bibinfo{author}{Nalbach, P.}, \bibinfo{author}{Mujica-Martinez, C.} \& \bibinfo{author}{Thorwart, M.}
\newblock \bibinfo{journal}{\bibinfo{title}{Vibronically coherent speed-up of the excitation energy transfer in the {Fenna-Matthews-Olson} complex}}.
\newblock {\emph{\JournalTitle{Phys. Rev. E}}} \textbf{\bibinfo{volume}{91}}, \bibinfo{pages}{022706} (\bibinfo{year}{2015}).

\bibitem{Fujihashi15}
\bibinfo{author}{Fujihashi, Y.}, \bibinfo{author}{Fleming, G.~R.} \& \bibinfo{author}{Ishizaki, A.}
\newblock \bibinfo{journal}{\bibinfo{title}{Impact of environmentally induced fluctuations on quantum mechanically mixed electronic and vibrational pigment states in photosynthetic energy transfer and 2d electronic spectra}}.
\newblock {\emph{\JournalTitle{J. Chem. Phys.}}} \textbf{\bibinfo{volume}{142}}, \bibinfo{pages}{212403} (\bibinfo{year}{2015}).

\bibitem{Engel07}
\bibinfo{author}{Engel, G.} \emph{et~al.}
\newblock \bibinfo{journal}{\bibinfo{title}{Evidence for wavelike energy transfer through quantum coherence in photosynthetic systems}}.
\newblock {\emph{\JournalTitle{Nature}}} \textbf{\bibinfo{volume}{446}}, \bibinfo{pages}{782--786} (\bibinfo{year}{2007}).

\bibitem{Christensson12}
\bibinfo{author}{Christensson, N.}, \bibinfo{author}{Kauffmann, H.~F.}, \bibinfo{author}{Pullerits, T.} \& \bibinfo{author}{Mancal, T.}
\newblock \bibinfo{journal}{\bibinfo{title}{Origin of long-lived coherences in light-harvesting complexes}}.
\newblock {\emph{\JournalTitle{J. Phys. Chem. B}}} \textbf{\bibinfo{volume}{116}}, \bibinfo{pages}{7449--7454} (\bibinfo{year}{2012}).

\bibitem{Duan17}
\bibinfo{author}{Duan, H.-G.} \emph{et~al.}
\newblock \bibinfo{journal}{\bibinfo{title}{Nature does not rely on long-lived electronic quantum coherence for photosynthetic energy transfer}}.
\newblock {\emph{\JournalTitle{Proc. Natl. Acad. Sci.}}} \textbf{\bibinfo{volume}{114}}, \bibinfo{pages}{8493--8498} (\bibinfo{year}{2017}).

\bibitem{Womick11}
\bibinfo{author}{Womick, J.~M.} \& \bibinfo{author}{Moran, A.~M.}
\newblock \bibinfo{journal}{\bibinfo{title}{Vibronic enhancement of exciton sizes and energy transport in photosynthetic complexes}}.
\newblock {\emph{\JournalTitle{J. Phys. Chem. B}}} \textbf{\bibinfo{volume}{115}}, \bibinfo{pages}{1347--1356} (\bibinfo{year}{2011}).

\bibitem{fuller2014vibronic}
\bibinfo{author}{Fuller, F.~D.} \emph{et~al.}
\newblock \bibinfo{journal}{\bibinfo{title}{Vibronic coherence in oxygenic photosynthesis}}.
\newblock {\emph{\JournalTitle{Nat. Chem.}}} \textbf{\bibinfo{volume}{6}}, \bibinfo{pages}{706--711} (\bibinfo{year}{2014}).

\bibitem{plenio2013origin}
\bibinfo{author}{Plenio, M.~B.}, \bibinfo{author}{Almeida, J.} \& \bibinfo{author}{Huelga, S.~F.}
\newblock \bibinfo{journal}{\bibinfo{title}{Origin of long-lived oscillations in {2D}-spectra of a quantum vibronic model: electronic versus vibrational coherence}}.
\newblock {\emph{\JournalTitle{J. Chem. Phys.}}} \textbf{\bibinfo{volume}{139}}, \bibinfo{pages}{12B614\_1} (\bibinfo{year}{2013}).

\bibitem{Tiwari13}
\bibinfo{author}{Tiwari, V.}, \bibinfo{author}{Peters, W.~K.} \& \bibinfo{author}{Jonas, D.~M.}
\newblock \bibinfo{journal}{\bibinfo{title}{Electronic resonance with anticorrelated pigment vibrations drives photosynthetic energy transfer outside the adiabatic framework}}.
\newblock {\emph{\JournalTitle{Proc. Natl. Acad. Sci.}}} \textbf{\bibinfo{volume}{110}}, \bibinfo{pages}{1203--1208} (\bibinfo{year}{2013}).

\bibitem{killoran2015enhancing}
\bibinfo{author}{Killoran, N.}, \bibinfo{author}{Huelga, S.~F.} \& \bibinfo{author}{Plenio, M.~B.}
\newblock \bibinfo{journal}{\bibinfo{title}{Enhancing light-harvesting power with coherent vibrational interactions: A quantum heat engine picture}}.
\newblock {\emph{\JournalTitle{J. Chem. Phys.}}} \textbf{\bibinfo{volume}{143}}, \bibinfo{pages}{10B614\_1} (\bibinfo{year}{2015}).

\bibitem{Li21}
\bibinfo{author}{Li, Z.-Z.}, \bibinfo{author}{Ko, L.}, \bibinfo{author}{Yang, Z.}, \bibinfo{author}{Sarovar, M.} \& \bibinfo{author}{Whaley, K.~B.}
\newblock \bibinfo{journal}{\bibinfo{title}{Unraveling excitation energy transfer assisted by collective behaviors of vibrations}}.
\newblock {\emph{\JournalTitle{New J. Phys.}}} \textbf{\bibinfo{volume}{23}}, \bibinfo{pages}{073012} (\bibinfo{year}{2021}).

\bibitem{Fassioli10}
\bibinfo{author}{Fassioli, F.}, \bibinfo{author}{Nazir, A.} \& \bibinfo{author}{Olaya-Castro, A.}
\newblock \bibinfo{journal}{\bibinfo{title}{Quantum state tuning of energy transfer in a correlated environment}}.
\newblock {\emph{\JournalTitle{J. Phys. Chem. Lett.}}} \textbf{\bibinfo{volume}{1}}, \bibinfo{pages}{2139--2143} (\bibinfo{year}{2010}).

\bibitem{Sarovar11}
\bibinfo{author}{Sarovar, M.}, \bibinfo{author}{Cheng, Y.-C.} \& \bibinfo{author}{Whaley, K.~B.}
\newblock \bibinfo{journal}{\bibinfo{title}{Environmental correlation effects on excitation energy transfer in photosynthetic light harvesting}}.
\newblock {\emph{\JournalTitle{Phys. Rev. E}}} \textbf{\bibinfo{volume}{83}}, \bibinfo{pages}{011906} (\bibinfo{year}{2011}).

\bibitem{Uchiyama18}
\bibinfo{author}{Uchiyama, C.}, \bibinfo{author}{Munro, W.~J.} \& \bibinfo{author}{Nemoto, K.}
\newblock \bibinfo{journal}{\bibinfo{title}{Environmental engineering for quantum energy transport}}.
\newblock {\emph{\JournalTitle{npj Quantum Inf.}}} \textbf{\bibinfo{volume}{4}}, \bibinfo{pages}{1--7} (\bibinfo{year}{2018}).

\bibitem{macdonell2023predicting}
\bibinfo{author}{MacDonell, R.~J.} \emph{et~al.}
\newblock \bibinfo{journal}{\bibinfo{title}{Predicting molecular vibronic spectra using time-domain analog quantum simulation}}.
\newblock {\emph{\JournalTitle{Chem. Sci.}}}  (\bibinfo{year}{2023}).

\bibitem{shen2018quantum}
\bibinfo{author}{Shen, Y.} \emph{et~al.}
\newblock \bibinfo{journal}{\bibinfo{title}{Quantum optical emulation of molecular vibronic spectroscopy using a trapped-ion device}}.
\newblock {\emph{\JournalTitle{Chem. Sci.}}} \textbf{\bibinfo{volume}{9}}, \bibinfo{pages}{836--840} (\bibinfo{year}{2018}).

\bibitem{rebentrost2009environment}
\bibinfo{author}{Rebentrost, P.}, \bibinfo{author}{Mohseni, M.}, \bibinfo{author}{Kassal, I.}, \bibinfo{author}{Lloyd, S.} \& \bibinfo{author}{Aspuru-Guzik, A.}
\newblock \bibinfo{journal}{\bibinfo{title}{Environment-assisted quantum transport}}.
\newblock {\emph{\JournalTitle{New J. Phys.}}} \textbf{\bibinfo{volume}{11}}, \bibinfo{pages}{033003} (\bibinfo{year}{2009}).

\bibitem{Potovcnik18}
\bibinfo{author}{Poto{\v{c}}nik, A.} \emph{et~al.}
\newblock \bibinfo{journal}{\bibinfo{title}{Studying light-harvesting models with superconducting circuits}}.
\newblock {\emph{\JournalTitle{Nat. Commun.}}} \textbf{\bibinfo{volume}{9}}, \bibinfo{pages}{904} (\bibinfo{year}{2018}).

\bibitem{berkelbach2012reduced}
\bibinfo{author}{Berkelbach, T.~C.}, \bibinfo{author}{Markland, T.~E.} \& \bibinfo{author}{Reichman, D.~R.}
\newblock \bibinfo{journal}{\bibinfo{title}{Reduced density matrix hybrid approach: Application to electronic energy transfer}}.
\newblock {\emph{\JournalTitle{J. Chem. Phys.}}} \textbf{\bibinfo{volume}{136}} (\bibinfo{year}{2012}).

\bibitem{montoya2015extending}
\bibinfo{author}{Montoya-Castillo, A.}, \bibinfo{author}{Berkelbach, T.~C.} \& \bibinfo{author}{Reichman, D.~R.}
\newblock \bibinfo{journal}{\bibinfo{title}{Extending the applicability of redfield theories into highly non-markovian regimes}}.
\newblock {\emph{\JournalTitle{J. Chem. Phys.}}} \textbf{\bibinfo{volume}{143}} (\bibinfo{year}{2015}).

\bibitem{Li22}
\bibinfo{author}{Li, Z.-Z.}, \bibinfo{author}{Ko, L.}, \bibinfo{author}{Yang, Z.}, \bibinfo{author}{Sarovar, M.} \& \bibinfo{author}{Whaley, K.~B.}
\newblock \bibinfo{journal}{\bibinfo{title}{Interplay of vibration-and environment-assisted energy transfer}}.
\newblock {\emph{\JournalTitle{New J. Phys.}}} \textbf{\bibinfo{volume}{24}}, \bibinfo{pages}{033032} (\bibinfo{year}{2022}).

\bibitem{etmarcus}
\bibinfo{author}{Marcus, R.~A.}
\newblock \bibinfo{journal}{\bibinfo{title}{Chemical and electrochemical electron-transfer theory}}.
\newblock {\emph{\JournalTitle{Annu. Rev. Phys. Chem.}}} \textbf{\bibinfo{volume}{15}} (\bibinfo{year}{1964}).

\bibitem{etdb}
\bibinfo{author}{Beratan, D.~N.}
\newblock \bibinfo{journal}{\bibinfo{title}{Why are dna and protein electron transfer so different?}}
\newblock {\emph{\JournalTitle{Annu. Rev. Phys. Chem.}}} \textbf{\bibinfo{volume}{70}} (\bibinfo{year}{2019}).

\bibitem{prytkova2007coupling}
\bibinfo{author}{Prytkova, T.~R.}, \bibinfo{author}{Kurnikov, I.~V.} \& \bibinfo{author}{Beratan, D.~N.}
\newblock \bibinfo{journal}{\bibinfo{title}{Coupling coherence distinguishes structure sensitivity in protein electron transfer}}.
\newblock {\emph{\JournalTitle{Science}}} \textbf{\bibinfo{volume}{315}}, \bibinfo{pages}{622--625} (\bibinfo{year}{2007}).

\bibitem{skourtis2004inelastic}
\bibinfo{author}{Skourtis, S.~S.}, \bibinfo{author}{Waldeck, D.~H.} \& \bibinfo{author}{Beratan, D.~N.}
\newblock \bibinfo{journal}{\bibinfo{title}{Inelastic electron tunneling erases coupling-pathway interferences}}.
\newblock {\emph{\JournalTitle{J. Phys. Chem. B}}} \textbf{\bibinfo{volume}{108}}, \bibinfo{pages}{15511--15518} (\bibinfo{year}{2004}).

\bibitem{goldsmith2006electron}
\bibinfo{author}{Goldsmith, R.~H.}, \bibinfo{author}{Wasielewski, M.~R.} \& \bibinfo{author}{Ratner, M.~A.}
\newblock \bibinfo{journal}{\bibinfo{title}{Electron transfer in multiply bridged donor- acceptor molecules: dephasing and quantum coherence}}.
\newblock {\emph{\JournalTitle{J. Phys. Chem. B}}} \textbf{\bibinfo{volume}{110}}, \bibinfo{pages}{20258--20262} (\bibinfo{year}{2006}).

\bibitem{zarea2013decoherence}
\bibinfo{author}{Zarea, M.}, \bibinfo{author}{Powell, D.}, \bibinfo{author}{Renaud, N.}, \bibinfo{author}{Wasielewski, M.~R.} \& \bibinfo{author}{Ratner, M.~A.}
\newblock \bibinfo{journal}{\bibinfo{title}{Decoherence and quantum interference in a four-site model system: Mechanisms and turnovers}}.
\newblock {\emph{\JournalTitle{J. Phys. Chem. B}}} \textbf{\bibinfo{volume}{117}}, \bibinfo{pages}{1010--1020} (\bibinfo{year}{2013}).

\bibitem{spiroscpl}
\bibinfo{author}{Skourtis, S.~S.}, \bibinfo{author}{Beratan, D.~N.}, \bibinfo{author}{Naaman, R.}, \bibinfo{author}{Nitzan, A.} \& \bibinfo{author}{Waldeck, D.~H.}
\newblock \bibinfo{journal}{\bibinfo{title}{Chiral control of electron transmission through molecules}}.
\newblock {\emph{\JournalTitle{Phys. Rev. Lett.}}} \textbf{\bibinfo{volume}{101}} (\bibinfo{year}{2008}).

\bibitem{xiao2009turning}
\bibinfo{author}{Xiao, D.}, \bibinfo{author}{Skourtis, S.~S.}, \bibinfo{author}{Rubtsov, I.~V.} \& \bibinfo{author}{Beratan, D.~N.}
\newblock \bibinfo{journal}{\bibinfo{title}{Turning charge transfer on and off in a molecular interferometer with vibronic pathways}}.
\newblock {\emph{\JournalTitle{Nano Lett.}}} \textbf{\bibinfo{volume}{9}}, \bibinfo{pages}{1818--1823} (\bibinfo{year}{2009}).

\bibitem{lin2009modulating}
\bibinfo{author}{Lin, Z.} \emph{et~al.}
\newblock \bibinfo{journal}{\bibinfo{title}{Modulating unimolecular charge transfer by exciting bridge vibrations}}.
\newblock {\emph{\JournalTitle{J. Am. Chem. Soc.}}} \textbf{\bibinfo{volume}{131}}, \bibinfo{pages}{18060--18062} (\bibinfo{year}{2009}).

\bibitem{evenson1992effective}
\bibinfo{author}{Evenson, J.~W.} \& \bibinfo{author}{Karplus, M.}
\newblock \bibinfo{journal}{\bibinfo{title}{Effective coupling in bridged electron transfer molecules: computational formulation and examples}}.
\newblock {\emph{\JournalTitle{J. Chem. Phys.}}} \textbf{\bibinfo{volume}{96}}, \bibinfo{pages}{5272--5278} (\bibinfo{year}{1992}).

\bibitem{skourtis1993two}
\bibinfo{author}{Skourtis, S.~S.}, \bibinfo{author}{Beratan, D.~N.} \& \bibinfo{author}{Onuchic, J.~N.}
\newblock \bibinfo{journal}{\bibinfo{title}{The two-state reduction for electron and hole transfer in bridge-mediated electron-transfer reactions}}.
\newblock {\emph{\JournalTitle{Chem. Phys.}}} \textbf{\bibinfo{volume}{176}}, \bibinfo{pages}{501--520} (\bibinfo{year}{1993}).

\bibitem{troisi2003rate}
\bibinfo{author}{Troisi, A.}, \bibinfo{author}{Nitzan, A.} \& \bibinfo{author}{Ratner, M.~A.}
\newblock \bibinfo{journal}{\bibinfo{title}{A rate constant expression for charge transfer through fluctuating bridges}}.
\newblock {\emph{\JournalTitle{J. Chem. Phys.}}} \textbf{\bibinfo{volume}{119}}, \bibinfo{pages}{5782--5788} (\bibinfo{year}{2003}).

\bibitem{sun2016exact}
\bibinfo{author}{Sun, X.} \& \bibinfo{author}{Geva, E.}
\newblock \bibinfo{journal}{\bibinfo{title}{Exact vs. asymptotic spectral densities in the {Garg-Onuchic-Ambegaokar} charge transfer model and its effect on {Fermi}’s golden rule rate constants}}.
\newblock {\emph{\JournalTitle{J. Chem. Phys.}}} \textbf{\bibinfo{volume}{144}}, \bibinfo{pages}{044106} (\bibinfo{year}{2016}).

\bibitem{Lemmer18}
\bibinfo{author}{Lemmer, A.} \emph{et~al.}
\newblock \bibinfo{journal}{\bibinfo{title}{A trapped-ion simulator for spin-boson models with structured environments}}.
\newblock {\emph{\JournalTitle{New J. Phys.}}} \textbf{\bibinfo{volume}{20}}, \bibinfo{pages}{073002} (\bibinfo{year}{2018}).

\bibitem{Tamascelli18}
\bibinfo{author}{Tamascelli, D.}, \bibinfo{author}{Smirne, A.}, \bibinfo{author}{Huelga, S.~F.} \& \bibinfo{author}{Plenio, M.~B.}
\newblock \bibinfo{journal}{\bibinfo{title}{Nonperturbative treatment of non-{Markovian} dynamics of open quantum systems}}.
\newblock {\emph{\JournalTitle{Phys. Rev. Lett.}}} \textbf{\bibinfo{volume}{120}}, \bibinfo{pages}{030402} (\bibinfo{year}{2018}).

\bibitem{onuchic1986some}
\bibinfo{author}{Onuchic, J.~N.}, \bibinfo{author}{Beratan, D.~N.} \& \bibinfo{author}{Hopfield, J.}
\newblock \bibinfo{journal}{\bibinfo{title}{Some aspects of electron-transfer reaction dynamics.}}
\newblock {\emph{\JournalTitle{J. Phys. Chem.}}} \textbf{\bibinfo{volume}{90}}, \bibinfo{pages}{3707--3721} (\bibinfo{year}{1986}).

\bibitem{caldeira1981influence}
\bibinfo{author}{Caldeira, A.~O.} \& \bibinfo{author}{Leggett, A.~J.}
\newblock \bibinfo{journal}{\bibinfo{title}{Influence of dissipation on quantum tunneling in macroscopic systems}}.
\newblock {\emph{\JournalTitle{Phys. Rev. Lett.}}} \textbf{\bibinfo{volume}{46}}, \bibinfo{pages}{211} (\bibinfo{year}{1981}).

\bibitem{leggett1984quantum}
\bibinfo{author}{Leggett, A.}
\newblock \bibinfo{journal}{\bibinfo{title}{Quantum tunneling in the presence of an arbitrary linear dissipation mechanism}}.
\newblock {\emph{\JournalTitle{Phys. Rev. B}}} \textbf{\bibinfo{volume}{30}}, \bibinfo{pages}{1208} (\bibinfo{year}{1984}).

\bibitem{leggett1987dynamics}
\bibinfo{author}{Leggett, A.~J.} \emph{et~al.}
\newblock \bibinfo{journal}{\bibinfo{title}{Dynamics of the dissipative two-state system}}.
\newblock {\emph{\JournalTitle{Rev. Mod. Phys.}}} \textbf{\bibinfo{volume}{59}}, \bibinfo{pages}{1} (\bibinfo{year}{1987}).

\bibitem{Schlawin21}
\bibinfo{author}{Schlawin, F.}, \bibinfo{author}{Gessner, M.}, \bibinfo{author}{Buchleitner, A.}, \bibinfo{author}{Sch{\"a}tz, T.} \& \bibinfo{author}{Skourtis, S.~S.}
\newblock \bibinfo{journal}{\bibinfo{title}{Continuously parametrized quantum simulation of molecular electron-transfer reactions}}.
\newblock {\emph{\JournalTitle{PRX Quantum}}} \textbf{\bibinfo{volume}{2}}, \bibinfo{pages}{010314} (\bibinfo{year}{2021}).

\bibitem{tanimura2020numerically}
\bibinfo{author}{Tanimura, Y.}
\newblock \bibinfo{journal}{\bibinfo{title}{Numerically “exact” approach to open quantum dynamics: The hierarchical equations of motion ({HEOM})}}.
\newblock {\emph{\JournalTitle{J. Chem. Phys.}}} \textbf{\bibinfo{volume}{153}}, \bibinfo{pages}{020901} (\bibinfo{year}{2020}).

\bibitem{zhao2022hierarchy}
\bibinfo{author}{Zhao, Y.}, \bibinfo{author}{Sun, K.}, \bibinfo{author}{Chen, L.} \& \bibinfo{author}{Gelin, M.}
\newblock \bibinfo{journal}{\bibinfo{title}{The hierarchy of {Davydov's Ans{\"a}tze} and its applications}}.
\newblock {\emph{\JournalTitle{Wiley Interdiscip. Rev. Comput. Mol. Sci.}}} \textbf{\bibinfo{volume}{12}}, \bibinfo{pages}{e1589} (\bibinfo{year}{2022}).

\bibitem{ye2016heom}
\bibinfo{author}{Ye, L.} \emph{et~al.}
\newblock \bibinfo{journal}{\bibinfo{title}{{HEOM-QUICK}: a program for accurate, efficient, and universal characterization of strongly correlated quantum impurity systems}}.
\newblock {\emph{\JournalTitle{Wiley Interdiscip. Rev. Comput. Mol. Sci.}}} \textbf{\bibinfo{volume}{6}}, \bibinfo{pages}{608--638} (\bibinfo{year}{2016}).

\bibitem{gelin2022equation}
\bibinfo{author}{Gelin, M.~F.}, \bibinfo{author}{Chen, L.} \& \bibinfo{author}{Domcke, W.}
\newblock \bibinfo{journal}{\bibinfo{title}{Equation-of-motion methods for the calculation of femtosecond time-resolved 4-wave-mixing and $n$-wave-mixing signals}}.
\newblock {\emph{\JournalTitle{Chem. Rev.}}}  (\bibinfo{year}{2022}).

\bibitem{ren2022time}
\bibinfo{author}{Ren, J.}, \bibinfo{author}{Li, W.}, \bibinfo{author}{Jiang, T.}, \bibinfo{author}{Wang, Y.} \& \bibinfo{author}{Shuai, Z.}
\newblock \bibinfo{journal}{\bibinfo{title}{Time-dependent density matrix renormalization group method for quantum dynamics in complex systems}}.
\newblock {\emph{\JournalTitle{Wiley Interdiscip. Rev. Comput. Mol. Sci.}}} \bibinfo{pages}{e1614} (\bibinfo{year}{2022}).

\bibitem{weimer2021simulation}
\bibinfo{author}{Weimer, H.}, \bibinfo{author}{Kshetrimayum, A.} \& \bibinfo{author}{Or{\'u}s, R.}
\newblock \bibinfo{journal}{\bibinfo{title}{Simulation methods for open quantum many-body systems}}.
\newblock {\emph{\JournalTitle{Rev. Mod. Phys.}}} \textbf{\bibinfo{volume}{93}}, \bibinfo{pages}{015008} (\bibinfo{year}{2021}).

\bibitem{de2017dynamics}
\bibinfo{author}{De~Vega, I.} \& \bibinfo{author}{Alonso, D.}
\newblock \bibinfo{journal}{\bibinfo{title}{Dynamics of non-{Markovian} open quantum systems}}.
\newblock {\emph{\JournalTitle{Rev. Mod. Phys.}}} \textbf{\bibinfo{volume}{89}}, \bibinfo{pages}{015001} (\bibinfo{year}{2017}).

\bibitem{thoss2004semiclassical}
\bibinfo{author}{Thoss, M.} \& \bibinfo{author}{Wang, H.}
\newblock \bibinfo{journal}{\bibinfo{title}{Semiclassical description of molecular dynamics based on initial-value representation methods}}.
\newblock {\emph{\JournalTitle{Annu. Rev. Phys. Chem.}}} \textbf{\bibinfo{volume}{55}}, \bibinfo{pages}{299} (\bibinfo{year}{2004}).

\bibitem{lee2016semiclassical}
\bibinfo{author}{Lee, M.~K.}, \bibinfo{author}{Huo, P.} \& \bibinfo{author}{Coker, D.~F.}
\newblock \bibinfo{journal}{\bibinfo{title}{Semiclassical path integral dynamics: Photosynthetic energy transfer with realistic environment interactions}}.
\newblock {\emph{\JournalTitle{Annu. Rev. Phys. Chem.}}} \textbf{\bibinfo{volume}{67}}, \bibinfo{pages}{639--668} (\bibinfo{year}{2016}).

\bibitem{mac2002surface}
\bibinfo{author}{Mac~Kernan, D.}, \bibinfo{author}{Ciccotti, G.} \& \bibinfo{author}{Kapral, R.}
\newblock \bibinfo{journal}{\bibinfo{title}{Surface-hopping dynamics of a spin-boson system}}.
\newblock {\emph{\JournalTitle{J. Chem. Phys.}}} \textbf{\bibinfo{volume}{116}}, \bibinfo{pages}{2346--2353} (\bibinfo{year}{2002}).

\bibitem{makri1991feynman}
\bibinfo{author}{Makri, N.}
\newblock \bibinfo{journal}{\bibinfo{title}{Feynman path integration in quantum dynamics}}.
\newblock {\emph{\JournalTitle{Comput. Phys. Commun.}}} \textbf{\bibinfo{volume}{63}}, \bibinfo{pages}{389--414} (\bibinfo{year}{1991}).

\bibitem{feynman1963}
\bibinfo{author}{Feynman, R.} \& \bibinfo{author}{Vernon, F.}
\newblock \bibinfo{journal}{\bibinfo{title}{The theory of a general quantum system interacting with a linear dissipative system}}.
\newblock {\emph{\JournalTitle{Ann. Phys.}}} \textbf{\bibinfo{volume}{24}}, \bibinfo{pages}{118--173} (\bibinfo{year}{1963}).

\bibitem{shi2004derivation}
\bibinfo{author}{Shi, Q.} \& \bibinfo{author}{Geva, E.}
\newblock \bibinfo{journal}{\bibinfo{title}{A derivation of the mixed quantum-classical {Liouville} equation from the influence functional formalism}}.
\newblock {\emph{\JournalTitle{J. Chem. Phys.}}} \textbf{\bibinfo{volume}{121}}, \bibinfo{pages}{3393--3404} (\bibinfo{year}{2004}).

\bibitem{may2011charge}
\bibinfo{author}{May, V.} \& \bibinfo{author}{K{\"u}hn, O.}
\newblock \emph{\bibinfo{title}{Charge and Energy Transfer Dynamics in Molecular Systems}} (\bibinfo{publisher}{John Wiley \& Sons}, \bibinfo{year}{2011}), \bibinfo{edition}{3} edn.

\bibitem{feynman1948space}
\bibinfo{author}{Feynman, R.~P.}
\newblock \bibinfo{journal}{\bibinfo{title}{Space-time approach to non-relativistic quantum mechanics}}.
\newblock {\emph{\JournalTitle{Rev. Mod. Phys.}}} \textbf{\bibinfo{volume}{20}}, \bibinfo{pages}{367} (\bibinfo{year}{1948}).

\bibitem{zobel2021surface}
\bibinfo{author}{Zobel, J.~P.}, \bibinfo{author}{Heindl, M.}, \bibinfo{author}{Plasser, F.}, \bibinfo{author}{Mai, S.} \& \bibinfo{author}{Gonz{\'a}lez, L.}
\newblock \bibinfo{journal}{\bibinfo{title}{Surface hopping dynamics on vibronic coupling models}}.
\newblock {\emph{\JournalTitle{Acc. Chem. Res.}}} \textbf{\bibinfo{volume}{54}}, \bibinfo{pages}{3760--3771} (\bibinfo{year}{2021}).

\bibitem{Harrington22}
\bibinfo{author}{Harrington, P.~M.}, \bibinfo{author}{Mueller, E.~J.} \& \bibinfo{author}{Murch, K.~W.}
\newblock \bibinfo{journal}{\bibinfo{title}{Engineered dissipation for quantum information science}}.
\newblock {\emph{\JournalTitle{Nat. Rev. Phys.}}} \textbf{\bibinfo{volume}{4}}, \bibinfo{pages}{660--671} (\bibinfo{year}{2022}).

\bibitem{Monroe95}
\bibinfo{author}{Monroe, C.} \emph{et~al.}
\newblock \bibinfo{journal}{\bibinfo{title}{Resolved-sideband raman cooling of a bound atom to the {3D} zero-point energy}}.
\newblock {\emph{\JournalTitle{Phys. Rev. Lett.}}} \textbf{\bibinfo{volume}{75}}, \bibinfo{pages}{4011} (\bibinfo{year}{1995}).

\bibitem{Larson86}
\bibinfo{author}{Larson, D.}, \bibinfo{author}{Bergquist, J.~C.}, \bibinfo{author}{Bollinger, J.~J.}, \bibinfo{author}{Itano, W.~M.} \& \bibinfo{author}{Wineland, D.~J.}
\newblock \bibinfo{journal}{\bibinfo{title}{Sympathetic cooling of trapped ions: {A} laser-cooled two-species nonneutral ion plasma}}.
\newblock {\emph{\JournalTitle{Phys. Rev. Lett.}}} \textbf{\bibinfo{volume}{57}}, \bibinfo{pages}{70} (\bibinfo{year}{1986}).

\bibitem{Blinov02}
\bibinfo{author}{Blinov, B.} \emph{et~al.}
\newblock \bibinfo{journal}{\bibinfo{title}{Sympathetic cooling of trapped {Cd}$^+$ isotopes}}.
\newblock {\emph{\JournalTitle{Phys. Rev. A}}} \textbf{\bibinfo{volume}{65}}, \bibinfo{pages}{040304} (\bibinfo{year}{2002}).

\bibitem{Barrett03}
\bibinfo{author}{Barrett, M.~D.} \emph{et~al.}
\newblock \bibinfo{journal}{\bibinfo{title}{Sympathetic cooling of $^9${Be}$^+$ and $^{24}${Mg}$^+$ for quantum logic}}.
\newblock {\emph{\JournalTitle{Phys. Rev. A}}} \textbf{\bibinfo{volume}{68}}, \bibinfo{pages}{042302} (\bibinfo{year}{2003}).

\bibitem{negnevitsky2018repeated}
\bibinfo{author}{Negnevitsky, V.} \emph{et~al.}
\newblock \bibinfo{journal}{\bibinfo{title}{Repeated multi-qubit readout and feedback with a mixed-species trapped-ion register}}.
\newblock {\emph{\JournalTitle{Nature}}} \textbf{\bibinfo{volume}{563}}, \bibinfo{pages}{527--531} (\bibinfo{year}{2018}).

\bibitem{Clark10}
\bibinfo{author}{Clark, C.~R.}, \bibinfo{author}{Goeders, J.~E.}, \bibinfo{author}{Dodia, Y.~K.}, \bibinfo{author}{Viteri, C.~R.} \& \bibinfo{author}{Brown, K.~R.}
\newblock \bibinfo{journal}{\bibinfo{title}{Detection of single-ion spectra by coulomb-crystal heating}}.
\newblock {\emph{\JournalTitle{Phys. Rev. A}}} \textbf{\bibinfo{volume}{81}}, \bibinfo{pages}{043428} (\bibinfo{year}{2010}).

\bibitem{Cai13}
\bibinfo{author}{Cai, Z.} \& \bibinfo{author}{Barthel, T.}
\newblock \bibinfo{journal}{\bibinfo{title}{Algebraic versus exponential decoherence in dissipative many-particle systems}}.
\newblock {\emph{\JournalTitle{Phys. Rev. Lett.}}} \textbf{\bibinfo{volume}{111}}, \bibinfo{pages}{150403} (\bibinfo{year}{2013}).

\bibitem{Chenu17}
\bibinfo{author}{Chenu, A.}, \bibinfo{author}{Beau, M.}, \bibinfo{author}{Cao, J.} \& \bibinfo{author}{del Campo, A.}
\newblock \bibinfo{journal}{\bibinfo{title}{Quantum simulation of generic many-body open system dynamics using classical noise}}.
\newblock {\emph{\JournalTitle{Phys. Rev. Lett.}}} \textbf{\bibinfo{volume}{118}}, \bibinfo{pages}{140403} (\bibinfo{year}{2017}).

\bibitem{marshall2017linear}
\bibinfo{author}{Marshall, K.} \& \bibinfo{author}{James, D.~F.}
\newblock \bibinfo{journal}{\bibinfo{title}{Linear mode-mixing of phonons with trapped ions}}.
\newblock {\emph{\JournalTitle{Appl. Phys. B}}} \textbf{\bibinfo{volume}{123}}, \bibinfo{pages}{26} (\bibinfo{year}{2017}).

\bibitem{gan2020hybrid}
\bibinfo{author}{Gan, H.}, \bibinfo{author}{Maslennikov, G.}, \bibinfo{author}{Tseng, K.-W.}, \bibinfo{author}{Nguyen, C.} \& \bibinfo{author}{Matsukevich, D.}
\newblock \bibinfo{journal}{\bibinfo{title}{Hybrid quantum computing with conditional beam splitter gate in trapped ion system}}.
\newblock {\emph{\JournalTitle{Phys. Rev. Lett.}}} \textbf{\bibinfo{volume}{124}}, \bibinfo{pages}{170502} (\bibinfo{year}{2020}).

\bibitem{chen2021quantum}
\bibinfo{author}{Chen, W.}, \bibinfo{author}{Gan, J.}, \bibinfo{author}{Zhang, J.-N.}, \bibinfo{author}{Matuskevich, D.} \& \bibinfo{author}{Kim, K.}
\newblock \bibinfo{journal}{\bibinfo{title}{Quantum computation and simulation with vibrational modes of trapped ions}}.
\newblock {\emph{\JournalTitle{Chin. Phys. B}}}  (\bibinfo{year}{2021}).

\bibitem{nguyen2021experimental}
\bibinfo{author}{Nguyen, C.-H.}, \bibinfo{author}{Tseng, K.-W.}, \bibinfo{author}{Maslennikov, G.}, \bibinfo{author}{Gan, H.} \& \bibinfo{author}{Matsukevich, D.}
\newblock \bibinfo{journal}{\bibinfo{title}{Experimental swap test of infinite dimensional quantum states}}.
\newblock {\emph{\JournalTitle{arXiv preprint arXiv:2103.10219}}}  (\bibinfo{year}{2021}).

\bibitem{chen2023scalable}
\bibinfo{author}{Chen, W.} \emph{et~al.}
\newblock \bibinfo{journal}{\bibinfo{title}{Scalable and programmable phononic network with trapped ions}}.
\newblock {\emph{\JournalTitle{Nat. Phys.}}} \bibinfo{pages}{1--7} (\bibinfo{year}{2023}).

\bibitem{Katz23}
\bibinfo{author}{Katz, O.} \& \bibinfo{author}{Monroe, C.}
\newblock \bibinfo{journal}{\bibinfo{title}{Programmable quantum simulations of bosonic systems with trapped ions}}.
\newblock {\emph{\JournalTitle{Phys. Rev. Lett.}}} \textbf{\bibinfo{volume}{131}}, \bibinfo{pages}{033604} (\bibinfo{year}{2023}).

\end{thebibliography}

\section*{Acknowledgments}
This material is based upon work supported by the Department of Energy under Grant No. DE-SC0019400 (to HN, SNC, JLY, JV, ZZ, and DNB), Grant No. 2127309 to the Computing Research Association for the CIF ellows 2021 Project (to JV), and Grant No. DE-SC0019449 (to KS and JW). In addition, the authors acknowledge the support of the National Science Foundation STAQ Project Phy-181891 (to JW and KRB) and the NSF Quantum Leap Challenge Institute for Robust Quantum Simulation Grant No. OMA-2120757 (to MK, KS, JW, and KRB). JLY acknowledges partial support from the Lewis-Sigler Institute for Integrative Genomics.

% \section*{Author contributions}
% MK, HN, SNC, and JLY led the overall project. MK, HN, and SNC performed the simulations presented. MK, HN, KS, JW, and JV prepared the figures. DNB and KRB developed the plan for the article and provided feedback. All authors contributed to discussions and writing the manuscript. 

% \section*{Competing interests}
% KRB is a scientific advisor for IonQ, Inc. and has a personal financial interest in the company. The remaining authors declare no competing interests. 

% \section*{Short summary} Analog-quantum simulation derived from tracking the evolution of trapped-ion systems holds the potential to simulate molecular quantum dynamics that is beyond the reach of classical-digital strategies. This Review explores the prospects for developing this quantum advantage.  

\end{document}